\documentclass[reprint, amsmath, amssymb,aps,prb,floatfix, eqsecnum]{revtex4-2}

\pdfoutput=1

\usepackage{amsmath,graphicx,natbib,bm,array,physics,hyperref}
\usepackage{diagbox,longtable}
\usepackage[caption=false]{subfig}
\hypersetup{colorlinks = true,allcolors = blue }
\usepackage{bbm}
\usepackage{xcolor}
\usepackage[normalem]{ulem} 
\usepackage{mathtools}

\newcolumntype{L}{>{$}l<{$}}
\newcolumntype{C}{>{$}c<{$}}
\newcolumntype{R}{>{$}r<{$}}

\DeclareMathOperator{\Arg}{Arg}
\DeclareMathOperator{\sgn}{sgn}
\DeclareMathOperator{\sinc}{sinc}

\begin{document}

\newcommand{\pd}{\partial}
\newcommand{\beq}{\begin{equation}}
\newcommand{\eeq}{\end{equation}}
\newcommand{\bseq}{\begin{subequations}}
\newcommand{\eseq}{\end{subequations}}
\newcommand{\bpmat}{\begin{pmatrix}}
\newcommand{\epmat}{\end{pmatrix}}
\newcommand{\bpl}{\boldsymbol{(}}
\newcommand{\bpr}{\boldsymbol{)}}

\newcommand{\param}{\lambda}
\newcommand{\lloc}{L_{\text{loc}}}

\newcommand{\phiup}{\phi_{\uparrow}}
\newcommand{\phidown}{\phi_{\downarrow}}

\newcommand{\rr}[1]{\textcolor{red}{#1}}
\newcommand{\ac}[1]{\textcolor{blue}{#1}}
\newcommand{\ps}[1]{\textcolor{green}{#1}}


\title{Scattering expansion for localization in one dimension: From disordered wires to quantum walks}

\author{Adrian B. Culver}
\email{adrianculver@physics.ucla.edu}
\author{Pratik Sathe}
\author{Rahul Roy}
\email{rroy@physics.ucla.edu}
\affiliation{Mani L. Bhaumik Institute for Theoretical Physics, Department of Physics and Astronomy, University of California, Los Angeles, Los Angeles, CA 90095}

\begin{abstract}
We present a perturbative approach to disordered systems in one spatial dimension that accesses the full range of phase disorder and clarifies the connection between localization and phase information.  We consider a long chain of identically disordered scatterers and expand in the reflection strength of any individual scatterer.  We apply this expansion to several examples, including the Anderson model, a general class of periodic-on-average-random potentials, and a two-component discrete-time quantum walk, showing analytically in the latter case that the localization length can depend non-monotonically on the strength of phase disorder (whereas expanding in weak disorder yields monotonic decrease).  More generally, we obtain to all orders in the expansion a particular non-separable form for the joint probability distribution of the transmission coefficient logarithm and reflection phase.  Furthermore, we show that for weak local reflection strength, a version of the scaling theory of localization holds: the joint distribution is determined by just three parameters.
\end{abstract}

\maketitle


\setcounter{footnote}{0} 
\footnotetext[0]{For clarification on how a wave interference effect such as Anderson localization (i.e., an effect not uniquely quantum) can be relevant to quantum computing, see Refs.~\cite{LloydQuantum1999,BhattacharyaImplementation2002,KnightQuantum2003,Perez-GarciaQuantum2015}.}

\section{Introduction}\label{sec:Introduction}

The localization of waves by disorder (Anderson localization) is a topic of enduring interest due to the wide range of settings in which it occurs, including electron transport, classical optics, acoustics, and Bose-Einstein condensates~\cite{Abrahams502010}.  Progress in the general theory of localization, independent of model details or of physical realization, can have similarly broad implications.  Another setting for localization, of recent interest as a potential quantum computing platform~\cite{ChildsUniversal2009,LovettUniversal2010,SinghUniversal2021}, is the quantum walk~\cite{KempeQuantum2003,Venegas-AndracaQuantum2012,KadianQuantum2021}, which is a quantum version of the classical random walk.  Localization has been demonstrated in quantum walks both experimentally and theoretically~\cite{PeretsRealization2008,SchreiberDecoherence2011,CrespiAnderson2013,VatnikAnderson2017,TormaLocalization2002,KeatingLocalization2007,YinQuantum2008,JoyeDynamical2010,AhlbrechtDisordered2011,ObuseTopological2011,VakulchykAnderson2017,DerevyankoAnderson2018} and could impact quantum computing proposals even in the idealized limit of no decoherence~\cite{KeatingLocalization2007,YinQuantum2008,ChandrashekarLocalized2015,Note0}.

A distinctive feature of localization in quantum walks is the prominent role of phase disorder.  Modern experimental platforms allow a high degree of control over a spatially varying phase which can be disordered~\cite{PeretsRealization2008,SchreiberDecoherence2011,CrespiAnderson2013,VatnikAnderson2017}.  Localization in what is perhaps the simplest quantum walk, a discrete-time quantum walk (DTQW) in one spatial dimension, has been experimentally realized both for strong phase disorder~\cite{SchreiberDecoherence2011} and for a controllable range of phase disorder from weak to strong~\cite{VatnikAnderson2017}.   However, existing analytical approaches seem to apply only in the limiting cases when phase disorder is either weak or strong~\cite{VakulchykAnderson2017,DerevyankoAnderson2018}.  Furthermore, there are several phases that can appear in the quantum ``coin'' (see below) of a DTQW~\cite{ChandrashekarOptimizing2008,*ChandrashekarErratum2010,VakulchykAnderson2017}, and a localization calculation that allows them all to be disordered simultaneously seems to be lacking in the literature.

A pillar of our general understanding of localization is the scaling theory initiated by Abrahams \textit{et al}. in Ref.~\cite{AbrahamsScaling1979} in the context of electron transport.  While localization may be characterized in several ways, including the absence of diffusion and the exponential decay of eigenstates, the scaling theory focuses on the suppression of the dimensionless conductance $g$ through a disordered sample (e.g., a cube of side length $L$).  The conductance has some probability distribution $P_L(g)$ over disorder realizations, and the scaling theory asserts that for sufficiently large $L$, all dependence of the function $P_L$ on $L$ and on microscopic details may be absorbed into a small (i.e., at least not growing with $L$) number of parameters $X_1(L),X_2(L),\dots$ that obey scaling equations of the form $d \bpl\ln X_i(L)\bpr/d(\ln L) = \beta_i\bpl X_1(L),X_2(L),\dots\bpr$ (see, e.g, Refs.~\cite{ShapiroScaling1987,CohenUniversal1988}).  The study of these scaling equations, and of the number of parameters needed in various cases, gives insight into disorder-driven phase transitions and has been a significant area of research in disordered systems.

In one spatial dimension, in which case localization is generally strongest, one can use a scattering setup~\cite{AndersonNew1980} based on the Landauer formula for the conductance: $g = \frac{e^2}{2\pi\hbar}T$, where $T$ is the transmission coefficient (we consider the case of a single scattering channel) of the sample \footnote{To be precise, we are considering here the two-terminal conductance rather than the four-terminal conductance $\frac{e^2}{2\pi\hbar}\frac{T}{R}$~\cite{DattaElectronic1995}, where $R=1-T$ is the reflection coefficient of the sample; however, the distinction does not matter in the localized regime (in which $R\to 1$) that is our focus throughout.  Also, we are ignoring spin degeneracy.  In the rest of the paper, we set $\hbar=1$.}.  The typical transmission coefficient $T$ of a long sample of length $L$ decays exponentially: $T_\text{typical}\sim e^{-2 L/\lloc}$, which defines the localization length $\lloc$.  In the transfer matrix approach, theorems for random matrices demonstrate that the probability distribution $P_L(-\ln T)$ over disorder realizations [which by the Landauer formula is simply related to $P_L(g)$] is Gaussian for large $L$~\cite{BougerolProducts1985}.  All dependence of $P_L(-\ln T)$ on $L$ and on disorder thus reduces to two parameters (the mean and variance).  A further reduction called single-parameter scaling (SPS), in which the two parameters reduce to one by an equation relating them, was originally obtained using an assumption of phase uniformity~\cite{AndersonNew1980}, but has since been shown to hold in certain limits even without this assumption~\cite{DeychSingle2000, DeychSingleparameter2001,SchraderPerturbative2004}.

In this paper, we present a perturbative approach to localization in one spatial dimension.  Our scattering-based approach accesses the full range of phase disorder, clarifies the connection between localization and phase information more broadly, and applies to a general class of systems that includes DTQWs~\cite{TarasinskiScattering2014}.  A central feature of our approach is the relation between the localization properties and the reflection phase of a disordered sample~\cite{AndersonNew1980,LambertPhase1982,LambertRandom1983}.  (This phase has been measured in a DTQW experiment~\cite{BarkhofenMeasuring2017}.)  We calculate the localization length and the probability distribution of the reflection phase, and we extend the scaling theory to include correlations between the reflection phase and the transmission coefficient.  For a Letter version of this paper, see Ref.~\cite{CulverScattering2024a}.

We now summarize our approach and results in more detail.  We consider a disordered sample consisting of many single-channel scatterers that are independently and identically disordered, and we expand in the magnitude of the reflection amplitude of any individual scatterer \footnote{As we comment on in the main body of this paper and in Appendix~\ref{sec:Schrader et al.'s formula}, some of the results of our expansion at leading order were obtained in an equivalent form by Schrader \textit{et al}. in Ref.~\cite{SchraderPerturbative2004} (see also Ref.~\cite{DrabkinGaussian2015}).  Also, in the context of light propagation in disordered metamaterials, a similar expansion called the weak scattering approximation (WSA) was done in Refs.~\cite{AsatryanSuppression2007,AsatryanAnderson2010} (see~\cite{GredeskulAnderson2012} for a review); however, the WSA has an uncontrolled aspect that is not present in our work (see Appendix~\ref{sec:Comparison with the weak scattering approximation}).}.  Our first main result is the expansion of the inverse localization length.  We construct this expansion recursively and show that all orders depend only on local averages (that is, disorder averages over any single site).  We obtain a similar expansion of the probability distribution of the reflection phase (finding that it is generally uniform only at the zeroth order), and indeed use this expansion in calculating the localization length.

We proceed to apply our first result to disordered wires and DTQWs.  In the former case, we consider both the Anderson model and a broad class of periodic-on-average random systems (PARS), verifying that our results agree with and extend results from the literature.  (For general background on PARS, see, e.g., Ref.~\cite{DeychStatistics1998} and references therein.)  We then calculate the localization length analytically as a function of arbitrary phase disorder in a two-component DTQW in one dimension.  We expect our calculation to apply to scattering setups~\cite{TarasinskiScattering2014} and beyond, and indeed we verify that our result interpolates between known results for weak and strong disorder that were calculated without reference to scattering~\cite{VakulchykAnderson2017,DerevyankoAnderson2018}.  We find that the localization length can depend non-monotonically on the strength of phase disorder (similar to behavior seen numerically in Refs.~\cite{VatnikAnderson2017,VakulchykAnderson2017}) \footnote{Non-monotonicity in disorder strength has been found previously, both analytically and numerically, in particular PARS problems~\cite{LambertRandom1983,MafiAnderson2015} (see Sec.~\ref{sec:Periodic-on-average random potential}).  Also, it has been shown that transmission through a disordered region can be non-monotonic in disorder strength when the incident wave is in an energy gap of the non-disordered system (see Ref.~\cite{HeinrichsEnhanced2008} and references therein); we do not consider this case.}.  Our expansion applies when the quantum ``coin'' is highly biased (see below), which is a regime of interest for optimizing quantum search~\cite{ChandrashekarOptimizing2008,*ChandrashekarErratum2010}.  Even if the coin is only moderately biased, we find favorable agreement with numerics using the first two non-vanishing orders of our expansion.

Our second main result concerns the joint probability distribution $P_L(-\ln T,\phi)$, where $\phi$ is the reflection phase of the disordered region.  We use an ansatz to find that for large $L$ and to all orders in the scattering expansion, $P_L(-\ln T,\phi)$ tends to a Gaussian function (of $-\ln T$) with mean, variance, and overall scale all depending on $\phi$ and all calculable order by order in terms of local averages.  We further show that at the leading order in the local reflection strength, a version of the scaling theory applies: the joint distribution is determined by three parameters, which we may take to be the mean of $-\ln T$ and the mean and variance of $\phi$.  The latter two reach constant values for large system size ($\beta_2=\beta_3=0$ in the notation used above).

Another contribution of this paper is to extend the perturbative derivation of SPS by Schrader \textit{et al}.~\cite{SchraderPerturbative2004}, while also explicitly connecting their result to a scattering setup.  As we review in the main text, Ref.~\cite{SchraderPerturbative2004} demonstrates SPS in a general setting at the leading order in a small parameter.  We point out that in a scattering setup, the local reflection strength is a valid choice for this parameter, which confirms (without assuming phase uniformity) the expectation in the literature that SPS should hold for weak local scattering~\cite{CohenUniversal1988}.  We also extend the SPS relation to another order in the local reflection strength.  (SPS cannot hold generally at the next order beyond, as Schrader \textit{et al}. have shown that the Anderson model is a counter-example~\cite{SchraderPerturbative2004}.)  Note then that both SPS (for the distribution of $-\ln T$) and the three-parameter scaling that we find for the joint distribution are obtained in the same regime (weak local scattering).

We now make several comments about the scope of this paper.  First, we focus throughout on the strongly localized regime ($L\gg \lloc$).  Second, we confine our attention to static (quenched) disorder that is independently and identically distributed (i.i.d.).  Finally, we assume that there is no decoherence.  In spite of the above restrictions, our work applies to a variety of problems that can be put into a scattering framework, including classical as well as quantum problems, since Anderson localization is ultimately a phenomenon of wave interference.  Also, our approach might extend to the quasi-one-dimensional case (which we discuss more in the Conclusion). 

The paper is organized as follows.  In Sec.~\ref{sec:Setup and review of prior work}, we present the basic setup and review the most relevant prior work.  In Sec.~\ref{sec:Scattering expansion of the localization length}, we present our results for the localization length and reflection phase distribution, and we apply them to disordered wires and discrete-time quantum walks.  (We also include our results on SPS in this section, though they partially depend on results from the section following it.)  In Sec.~\ref{sec:Joint probability distribution}, we present our results for the joint probability distribution.  We summarize and discuss possible future work in Sec.~\ref{sec:Conclusion}.

\section{Setup and review of prior work}\label{sec:Setup and review of prior work}
\subsection{Setup}\label{sec:Setup and review of prior work, Setup}
\begin{figure}[t]
    \includegraphics[scale=0.3]{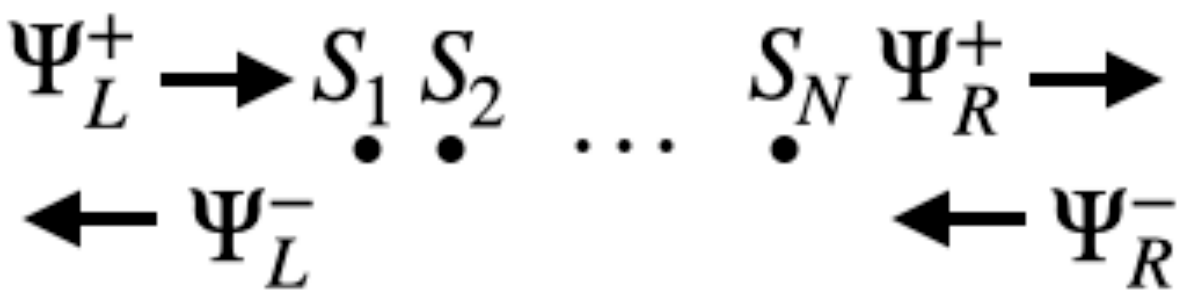}
    \caption{Schematic of our setup.}\label{fig:setup_diagram}
\end{figure}

We consider a one-dimensional ``sample'' consisting of $N$ sites labeled by $n=1,\dots,N$ (Fig.~\ref{fig:setup_diagram}).
The sample is attached to infinite ``leads'' and is a scattering region for incoming and outgoing waves in the leads.  Each site in the sample is associated with an array $\mathbf{D}_n$ of disorder variables.  For instance, in the Anderson model with diagonal disorder, $\mathbf{D}_n$ would be the onsite potential, i.e., an array of length $1$.  The disorder is assumed to be i.i.d. with some normalized probability distribution $\rho(\mathbf{D}_n)$, and we write disorder averages of any quantity $X$ as $\langle X \rangle_n \equiv \int d \mathbf{D}_n\  \rho(\mathbf{D}_n) X$ (any site $n$) and $\langle X \rangle_{1\dots N}\equiv\int \left[ \prod_{n=1}^N d \mathbf{D}_n\  \rho(\mathbf{D}_n)\right] X$ (the average over all $N$ sites).

We assume that each site has a $2\times2$ disordered $S$ matrix $\mathcal{S}_n$ which we parametrize by
\bseq
\begin{align}
    \mathcal{S}_n  &=
    \bpmat
        t_n & r_n' \\
        r_n & t_n'
    \epmat\label{eq:Sn general parametrization amplitudes}
    \\
    &= \bpmat
        \sqrt{T_n} \exp[i \phi_{t_n}] & \sqrt{R_n}\exp[i \phi_{r_n'}] \\
        \sqrt{R_n} \exp[i \phi_{r_n}] & \sqrt{T_n} \exp[i \phi_{t_n'}]
    \epmat,\label{eq:Sn general parametrization coeffs and phases}
\end{align}
\eseq
where $t_n$ and $t_n'$ ($r_n$ and $r_n'$) are the local transmission (reflection) amplitudes, $T_n = |t_n|^2 = |t_n'|^2$ ($R_n = |r_n|^2 = |r_n'|^2$) are the local transmission (reflection) coefficients, and $\phi_{t_n}$ and $\phi_{t_n'}$ ($\phi_{r_n}$ and $\phi_{r_n'}$) are the transmission (reflection) phases, which unitarity constrains by $R_n+T_n=1$ and $\phi_{t_n} +\phi_{t_n'} -  \phi_{r_n} -\phi_{r_n'} = \pi$ (modulo $2\pi$).  Here and throughout, the subscript $n$ indicates dependence on the disorder variables $\mathbf{D}_n$ of site $n$.  We consider only the single channel case, i.e., the amplitudes are complex numbers and not matrices.

We consider the scattering problem of the $N$-site sample.  The $S$ matrix for the sample, denoted $\mathcal{S}_{1\dots N}$, relates incoming and outgoing scattering amplitudes in the leads by
\beq
    \bpmat
        \Psi_R^+\\
        \Psi_L^-
    \epmat
    =
    \mathcal{S}_{1\dots N}
    \bpmat
        \Psi_L^+\\
        \Psi_R^-
    \epmat.
\eeq
The $S$ matrix is obtained in the usual way by multiplying the scattering transfer matrices of the $N$ sites and converting the resulting scattering transfer matrix back to an $S$ matrix (see Appendix~\ref{sec:S matrices and scattering transfer matrices}).  We write
\beq
    \mathcal{S}_{1\dots N} = \bpmat
        \sqrt{T_{1\dots N}} \exp[i \phi_{t_{1\dots N} }] & \sqrt{R_{1\dots N} }\exp[i \phi_{r_{1\dots N}'}] \\
        \sqrt{R_{1\dots N}} \exp[i \phi_{r_{1\dots N}}] & \sqrt{T_{1\dots N}} \exp[i \phi_{t_{1\dots N}'}]
    \epmat,\label{eq:Ssample parametrization}
\eeq
in which the parameters are constrained by unitarity to satisfy the same equations mentioned below Eq.~\eqref{eq:Sn general parametrization coeffs and phases}.  No further discussion of the leads is necessary at this point; the problem is to study the sample $S$ matrix~\eqref{eq:Ssample parametrization} given the local $S$ matrix~\eqref{eq:Sn general parametrization amplitudes}, which in any particular problem depends on the properties of the sample and leads.

The central object of our study is the joint probability distribution of the minus logarithm of the transmission coefficient (denoted $s_{1\dots N}\equiv -\ln T_{1\dots N}$) and of either one of the reflection phases (we choose the right-to-left phase, $\phi_{r_{1\dots N}'}$).  We focus at first on the localization length and the marginal distributions of $s$ and $\phi_{r'}$ (which indeed are determined by the joint distribution).  We write the joint and marginal distributions as
\bseq
\begin{align}
    P_{N}(s',\phi_{r'}') &\equiv \langle \delta( s_{1\dots N}-s')\delta(  \phi_{r_{1\dots N}'}-\phi_{r'}')\rangle_{1\dots N},\label{eq:joint prob dist def}\\
    P_{N}(s) &\equiv \int_{-\pi}^\pi d\phi_{r'}\ P_{N}(s,\phi_{r'}),\\
    p_{N}(\phi_{r'}) &\equiv \int_0^\infty ds\ P_{N}(s,\phi_{r'}),
\end{align}
\eseq
where the subscript $N$ on the left-hand side, in departure from our usual convention, indicates dependence on the number $N$ rather than on $\mathbf{D}_N$ \footnote{Explicitly, we have (e.g.) $p_N(\phi_{r'}') = \int d\mathbf{D}_1\dots d\mathbf{D}_N\ \rho(\mathbf{D}_1)\dots\rho(\mathbf{D}_N) \delta[ \phi_{r'}(\mathbf{D}_1,\dots,\mathbf{D}_N)- \phi_{r'}']$, where $\phi_{r'}(\mathbf{D}_1,\dots,\mathbf{D}_N)\equiv \phi_{r_{1\dots N}'}$.  Due to the disorder being i.i.d., the probability distributions depend only on the number of consecutive sites $N$ and not on their location.  Also, to make contact with the notation used in the Introduction, we take the spacing between sites to be the unit of distance; then the length of the sample is $L=N$.}.  As discussed in the Introduction, for a large sample ($N \gg 1$) the probability distribution of $s$ is Gaussian with the mean growing linearly with $N$ (although the Gaussian form may not accurately calculate moments of $T$ and $T^{-1}$~\cite{VagnerCoherent2003}).  The localization length $\lloc$ is then determined by
\beq
    \langle s_{1\dots N} \rangle_{1\dots N} = \frac{2}{\lloc} N + O(N^0).
\eeq
The variance [denoted $\sigma(N)^2$] of $P_{N}(s)$ also grows linearly with $N$, and SPS holds when the mean and variance are related by an equation (see below).

\subsection{Review of prior work}\label{sec:Review of prior work}
In this discussion, we present results from the literature as they apply to our setup, using our notation throughout.  We focus on prior work most relevant to our work and do not attempt a comprehensive review.  We first review explicit, model-independent results for the localization length and for SPS from the literature, and then we discuss the relation of our work to the literature on the joint distribution.

In Ref.~\cite{AndersonNew1980}, Anderson \textit{et al}. proposed that there would be a length scale $\ell_p$ beyond which complete phase randomization would occur.  According to this uniform phase hypothesis, for $N\gg \ell_p$ the phase of $r_{1\dots N}'r_{N+1}$, which we write as $\nu_{1\dots N} \equiv \phi_{r_{1\dots N}'} + \phi_{r_{N+1}}$, is distributed uniformly in $(-\pi,\pi)$ and independently of $T_{N+1}$.  [We will use ``uniform'' to refer to a phase distribution being uniform over $(-\pi,\pi)$, unless otherwise noted.]  This hypothesis yields an explicit formula for the localization length in terms of a local average~\cite{AndersonNew1980}:
\beq
    \frac{2}{\lloc} = \langle -\ln T_n \rangle_n \qquad (\text{any site } n).\label{eq:lloc AndersonNew1980}
\eeq
Furthermore, the same hypothesis yields SPS by relating the mean and variance through~\cite{AndersonNew1980}
\beq
    \lim_{N\to\infty}\frac{\sigma(N)^2}{2N}= \frac{2}{\lloc}.\label{eq:SPS relation in terms of lloc}
\eeq
However, the uniform phase hypothesis was shown to fail in a number of cases including strong disorder (see~\cite{PendrySymmetry1994} and references therein, as well as~\cite{JayannavarPhase1990,JayannavarScaling1991,HeinrichsRelation2002,PradhanPhase2021}), multiple scattering channels~\cite{DorokhovLocalization1982}, and anomalies in the Anderson model~\cite{SchomerusBandcenter2003,TessieriAnomalies2018}.  In Ref.~\cite{LambertPhase1982}, though, Lambert and Thorpe obtained a more general formula for the localization length that accounts for deviations from phase uniformity:
\begin{multline}
    \frac{2}{\lloc} = \langle - \ln T_n \rangle_n  \\
    + \langle \ln \left[1 +R_{N+1} + 2 \sqrt{R_{N+1}} \cos(\nu_{1\dots N}) \right]\rangle_{1\dots N+1},\label{eq:lloc LambertPhase1982}
\end{multline}
in which $N\gg1$ and $n$ is any site.  Equation~\eqref{eq:lloc LambertPhase1982} recovers the uniform phase result~\eqref{eq:lloc AndersonNew1980} when $\nu_{1\dots N}$ is uniformly distributed independently of $R_{N+1}$ [since $\int_{-\pi}^\pi d\nu\ \ln (1 +R + 2 \sqrt{R} \cos\nu)=0$].  In Refs.~\cite{LambertPhase1982, LambertRandom1983}, Lambert and Thorpe furthermore showed that, in a model consisting of randomly spaced delta function scatterers of equal strength, the probability distribution of $\nu_{1\dots N}$ is generically non-uniform even for an arbitrarily long chain.  The second term in~\eqref{eq:lloc LambertPhase1982} makes a substantial correction to the first, particularly for weak disorder.  In a more general setting, an expression for the probability distribution of $\nu_{1\dots N}$ at the zeroth order in disorder strength (with the thermodynamic limit taken before the limit of zero disorder) was derived by Lambert \textit{et al}. in Ref.~\cite{LambertPhase1983}; the authors then used this probability distribution in Eq.~\eqref{eq:lloc LambertPhase1982} to find that the localization length is finite in band gaps and infinite outside of band gaps, as expected~\cite{LambertPhase1983}.

The significance of the delta-function model considered by Refs.~\cite{LambertPhase1982,LambertRandom1983} was challenged by Stone \textit{et al}. in Ref.~\cite{StonePhase1983}, in which the authors showed numerically that in the Anderson model with weak, diagonal disorder with vanishing mean onsite energy, the probability distribution of $\nu_{1\dots N}$ is indeed uniform for large $N$.  Stone \textit{et al}. obtained a formula analogous to~\eqref{eq:lloc LambertPhase1982} for the variance $\sigma(N)^2$; using this formula, Eq.~\eqref{eq:lloc LambertPhase1982}, and the uniformity of the distribution of $\nu_{1\dots N}$, they obtained both the known weak disorder expansion for the inverse localization length and the SPS relation~\eqref{eq:SPS relation in terms of lloc}.  The authors further argued that the delta-function model considered by Lambert and Thorpe is a special case and that any model with the potential being positive as often as negative would yield results similar to those in the Anderson model.

A number of works (including some that we have cited above) have studied the distribution of the reflection phase $\phi_{r_{1\dots N}'}$ rather than of $\nu_{1\dots N}$.  We can understand this as follows.  By construction, $\nu_{1\dots N}$ is distributed uniformly provided that either or both of $\phi_{r_{1\dots N}'}$ and $\phi_{r_{N+1}}$ are distributed uniformly.  The challenging case, then, is what happens when the local phase $\phi_{r_{N+1}}$ is non-uniformly distributed: Does the distribution of the sample phase $\phi_{r_{1\dots N}'}$ become uniform for large $N$, or not?  From here on, we mean by the ``uniform phase hypothesis'' the assertion that either or both of the sample reflection phases, $\phi_{r_{1\dots N}'}$ and $\phi_{r_{1\dots N}}$, are distributed uniformly for large $N$.

Another approach to the SPS relation was provided by Deych \textit{et al}. in Refs.~\cite{DeychSingle2000, DeychSingleparameter2001}.  Deych \textit{et al}. argue that the SPS relation holds in the regime $\lloc \gg \ell_s$, where $\ell_s$ is a new length scale that they identify, without appeal to any uniform phase hypothesis.  In a setup encompassing both the Anderson model with diagonal disorder and the case of equally spaced delta functions with random strengths~\cite{StonePhase1983, DeychSingle2000, DeychSingleparameter2001}, Deych \textit{et al}. provide analytical results in the Lloyd model (i.e., disorder following a Cauchy distribution).  They obtain the SPS relation, albeit with an extra prefactor of $2$ that they attribute to the peculiarity of the Cauchy distribution; they also present numerical evidence for the SPS relation following from the condition $\lloc \gg \ell_s$ in a more generic model (an array of square well potentials with random widths)~\cite{DeychSingle2000,DeychSingleparameter2001}.

The work with most overlap with ours is that of Schrader \textit{et al}. in Ref.~\cite{SchraderPerturbative2004} (see also Ref.~\cite{DrabkinGaussian2015}), which presents an explicit formula for the inverse localization length and variance and a proof of the SPS relation, all at the leading order in a small parameter.   In particular, Schrader \textit{et al}. proved $2/\lloc =D \lambda'^2 +O(\lambda'^3)$ and $\lim_{N\to\infty}\sigma(N)^2/(2N)= 2 /\lloc + O(\lambda'^3)$, where $D$ is given in terms of averages over local quantities [unlike Eq.~\eqref{eq:lloc LambertPhase1982}, which includes an average over the whole sample] and where $\lambda'$ is a small parameter that Schrader \textit{et al}. identified as ``the effective size of the randomness''~\cite{SchraderPerturbative2004}.  The quantity $D$ consists of a sum of two terms, one that recovers the uniform phase result~\eqref{eq:lloc AndersonNew1980} at leading order in $\lambda'$ and a second that contributes due to departures from the uniform phase hypothesis~\cite{SchraderPerturbative2004}.  Schrader \textit{et al}. point out that although the second term vanishes in the Anderson model (with diagonal disorder and vanishing mean onsite energy), it is in general \emph{not} a small correction to the first term (for instance, in the random polymer model~\cite{JitomirskayaDelocalization2003}).   Schrader \textit{et al}. also calculate $2/\lloc$ and $\lim_{N\to\infty}\sigma(N)^2/(2N)$ to the next non-vanishing order in the Anderson model ($\lambda'^4$), showing that both the uniform phase formula~\eqref{eq:lloc AndersonNew1980} and the SPS relation~\eqref{eq:SPS relation in terms of lloc} break down at this order~\cite{SchraderPerturbative2004}.

Let us next review prior work on the joint distribution of $s$ and $\phi_{r'}$.  Given a maximum entropy ansatz, Ref.~\cite{MelloMacroscopic1987} found that the probability distribution of $s$ satisfies a Fokker-Planck equation (which in turn yields a Gaussian distribution for $s$) \footnote{Reference~\cite{MelloMacroscopic1987} obtains a Fokker-Planck equation for the probability distribution of $\rho \equiv  R/T$.  In the localized regime that is our focus, Ref.~\cite{MelloMacroscopic1987} uses this Fokker-Planck equation to obtain a Gaussian distribution for $s$ (there denoted $z$), as we have stated in the main text.  The phase variables $\mu,\nu$ in Ref.~\cite{MelloMacroscopic1987}, which are distributed uniformly [c.f. Eq. (2.6) there)], are related to the reflection phases by $\phi_{r'}=-2\mu$ and $\phi_r= -2 \nu -\pi$ [c.f. Eqs. (2.3a) and (2.3b) there].}.  The maximum entropy ansatz yields a factorization of the joint distribution of $s$ and $\phi_{r'}$ into a product of the two marginal distributions, with the marginal distribution of $\phi_{r'}$ being uniform.  This calculation applies when either or both of the sample reflection phase and local reflection phase are distributed uniformly.

While our work focuses on the case of a single scattering channel (i.e., $2\times2$ S matrices), it is useful for further comparison with the literature to consider the case of an arbitrary number $N_c$ of channels ($2N_c\times 2N_c$ S matrices). In this more general case, the $S$ matrix may be parametrized by $N_c$ transmission coefficients and $4N_c^2 -N_c$ phases~\cite{MelloMacroscopic1988, MelloSymmetries1991}.  The generalization of the uniform phase hypothesis to this case is known as the isotropy assumption and is believed to hold when there are many channels ($N_c\gg 1$) with sufficient coupling between them (see Ref.~\cite{XieAnderson2012} and references therein).  Under this assumption, the joint probability distribution of the $N_c$ transmission coefficients satisfies a version of the Fokker-Planck equation known as the Dorokhov-Mello-Pereyra-Kumar (DMPK) equation~\cite{DorokhovLocalization1982, MelloMacroscopic1988}. For the case $N_c = 2$, it has been shown in particular models that the isotropy assumption breaks down and that the transmission coefficients are entangled with the phases in the Fokker-Planck equation~\cite{DorokhovLocalization1982,XieAnderson2012}.

For the case $N_c=1$, which is our focus, general and analytical results for the joint distribution of $(s,\phi_{r'},\phi_t)$ were obtained by Roberts in Ref.~\cite{RobertsJoint1992}; see also Ref.~\cite{PendrySymmetry1994}.  These works obtained factorization of the joint distribution of $(s,\phi_{r'})$ into the product of separate distributions for $s$ and $\phi_{r'}$, apparently in conflict with our result.  In Sec.~\ref{sec:Joint probability distribution}, we support our result with analytical calculations and numerical checks, and we provide an explanation for the discrepancy with Refs.~\cite{RobertsJoint1992,PendrySymmetry1994}.

\section{Scattering expansion of the localization length}\label{sec:Scattering expansion of the localization length}
\subsection{Overview}\label{sec:Overview}
Our first general result in this paper is a series expansion of the inverse localization length with the local reflection strength $|r_n|$ as a small parameter.  In the course of this calculation, we also present results on SPS and on the marginal distribution of the reflection phase.

Our scattering expansion is defined by rescaling $r_n \to \param r_n$ and $r_n' \to \param r_n'$ in the local $S$ matrix~\eqref{eq:Sn general parametrization amplitudes} (with $t_n$ and $t_n'$ also rescaled to maintain unitarity), and, roughly speaking, expanding in the real and positive parameter $\param$.  (As we discuss below, it is necessary in the calculation to consider the order of limits $N\to\infty$ and $\param\to0^+$ in a particular way to maintain the system always in the localized regime.  However, this subtlety does not appear in the final results, which may be understood as straightforward expansions in $\param$.)  For convenience, we let the dependence on $\param$ be implicit in our notation and refer informally to an expansion in $|r_n|$.  Thus, the reflection amplitudes $r_n$ and $r_n'$ are both first order and the reflection coefficient $R_n$ is second order.  While we assume for now that $|r_n|$ is linear in the small parameter $\lambda$, we explain below the simple modifications to our results in the more general case of $|r_n|$ starting at linear order but also having higher-order corrections.

Before presenting our calculation, we first note that the result of Schrader \textit{et al}.~\cite{SchraderPerturbative2004} mentioned in the previous section has the following corollary: our parameter $\param$ is a valid choice for their parameter $\lambda'$, and their formula for the localization length and variance may be written in terms of local averages at any site $n$ as (see Appendix~\ref{sec:Schrader et al.'s formula} for the calculation) \footnote{In the case of $t_n=t_n'$ and $r_n=r_n'$, a result related to Eq.~\eqref{eq:lloc to 2nd order} has also been obtained in Refs.~\cite{AsatryanSuppression2007,AsatryanAnderson2010,GredeskulAnderson2012} using the WSA approach developed there.  See Appendix~\ref{sec:Comparison with the weak scattering approximation} for details.}
\beq
    \frac{2}{\lloc} = \langle R_n\rangle_n - 2 \text{Re}\left[ \frac{\langle r_n \rangle_n \langle r_n'\rangle_n}{1+ \langle r_n r_n'/R_n\rangle_n } \right]+O(|r_n|^4),\label{eq:lloc to 2nd order}
\eeq
and
\beq
    \lim_{N\to\infty}\frac{\sigma(N)^2}{2N} =\frac{2}{\lloc} +O(|r_n|^4),\label{eq:sigma to 2nd order}
\eeq
where we have also improved their results from $O(|r_n|^3)$ error to $O(|r_n|^4)$ error (see next paragraph).  The first term on the right-hand side of Eq.~\eqref{eq:lloc to 2nd order} is the small $|r_n|$ expansion of the uniform phase result~\eqref{eq:lloc AndersonNew1980}, while the second term is the contribution from deviations from the uniform phase hypothesis.  We emphasize again that the second term is not, in general, a small correction to the first; indeed, both terms are of the same order in the small parameter.  We also note that unlike the Lambert and Thorpe formula~\eqref{eq:lloc LambertPhase1982}, Eq.~\eqref{eq:lloc to 2nd order} involves only local averages and no averaging over the whole chain.

In writing Eqs.~\eqref{eq:lloc to 2nd order} and~\eqref{eq:sigma to 2nd order}, we have in fact gone beyond the immediate corollary of the result of Schrader \textit{et al}., which would yield the leading order contributions as given above but with $O(|r_n|^3)$ error instead of $O(|r_n|^4)$ \footnote{In the particular case of the Anderson model with onsite disorder with vanishing mean, Ref.~\cite{SchraderPerturbative2004} calculates the third and fourth order terms and finds that the third order terms indeed vanish.}. The third order terms, and indeed all odd orders, vanish due to the following symmetry argument.  The localization length and variance must be invariant under phase redefinitions of the incoming and outgoing scattering amplitudes; in particular, we can redefine $\Psi_\alpha^\pm\to e^{\pm i\phi/2}\Psi_\alpha^\pm$ ($\alpha=R,L$), which sends $r_n\to e^{-i\phi}r_n$ and $r_n'\to e^{i\phi}r_n'$.  It follows that $r_n$ and $r_n'$ must appear in equal numbers in each term (hence there are no terms of the order of an odd power of $|r_n|$).  This argument relies on the existence of series expressions for $2/\lloc$ and $\lim_{N\to\infty}\sigma(N)^2/(2N)$ with coefficients involving only local averages of integer powers of $r_n$, $r_n'$, $v_n$, and $R_n$ (note that the latter two are invariant under the phase re-definition).  In this section we show that such a series exists for $2/\lloc$, and in Sec.~\ref{sec:Joint probability distribution} we show it for  $\lim_{N\to\infty}\sigma(N)^2/(2N)$. 

We have thus explicitly connected SPS to the weakness of the local reflection strength, and we have extended the SPS relation to one more order [Eq.~\eqref{eq:sigma to 2nd order}].  Our next task is to apply the scattering expansion to the inverse localization length, recovering Eq.~\eqref{eq:lloc to 2nd order} and providing a recursive calculation of higher orders.  We show that all orders of the series depend, as in Eq.~\eqref{eq:lloc to 2nd order}, only on averages involving the local reflection amplitudes $r_n$ and $r_n'$.  We explicitly verify the vanishing of the third order and present the next non-vanishing order ($|r_n|^4$).

A key ingredient in our calculation is the limiting form as $N\to\infty$ of the probability distribution of the reflection phase, which we write as
\beq
    p_{\infty}(\phi_{r'}) \equiv \lim_{N\to\infty}p_{N}(\phi_{r'}). 
\eeq
(In Appendix~\ref{sec:Existence of the limiting distribution of the reflection phase}, we show that this limit exists given the assumption that localization occurs.)  Our approach is based on the following series formula, which we derive below, that expresses the inverse localization length in terms of the Fourier coefficients $p_{\infty,\ell}\equiv \int_{-\pi}^\pi \frac{d\phi_{r'}}{2\pi}e^{-i\ell\phi_{r'}}p_{\infty}(\phi_{r'})$ and the moments of $r_n$:
\beq
    \frac{2}{\lloc} = \langle -\ln T_n\rangle_n - 4\pi \text{Re}\left[ \sum_{\ell=1}^\infty \frac{1}{\ell}p_{\infty,-\ell} \langle r_n^\ell \rangle_n\right].\label{eq:lloc in terms of Fourier coefficients}
\eeq
Equation~\eqref{eq:lloc in terms of Fourier coefficients} is related to the Lambert and Thorpe formula~\eqref{eq:lloc LambertPhase1982}, with the essential difference being that we focus on the distribution of $\phi_{r'}$ rather than $\nu$ and work in frequency space.  Note that the uniform phase formula~\eqref{eq:lloc AndersonNew1980} is recovered from Eq.~\eqref{eq:lloc in terms of Fourier coefficients} in two non-exclusive special cases: (i) the local reflection phase is uniformly distributed independently of the local reflection coefficient (then  $\langle r_n^\ell\rangle_n =0$ for $\ell>0$), or (ii) the reflection phase distribution of the sample is uniform.  Case (i) is an example of strong phase disorder.  The difficulty of applying Eq.~\eqref{eq:lloc in terms of Fourier coefficients}, in the case that (i) does not hold, is that it has been shown in many examples that the reflection phase distribution $p_{\infty}(\phi_{r'})$ can be non-uniform (see references cited in Sec.~\ref{sec:Review of prior work}) and in general the distribution is only known numerically (although Schrader \textit{et al}.~\cite{SchraderPerturbative2004} calculated $p_{\infty,\pm1}$ in an equivalent form).

In order to use Eq.~\eqref{eq:lloc in terms of Fourier coefficients}, we calculate the probability distribution of the reflection phase order by order in the scattering expansion.  We show that the zeroth-order contribution is (generically) the uniform distribution, and that at any fixed order of the expansion, only finitely many of the Fourier coefficients $p_{\infty,\ell}$ are non-vanishing.  We further show that the expansion coefficients are local averages that can be calculated recursively.  Together with Eq.~\eqref{eq:lloc in terms of Fourier coefficients}, this recursive expansion of the Fourier coefficients $p_{\infty,\ell}$ yields our scattering expansion of the inverse localization length.  

In the remainder of this section, we derive the above results (Sec.~\ref{sec:General calculation}), then apply them to disordered wires (Sec.~\ref{sec:Application to disordered wires}) and to quantum walks (Sec.~\ref{sec:Application to discrete-time quantum walks}).  We note here that our calculation in this section only concerns the localization length and not the variance; however, we do obtain Eq.~\eqref{eq:sigma to 2nd order} by a different approach in Sec.~\ref{sec:Joint probability distribution}, and it is convenient to apply this result in this section to comment on SPS in a broad class of PARS.

\subsection{General calculation}\label{sec:General calculation}

We compare a sample of size $N$ with a sample of size $N+1$ using the following exact relations between the transmission coefficients and reflection phases (for completeness we derive them in Appendix~\ref{sec:S matrices and scattering transfer matrices}):
\bseq
\begin{align}
    T_{1\dots N+1} &= \frac{T_{1\dots N} T_{N+1} }{ \left| 1- \sqrt{R_{1\dots N} }e^{i \phi_{r_{1\dots N}'}} r_{N+1}  \right|^2},
    \label{eq:transmission coeff recursion}\\
    \phi_{r_{1\dots N+1}'} &= \phi_{r_{1\dots N}'} + \pi + \phi_{r_{N+1}} + \phi_{r_{N+1}'} \notag\\
    &\ -\Arg\left[ 1 -  \sqrt{R_{1\dots N}}r_{N+1}e^{i\phi_{r_{1\dots N}'} }\right]\notag\\
    &\ - \Arg\left[ 1-   r_{N+1}e^{i\phi_{r_{1\dots N}'} }/\sqrt{R_{1\dots N}} \right]\notag\\
    &\qquad\qquad\qquad\qquad\qquad(\text{modulo } 2\pi). \label{eq:reflection phase recursion}
\end{align}
\eseq

A basic assumption of our calculation is that localization occurs: that is, for large $N$ the sample reflection coefficient $R_{1\dots N}\approx 1$ in all disorder realizations \footnote{Whenever we require a property to hold for all disorder realizations, we expect that it would also suffice for that property to hold for almost all realizations (i.e., all except a set of measure zero).}.  We can then replace $R_{1\dots N}\to 1$ in the logarithm of Eq.~\eqref{eq:transmission coeff recursion} and in Eq.~\eqref{eq:reflection phase recursion}; in other words, we keep $\ln T_{1\dots N}$ and constant terms, but neglect terms linear or higher in $T_{1\dots N}= 1 -R_{1\dots N}$.  We thus obtain
\bseq
\begin{align}
    s_{1\dots N+1} &= s_{1\dots N} + g_{N+1}(\phi_{r_{1\dots N}'}),\label{eq:s recursion Rto1}\\
    \phi_{r_{1\dots N+1}'}&= \phi_{r_{1\dots N}'} + h_{N+1}(\phi_{r_{1\dots N}'}),\label{eq:phirp recursion Rto1}
\end{align}
\eseq
where
\bseq
\begin{align}
    g_n(\phi) &= -\ln T_n + \ln(1- r_n e^{i\phi} - r_n^* e^{-i \phi} +R_n),\label{eq:gn}\\
    h_n(\phi)&= \pi +\phi_{r_n} + \phi_{r_n'} +i\ln \frac{1-r_n e^{i\phi} }{1-r_n^* e^{-i \phi}}.\label{eq:hn}
\end{align}
\eseq

Let us pause to present an intuitive picture, in which we temporarily think of the increasing sample size $N$ as representing time.  Then, Eq.~\eqref{eq:phirp recursion Rto1} indicates that the reflection phase performs a random walk on the circle; furthermore, the probability distribution of the step size depends on the current position that the walk has reached.  From this point of view, it is clear that the long-time distribution of the phase should generally be non-uniform, since the walk will spend more time in regions where the step size is smaller.  From Eq.~\eqref{eq:s recursion Rto1}, we see that the variable $s$ performs a random walk on the half-line with a step size that depends on the current position of the phase walk, though not on the current position of $s$.  While the average position reached by $s$ after a long time may be obtained by treating the phase as simply another disorder variable in the step size (i.e., sampling the phase from its long-time distribution), the variance of $s$ and the correlations between $s$ and the phase seem to require a more careful treatment (Sec.~\ref{sec:Joint probability distribution}). 

To clarify a technical point about our expansion, we temporarily restore the true small parameter $\param$.  While we are interested in the regime of small $\param$, we cannot simply expand in $\param$ for fixed $N$, since the point $\param=0$ corresponds to a sample transmission coefficient of unity, inconsistent with our replacing $R_{1\dots N} \to 1$.  Instead, we must suppose that for any fixed $\param >0$, there is some sufficiently large $N_0(\param)$ for which the sample transmission coefficient is negligible for any $N\ge N_0(\param)$ in every disorder realization~\cite{Note8}.  [$N_0(\param)$ is of order of the localization length, so in view of our results below, it will turn out that $N_0(\param) \sim 1/\param^2$.]  All expansions in powers of $\param$ in our calculations below must be understood as occurring in the regime of $\param>0$ with $N\ge N_0(\param)$ \footnote{A similar point is mentioned in Ref.~\cite{CohenUniversal1988} (footnote 24).}.  This subtlety need not be considered in our final answer for  $2/\lloc$, and we have there a simple expansion in $\param$.

We make a few comments now about our conventions.  Here and throughout we use a superscript to indicate the order in $|r_n|$ (i.e., $X^{(j)} \sim |r_n|^j$ for any quantity $X$).  Also, since the disorder is i.i.d., we frequently replace the disorder average over site $N+1$ by an average over any site $n$.  Finally, all phases are constrained to be in $[-\pi,\pi)$, and any equations between non-exponentiated phases are to be understood as holding modulo $2\pi$. 

Next, we develop the scattering expansion for the Fourier coefficients $p_{\infty,\ell}$.  A recursion relation for $p_{N}(\phi_{r'})$ is determined by $p_{N+1}(\phi_{r'}) = \langle \delta(\phi_{r_{1\dots N+1}'}-\phi_{r'})\rangle_{1\dots N+1}$, from which Eq.~\eqref{eq:phirp recursion Rto1} then yields
\begin{multline}
    p_{N+1}(\phi_{r'})=  \int_{-\pi}^\pi d\phi_{r'}'\ p_{N}(\phi_{r'}')\\
    \times \langle \delta\bpl\phi_{r'}' + h_n(\phi_{r'}')-\phi_{r'}\bpr \rangle_n.\label{eq:p recursion finite N}
\end{multline}
We then send $N\to\infty$ and take the Fourier transform of both sides to obtain [noting Eq.~\eqref{eq:hn}]
\beq
    p_{\infty,\ell}=(-1)^\ell  \langle v_n^{-\ell} \int_{-\pi}^\pi \frac{d\phi}{2\pi}\ e^{-i\ell\phi}p_{\infty}(\phi)A_{n,\ell}(\phi)\rangle_n,\label{eq:p first eqn with A}
\eeq
where we have defined
\bseq
\begin{align}
    v_n &= \frac{r_n r_n'}{R_n},\\
    A_{n,\ell}(\phi) &= \left(\frac{1-r_n e^{i\phi} }{1-r_n^* e^{-i\phi}} \right)^\ell.\label{eq:A}
\end{align}
\eseq
Eq.~\eqref{eq:p first eqn with A} can be solved order by order in the scattering expansion, as we now show.  

We write the expansions of $p_{\infty}(\phi)$ and $A_{n,\ell}(\phi)$ as $p_{\infty}(\phi) = \sum_{j=0}^\infty p_{\infty}^{(j)}(\phi)$ and $A_{n,\ell}(\phi) = \sum_{j=0}^\infty A_{n,\ell}^{(j)}(\phi)$, with corresponding expansions for the Fourier coefficients.  Note that by the normalization condition, $\int_{-\pi}^\pi d\phi\ p_{\infty}(\phi) =1$, we have $p_{\infty,0}^{(j)} =0$ for all $j\ge 1$.

We now equate terms of the same order on both sides of Eq.~\eqref{eq:p first eqn with A}.  Since $A_{n,\ell}^{(0)}=1$, the quantity $p_{\infty,\ell}^{(q)}$ appears on both sides of the order $q$ equation (for any $q\ge 0$).  This has two immediate consequences.  First, the zeroth-order part of Eq.~\eqref{eq:p first eqn with A} is
\beq
    p_{\infty,\ell}^{(0)} = (-1)^\ell \langle v_n^{-\ell}\rangle_n p_{\infty,\ell}^{(0)},\label{eq:p0 equation}
\eeq
and second, for any $q\ge1$, the order $q$ part of Eq.~\eqref{eq:p first eqn with A} for $\ell\ne0$ can be re-arranged to
\begin{multline}
    p_{\infty,\ell}^{(q)} = \frac{(-1)^\ell}{1-(-1)^\ell \langle v_n^{-\ell}\rangle_n} \langle v_n^{-\ell} \sum_{j=0}^{q-1} \sum_{\ell'=-\infty}^\infty p_{\infty,-\ell'}^{(j)}\\
    \times A_{n,\ell;\ell+\ell'}^{(q-j)}\rangle_n,\label{eq:p recursion sum over all ellprime} 
\end{multline}
where $A_{n,\ell;\ell'}= \int_{-\pi}^\pi\frac{d\phi}{2\pi}\ e^{-i\ell'\phi}A_{n,\ell}(\phi)$ [and where we have written $\int_{-\pi}^\pi \frac{d\phi}{2\pi}\ e^{-i\ell\phi} p_{\infty}^{(j)}(\phi)A_{n,\ell}^{(q-j)}(\phi)$ as a sum over Fourier coefficients].

Equation~\eqref{eq:p0 equation} implies (see next paragraph) that $p_{\infty,\ell}^{(0)}=0$ for all $\ell\ne0$.  The $\ell=0$ Fourier coefficient is fixed by the normalization of the probability distribution, so we find that the zeroth-order distribution is uniform:
\beq
    p_{\infty}^{(0)}(\phi_{r'})=\frac{1}{2\pi}.\label{eq:uniform distribution}
\eeq

Equations~\eqref{eq:p recursion sum over all ellprime} and~\eqref{eq:uniform distribution} rely on assumptions of ``reasonable'' disorder and ``generic'' model parameters, as we now explain.  For instance, the limit of no disorder at all technically meets our definitions so far, but must be excluded.  Also, in applying our approach to any particular model, we must consider the model as part of some family of models (e.g., with parameters taking a continuous range of values), since our results are expected to hold generically but may break down at some special values of parameters (the anomalous momenta of the Anderson model are an example).  A precise statement of these assumptions is (a) that localization occurs (an assumption that we have used already), and (b) that the following inequality holds for all integers $\ell\ne 0$:
\beq
    \langle e^{i\ell(\phi_{r_n} + \phi_{r_n'} + \pi)} \rangle_n \ne 1,\label{eq:generic condition}
\eeq
where $n$ is any site.  We expect that any choice of disorder distribution and model parameters that violates~\eqref{eq:generic condition} is ``non-generic'' (since a particular alignment of reflection phases across disorder realizations is required).  [In Appendix~\ref{sec:Partial extension of our results to anomalies}, we consider the case that~\eqref{eq:generic condition} is only assumed for $0<|\ell| \le \ell_\text{max}$ for some finite $\ell_\text{max}$.]  Noting that $v_n=e^{i(\phi_{r_n}+\phi_{r_n'})}$, we see then that for $\ell\ne0$, we have $(-1)^\ell\langle v_n^{-\ell}\rangle_n\ne 1$, so indeed Eq.~\eqref{eq:p0 equation} implies $p_{\infty,\ell}^{(0)}=0$ and the denominator in Eq.~\eqref{eq:p recursion sum over all ellprime} is non-vanishing.

Equation~\eqref{eq:p recursion sum over all ellprime} is almost in a form in which we can calculate the reflection phase distribution to many orders.  The point is that the right-hand side involves only the reflection phase distribution at lower orders than the order $q$ on the left-hand side; thus, in principle we can start with Eq.~\eqref{eq:uniform distribution} and iterate Eq.~\eqref{eq:p recursion sum over all ellprime} to calculate higher orders.  To make this process more efficient, we show next that only finitely many Fourier coefficients are non-vanishing at any given order.  In particular, we show that for any $q\ge0$,
\beq
    p_{\infty,\ell}^{(q)}=0 \qquad (\text{for }|\ell|>q),\label{eq:vanishing of Fourier coefficients}
\eeq
and we show that the sum over $\ell'$ on the right-hand side of Eq. ~\eqref{eq:p recursion sum over all ellprime} can be truncated to a finite sum:
\begin{widetext}
\beq
    p_{\infty,\ell}^{(q)} = \frac{(-1)^\ell}{1-(-1)^\ell \langle v_n^{-\ell}\rangle_n} \langle v_n^{-\ell} \sum_{j=0}^{q-1} \sum_{\ell'=0}^j p_{\infty,-\ell'}^{(j)}A_{n,\ell;\ell+\ell'}^{(q-j)}\rangle_n \qquad (-q\le \ell <0).\label{eq:p recursion}
\eeq
\end{widetext}
The Fourier coefficients $A_{n,\ell;\ell'}$ that we need are easily calculated to many orders from the definition~\eqref{eq:A}, and the positive frequency coefficients ($\ell >0$) can be obtained from $p_{\infty,-\ell}=p_{\infty,\ell}^*$.

To derive Eqs.~\eqref{eq:vanishing of Fourier coefficients} and~\eqref{eq:p recursion}, we use some simple properties of the function $A$.  We start by noting that $A_{n,\ell;\ell'}^{(j)}=0$ for $|\ell'|>j$.  Suppose now that for some $q\ge 1$, $p_{\infty,\ell'}^{(j)}=0$ when $0\le j\le q-1$ and $|\ell'| > j$ [this holds for $q=1$ by Eq.~\eqref{eq:uniform distribution}].  Then the only terms on the right-hand side of Eq.~\eqref{eq:p recursion sum over all ellprime} that can be non-zero are those with $|\ell'|\le j$ and $|\ell + \ell'|\le q-j$, which implies $|\ell| \le q$ for these terms.  Thus, Eq.~\eqref{eq:vanishing of Fourier coefficients} holds by induction.  The sum over $\ell'$ in Eq.~\eqref{eq:p recursion sum over all ellprime} can therefore be truncated to $\ell'=-j$ to $j$.  Fixing $\ell<0$, we can see that the negative $\ell'$ terms can be dropped since $A_{n,\ell;\ell'}=0$ for $\ell' < \ell$; thus we obtain Eq.~\eqref{eq:p recursion}.  Further improvement (dropping terms in the sum that must be zero) is possible, but Eq.~\eqref{eq:p recursion} suffices for our purposes.

We can now calculate higher orders of the reflection phase distribution in a completely mechanical, recursive fashion using Eq.~\eqref{eq:p recursion}, starting from $p_{\infty,\ell}^{(0)}=\delta_{\ell,0}/(2\pi)$.  To present compactly the first few terms of these expansions, we use the following notation:
\bseq
\begin{align}
    \alpha_j &= \frac{1}{1 -(-1)^j \langle v_n^j\rangle_n },\label{eq:alpha}\\
    \gamma^{(1)} &= \alpha_1 \langle r_n'\rangle_n,\label{eq:gamma1}\\
    \gamma^{(2)} &= \alpha_2 \langle r_n' (r_n'-2\gamma^{(1)}v_n)\rangle_n,\label{eq:gamma2}\\
    \gamma_1^{(3)} & =\alpha_1 \langle r_n  (\gamma^{(1)}r_n' - \gamma^{(2)}v_n)\rangle_n,\label{eq:gamma13}\\
    \gamma_3^{(3)} &= \alpha_3 \langle r_n' ({r_n'}^2- 3\gamma^{(1)}r_n'v_n +3 \gamma^{(2)}v_n^2 )\rangle_n.\label{eq:gamma33}
\end{align}
\eseq
Then we obtain, transforming back from frequency space (see Appendix~\ref{sec:Partial extension of our results to anomalies} for further detail)
\begin{multline}
    2\pi p_{\infty}(\phi_{r'}) =  1 + 2\text{Re}\biggr[ \left( \gamma^{(1)} + \gamma_1^{(3)}  \right)e^{-i\phi_{r'}}\\
    + \gamma^{(2)} e^{-2i\phi_{r'}}
    +\gamma_3^{(3)} e^{-3i \phi_{r'}}
    \biggr]  + O(|r_n|^4),\label{eq:p to 3rd order}
\end{multline}
which shows explicitly the corrections to the uniform distribution that is obtained at zeroth order.  Let us also note that interchanging $r_n\leftrightarrow r_n'$ in Eqs.~\eqref{eq:gamma1}--\eqref{eq:gamma33} yields the corresponding expression for the limiting distribution of the left-to-right reflection phase ($\phi_{r_{1\dots N}}$).

We proceed to use the reflection phase distribution to calculate the inverse localization length in the scattering expansion.  We start by deriving the relation~\eqref{eq:lloc in terms of Fourier coefficients} between the localization length and the Fourier coefficients $p_{\infty,\ell}$.  From the recursion relation~\eqref{eq:s recursion Rto1} for $s$, it follows that for sufficiently large $N$, $\langle s_{1\dots N}\rangle_{1\dots N}$ increases by the same constant amount (which by definition is $2/\lloc$) each time $N$ is increased by one:
\begin{multline}
    \langle s_{1\dots N+1} \rangle_{1\dots N+1} - \langle s_{1\dots N} \rangle_{1\dots N} \\= \int_{-\pi}^{\pi} d\phi_{r'}\ p_{\infty}(\phi_{r'}) \langle g_{N+1}(\phi_{r'}) \rangle_{N+1}.\label{eq:s average increase}
\end{multline}
The site $N+1$ on the right-hand side can be replaced by any site $n$, since the disorder is by assumption distributed identically across the sites.  We thus obtain an expression for the inverse localization length in terms of the reflection phase distribution:
\bseq
\begin{align}
    \frac{2}{\lloc} &= \langle \int_{-\pi}^{\pi} d\phi\ p_{\infty}(\phi_{r'}) g_n(\phi_{r'}) \rangle_n\label{eq:lloc as integral} \\
    &= \langle -\ln T_n\rangle_n
    + \int_{-\pi}^\pi d\phi_{r'}\  p_{\infty}(\phi_{r'})\notag\\
    &\qquad \times \langle \ln(1- r_n e^{i\phi_{r'}} - r_n^* e^{-i \phi_{r'}} +R_n)\rangle_n.\label{eq:lloc as integral explicit}
\end{align}
\eseq
Equation~\eqref{eq:lloc as integral explicit} can also be obtained directly from the Lambert and Thorpe expression [Eq.~\eqref{eq:lloc LambertPhase1982}], or (using the correspondence discussed in Appendix~\ref{sec:Schrader et al.'s formula}) from the Furstenberg formula quoted in Ref.~\cite{SchraderPerturbative2004}.  Expressing the second term in frequency space is straightforward~\cite{RauhAnalytical2009} (see Appendix~\ref{sec:Evaluation of a Fourier integral}) and yields Eq.~\eqref{eq:lloc in terms of Fourier coefficients}.

Since the odd orders vanish due to the symmetry argument presented below Eq.~\eqref{eq:sigma to 2nd order}, we may write
\beq
    \frac{2}{\lloc} = \sum_{j=1}^\infty \left(\frac{2}{\lloc}\right)^{(2j)}, 
\eeq
where the superscript indicates the order in the expansion.  Furthermore, for any $j\ge 1$, Eq.~\eqref{eq:lloc in terms of Fourier coefficients} yields
\beq
    \left(\frac{2}{\lloc}\right)^{(2j)} = \frac{1}{j} \langle R_n^j \rangle_n - 4\pi \text{Re}\left[\sum_{\ell=1}^j \frac{1}{\ell}p_{\infty,-\ell}^{(2j-\ell)} \langle r_n^\ell\rangle_n \right],\label{eq:lloc order 2j}
\eeq
where we noted that Eq.~\eqref{eq:vanishing of Fourier coefficients} allows us to drop the terms $\ell=j+1$ to $2j$ in the sum.  Reading off the Fourier coefficients from Eq.~\eqref{eq:p to 3rd order}, we then calculate the first two non-vanishing orders of the scattering expansion for the inverse localization length (the main result of this section):
\begin{widetext}
\begin{multline}
    \frac{2}{\lloc} = \langle R_n\rangle_n - 2 \text{Re}\left[ \frac{\langle r_n \rangle_n \langle r_n'\rangle_n}{1+ \langle r_n r_n' / R_n\rangle_n } \right]
    + \frac{1}{2}\langle R_n^2\rangle_n  \\- \text{Re}\left[\alpha_2 ( \langle r_n^2\rangle_n -2\alpha_1\langle r_n\rangle_n \langle r_n v_n\rangle_n) ( \langle r_n'^2\rangle_n -2\alpha_1\langle r_n'\rangle_n \langle r_n' v_n\rangle_n)
    + 2\alpha_1^2\langle r_n\rangle_n \langle r_n'\rangle_n \langle r_n r_n'\rangle_n\right] + O(|r_n|^6).\label{eq:lloc to 4th order}
\end{multline}
\end{widetext}
On the right-hand side, the first two terms are second order and recover Eq.~\eqref{eq:lloc to 2nd order}.  The remaining terms are fourth order.  We have explicitly checked [using Eq.~\eqref{eq:lloc in terms of Fourier coefficients}] that the third- and fifth-order contributions vanish, consistent with  the general symmetry argument.  Our recursive formula~\eqref{eq:p recursion} can straightforwardly generate still higher orders of Fourier coefficients $p_{\infty,\ell}$, and thus of $2/\lloc$.

A final technical point is that we have so far assumed that $|r_n|$ is strictly linear in the small parameter $\param$.  However, it may be the case in a particular problem that $|r_n|$ starts at linear order in some parameter $\param$ but also has higher-order corrections in $\param$ (for instance, this occurs in the Anderson model, where $\param$ is essentially the disorder strength).  There are two straightforward steps to adapt the results of this section to this more general case.  First, the condition~\eqref{eq:generic condition} is required to hold at the zeroth order in $\param$ as $\param\to0$.  Second, the series for $2/\lloc$ is to be truncated to a fixed order in $\param$, which implies in particular that each term at a given order $j$ in the $|r_n|$ expansion contributes generally at all orders $j'\ge j$ in $\param$. 

\subsection{Application to disordered wires}\label{sec:Application to disordered wires}
In this section, we apply our general results to the Anderson model and to a broad class of PARS.  In the Anderson model, we consider the case of diagonal disorder (i.e., disorder in the onsite energies but not the tunneling) and show that our next-to-leading order result~\eqref{eq:lloc to 4th order} agrees with and extends a known formula for the weak disorder expansion of the inverse localization length.  We also verify our result for the probability distribution of the reflection phase with the literature, and we discuss the ``anomalous'' values of the momentum.  In the PARS case, we use the Born series to show that Eqs.~\eqref{eq:lloc to 2nd order} and~\eqref{eq:sigma to 2nd order} demonstrate the SPS relation to the first two orders in the potential strength.  As a special case, we obtain the SPS relation in the same square well array model considered by Deych \textit{et al}. (see Sec.~\ref{sec:Setup and review of prior work}).  We also verify our higher-order formula~\eqref{eq:lloc to 4th order} for the inverse localization length by comparing to another special case in the literature.

\subsubsection{Anderson model with diagonal disorder}\label{sec:Anderson model with diagonal disorder}
As a first application and check of our general results, we consider the Anderson model: a one-dimensional tight-binding chain with disordered onsite energies.  The well-known scattering setup for this problem is the following:
\beq
    H = -V \sum_n (\ket{n+1}\bra{n} + \text{H.c.}) + \sum_n \epsilon_n \ket{n}\bra{n},
\eeq
where the onsite potentials $\epsilon_n$ are i.i.d. for $n$ inside the ``sample'' region (defined as the sites $n=1,\dots, N$) and zero otherwise ($-\infty< n \le 0$ and $N+1 \le n <\infty$).  We take $V>0$.  The Schr\"{o}dinger equation for an eigenstate $\ket{\Psi} = \sum_n \Psi(n) \ket{n}$ with energy $E$ can be written in transfer matrix form:
\beq
    \bpmat
        \Psi(n+1)\\
        \Psi(n)
    \epmat
    =
    \mathcal{M}_n
    \bpmat
        \Psi(n)\\
        \Psi(n-1)
    \epmat,
\eeq
where
\beq
    \mathcal{M}_n = 
    \bpmat
        \frac{\epsilon_n-E}{V} & -1 \\
        1 & 0
    \epmat.
\eeq
The clean Hamiltonian has spectrum $E = -2 V \cos k$, and given a fixed $k\in(0,\pi)$ we can write a scattering state with energy $E$ as
\beq
    \Psi(n) = 
    \begin{cases}
        \Psi_L^+ e^{i k n} + \Psi_L^- e^{-i k n} & n \le 1,\\
        \Psi_R^+ e^{i k(n-N)} + \Psi_R^- e^{-i k(n-N)} & n \ge N.
    \end{cases}
\eeq
The phase convention for the amplitudes on the right (i.e., the factors of $e^{\pm i k N}$~\cite{SulemTotal1973}) is chosen so that the local reflection phase takes the same form at every site (i.e., it depends on $\epsilon_n$ but not on the integer value of the site index $n$), which in turn results in the existence of a limiting distribution of the reflection phase as $N\to\infty$.  Were we to omit the factors of $e^{\pm i k N}$, the distribution would instead tend to a fixed function evaluated with an $N$-dependent shift of its argument and hence would not strictly have any $N\to\infty$ limit.

To convert position-space amplitudes to scattering amplitudes, define a matrix $\Lambda$ as
\beq
    \Lambda =
    \bpmat
    e^{i k} & e^{-i k}\\
    1 & 1
    \epmat.
\eeq
Then,
\beq
    \bpmat
        \Psi(1)\\
        \Psi(0)
    \epmat
    = \Lambda
    \bpmat
        \Psi_L^+\\
        \Psi_L^-
    \epmat,\ 
    \bpmat
        \Psi(N+1)\\
        \Psi(N)
    \epmat
    = \Lambda 
    \bpmat
        \Psi_R^+\\
        \Psi_R^-
    \epmat.
\eeq
Chaining transfer matrices then yields the desired product form of the scattering transfer matrix of the sample:
\beq
    \mathcal{T}_{1\dots N} = \mathcal{T}_N \dots \mathcal{T}_1,
\eeq
where
\bseq
\begin{align}
    \mathcal{T}_n &= \Lambda^{-1} \mathcal{M}_n \Lambda\\
    &= 
    \bpmat
        \left( 1 - ie_n \right) e^{i k } & -ie_n e^{-i k}\\
        i e_n e^{i k}& \left( 1 + ie_n \right) e^{-i k } 
    \epmat,
\end{align}
\eseq
where we have defined a dimensionless onsite energy:
\beq
    e_n = \frac{\epsilon_n}{2V\sin k}= \frac{\epsilon_n}{\sqrt{4V^2-E^2}}.
\eeq
The corresponding $S$ matrix at site $n$ is
\beq
    \mathcal{S}_n = 
    \frac{e^{i k}}{1+ ie_n }
    \bpmat
        1 & - ie_ne^{- i k}\\
        - ie_ne^{i k} & 1
    \epmat,
\eeq
from which we read off the reflection amplitudes
\beq
    r_n'=e^{-2ik}r_n= \frac{-ie_n}{1+ie_n}.\label{eq:reflection amplitudes Anderson}
\eeq

Thus, our scattering expansion applies in the regime of weak onsite energies.  We assume that almost all disorder realizations have $|\epsilon_n| < \epsilon_\text{max}$, where $\epsilon_\text{max}$ is a fixed constant independent of the realization.  Note in particular that we are not considering any long-range probability distribution (such as the Cauchy distribution considered by Deych \textit{et al}. in Refs.~\cite{DeychSingle2000,DeychSingleparameter2001}).  The parameter $\param$ that controls the reflection strength $|r_n|$ can be taken to be $\param = \epsilon_\text{max} / V$, and we refer informally to an expansion in $e_n$.

Our main result for the Anderson model is obtained by substituting the reflection amplitudes~\eqref{eq:reflection amplitudes Anderson} into Eq.~\eqref{eq:lloc to 4th order}.  With no loss of generality (see below), we take the mean onsite energy to be zero ($\langle \epsilon_n\rangle_n=0$), yielding
\begin{multline}
    \frac{2}{\lloc} = \langle e_n^2\rangle_n + \frac{3}{2} \langle e_n^2\rangle_n^2 -\frac{1}{2} \langle  e_n^4\rangle_n\\
    + 4\frac{E^2 - V^2}{E\sqrt{4V^2-E^2}} \langle e_n^2\rangle_n\langle e_n^3\rangle_n 
    + O(e_n^6). \label{eq:lloc to 5th order Anderson}
\end{multline}

The terms in the first line of~\eqref{eq:lloc to 5th order Anderson} recover the known expansion to fourth order \footnote{See, e.g., the second unnumbered equation below Eq. (36) in Ref.~\cite{SchraderPerturbative2004} (which cites earlier work).  Our Appendix~\ref{sec:Schrader et al.'s formula} provides some details for converting the notation of Ref.~\cite{SchraderPerturbative2004} into our notation.}.  The fifth-order term in the second line is the leading dependence on the asymmetry of the probability distribution of the onsite energy, in particular its skewness (i.e., third moment).  We are able to calculate the fifth order term here because there is no term of fifth order in the reflection strength in Eq.~\eqref{eq:lloc to 4th order}.

We note that Eq.~\eqref{eq:lloc to 5th order Anderson} is consistent with the work of Ref.~\cite{FengLocalization2019}, which considers an asymmetric distribution with (in our notation) $\langle e_n\rangle_n=0$, $\langle e_n^2\rangle_n \ne 0$, and a small $\langle e_n^3\rangle_n \ne 0$.  Reference~\cite{FengLocalization2019} finds that the non-vanishing skewness has no effect at leading order on the inverse localization length.  

The result~\eqref{eq:lloc to 5th order Anderson} might be expected to hold more generally, in particular with periodic or open boundary conditions instead of the scattering setup that we used.  As a consistency check of this expectation, we consider the localization length as a function of the eigenstate energy $E$ and the average onsite energy: $\lloc(E,\epsilon_\text{avg})$, where $\epsilon_\text{avg}=\langle\epsilon_n\rangle_n$.  With periodic or open boundary conditions, the average onsite energy is just an additive constant in the Hamiltonian, so we must have the symmetry property $1/\lloc(E+\epsilon_\text{avg},\epsilon_\text{avg})=1/\lloc(E,0)$.  We have verified that our scattering calculation also has this symmetry property [by evaluating Eq.~\eqref{eq:lloc to 4th order} with general $\epsilon_\text{avg}$ and comparing to Eq.~\eqref{eq:lloc to 5th order Anderson}] up to error $O(\param^6)$.  Note that $\epsilon_\text{avg}=O(\param)$ here.

We emphasize that the equivalence of $\lloc$ across different boundary conditions must break down if we allow $\epsilon_n$ to be sufficiently large in sufficiently many disorder realizations (e.g., by taking $\epsilon_\text{avg}$ to be large or by having large variations about $\epsilon_\text{avg}$), for then the scattering setup has an attenuation effect (see next paragraph) that is not present in the problems with periodic or open boundary conditions.  This regime is outside the scope of our calculation, since the reflection strength approaches unity.  We do not know if the difference between boundary conditions can be detected at sufficiently high order in the scattering expansion, but due to the symmetry property mentioned in the previous paragraph, we expect that Eq.~\eqref{eq:lloc to 5th order Anderson} holds also for periodic or open boundary conditions.

To explain this point in more detail, we consider first the clean limit with strong onsite energy (all $\epsilon_n \equiv \epsilon_\text{avg} \gg V, |E|$).  In the case of periodic or open boundary conditions, $\epsilon_\text{avg}$ is of no significance and the localization length is infinite.  However, with scattering boundary conditions, the transmission coefficient $T_{1\dots N}$ decays exponentially in $N$ purely because of \emph{attenuation}.  More generally, in the course of a strong disorder calculation, Ref.~\cite{AbrahamsResistance1980} shows that the transmission coefficient (with any disorder such that all $\epsilon_n$ are large) is simply proportional to the product of $1/\epsilon_n$ over all sites $n$, from which it follows that $2/\lloc = \langle \ln(\epsilon_n/V)\rangle_n$ in the scattering setup; however, this cannot be the answer in the periodic or open setups because setting all $\epsilon_n$ to be equal yields a non-zero $\lloc$.

Next, we calculate the probability distribution of the reflection phase up to errors of third order in $e_n$, verifying our answer with the literature in the special case of $\epsilon_\text{avg}=0$.  Here we do not see any obvious symmetry relation between the problem with $\epsilon_\text{avg}=0$ and the problem with small, non-zero $\epsilon_\text{avg}$.

For convenience, we define
\beq
    \Delta e_n = e_n - \langle e_n \rangle_n.
\eeq
We then obtain
\bseq
\begin{align}
    \gamma^{(1)} &= \frac{e^{-ik}}{2\sin k} \left[-i \langle(\Delta e_n)^2\rangle_n + \left(1+ \langle e_n\rangle_n\cot k  \right)\langle e_n\rangle_n\right]\notag\\
    &\qquad + O(e_n^3),\label{eq:gamma1 Anderson}\\
    \gamma^{(2)} &= -\frac{ie^{-2ik}}{2\sin(2k)}\langle(\Delta e_n)^2\rangle_n + \frac{e^{-2ik}}{4\sin^2k}\langle e_n\rangle_n^2\notag\\
    &\qquad + O(e_n^3)\label{eq:gamma2 Anderson},
\end{align}
\eseq
which we substitute into Eq.~\eqref{eq:p to 3rd order} to obtain the probability distribution of the reflection phase up to second order (the third-order term may be obtained similarly).  In particular, the leading correction to phase uniformity is
\beq
    2\pi p_\infty(\phi_{r'}) = 1 + \frac{\cos(k+\phi_{r'})}{\sin k}\langle e_n\rangle_n + O(e_n^2).\label{eq:reflection phase dist to 1st order Anderson}
\eeq
For $\epsilon_\text{avg}=0$, the first-order term vanishes, and the leading correction is second order:
\begin{multline}
    2\pi p_{\infty}(\phi_{r'}) =1\\ - \left[ \frac{\sin(k+\phi_{r'})}{\sin k} + \frac{\sin\bpl2(k+\phi_{r'})\bpr}{\sin 2k}  \right] \langle e_n^2\rangle_n + O(e_n^4),\label{eq:reflection phase dist to 2nd order Anderson}
\end{multline}
in agreement with Ref.~\cite{Barnesdistribution1990}, where a symmetric exponential distribution for $\epsilon_n$ is considered.  Equation~\eqref{eq:reflection phase dist to 2nd order Anderson} also agrees with Ref.~\cite{RobertsJoint1992}, which considers an arbitrary disorder distribution symmetric about zero, except for the sign of the $\sin\bpl2(k+\phi_{r'})\bpr$ term (and our sign agrees with Refs.~\cite{Barnesdistribution1990} and with numerical checks) \footnote{Here we explain the comparison with Refs.~\cite{Barnesdistribution1990} and~\cite{RobertsJoint1992} in more detail.  From Eq. (4.21) of Ref.~\cite{Barnesdistribution1990}, we see that the coefficients $c_n$ there are related to our notation by $c_\ell^* = (-1)^{\ell-1}2\pi p_{\infty,-\ell}$.  We specialize to the symmetric exponential distribution with width $W$ by setting $\langle \epsilon_n^2\rangle_n= 2 W^2$, yielding agreement with Eq. (4.23) there.  To compare with Ref.~\cite{RobertsJoint1992}, we note from Eq. (54) there that $\sigma^{(n')} = \langle e_n^{2n'}\rangle_n$ in our notation, and we integrate over $\phi_t$ in Eq. (55) there to obtain $p_\infty^{[\phi_r]}(\phi)$, i.e., the distribution of $\phi_r$.  By Eq.~\eqref{eq:reflection amplitudes Anderson}, we have $p_\infty(\phi) = p_\infty^{[\phi_r]}(\phi+2k)$, resulting in the sign discrepancy mentioned in the main text.}.  Since Eq.~\eqref{eq:reflection phase dist to 1st order Anderson} contains a term linear in $\epsilon_\text{avg}$, it cannot be recovered from Eq.~\eqref{eq:reflection phase dist to 2nd order Anderson} by a simple shift of energy.

We conclude this section by discussing the special case of anomalous momenta.  So far, we have assumed that the momentum $k$ takes a ``generic'' value within $(0,\pi)$  (recall that the lattice spacing is the unit of distance).  In particular, the condition~\eqref{eq:generic condition} at the zeroth order in $\param$ [see the discussion below~\eqref{eq:generic condition}] is
\beq
    e^{2ik\ell} \ne 1 \qquad (\text{all integers }\ell\ne 0),\label{eq:generic condition Anderson model}
\eeq
which excludes $k=(p/q)\pi$ for some integers $p,q$.  These special points are known in the literature as anomalous (see Ref.~\cite{TessieriAnomalies2018} and references therein).

Let us consider $k=(p/q)\pi$ with $p/q$ reduced to simplest form and with $p\ne0$.  Then~\eqref{eq:generic condition Anderson model} holds for $|\ell| \le q-1$, but not for $\ell=q$.  The calculation in Appendix~\ref{sec:Partial extension of our results to anomalies} then shows that Eq.~\eqref{eq:lloc to 5th order Anderson} holds provided that $q\ge 6$.  More generally, our approach can be used to calculate the inverse localization length up to and including order $q-1$ in $\param$. (The increase, with $q$, in the expansion order affected by the anomaly has already been discussed in Refs.~\cite{LambertAnomalies1984, DerridaLyapounov1984,BovierWeak1988}.)  As a check, we note that at the band center anomaly~\cite{KappusAnomaly1981} ($k=\pi/2$ and hence $q=2$), the scattering expansion does not yield even the leading order ($\lambda^2$); this is consistent with the known result that the leading order formula $2/\lloc = \langle (\Delta e_n)^2\rangle_n$ is modified at the band center~\cite{KappusAnomaly1981}.  Furthermore, Ref.~\cite{TessieriAnomalies2018} finds that the leading order formula holds at all of the anomalies except the band center and band edge, which is consistent with what we find above (since $q\ge 3$ except at the band center and edge).

The reflection phase distribution is also affected at anomalous momenta (see Appendix~\ref{sec:Partial extension of our results to anomalies}).  The higher-order Fourier components that appear may be related to the higher-order terms found in the invariant measure in the real space approaches of Refs.~\cite{BovierWeak1988,TessieriAnomalies2018}. 

\subsubsection{Periodic-on-average random potential}\label{sec:Periodic-on-average random potential}
\paragraph{Setup.} We consider the problem of a free particle scattering off $N$ identically and independently disordered potentials with disorder also in the separations between the potentials.  Since the separations are not necessarily constant (though we may choose them to be as a special case), the scattering region is in general periodic-on-average rather than periodic.

\begin{figure}[t]
\subfloat[\label{subfig:PARS_diagram_ucorrelated}]{%
  \includegraphics[width=\columnwidth]{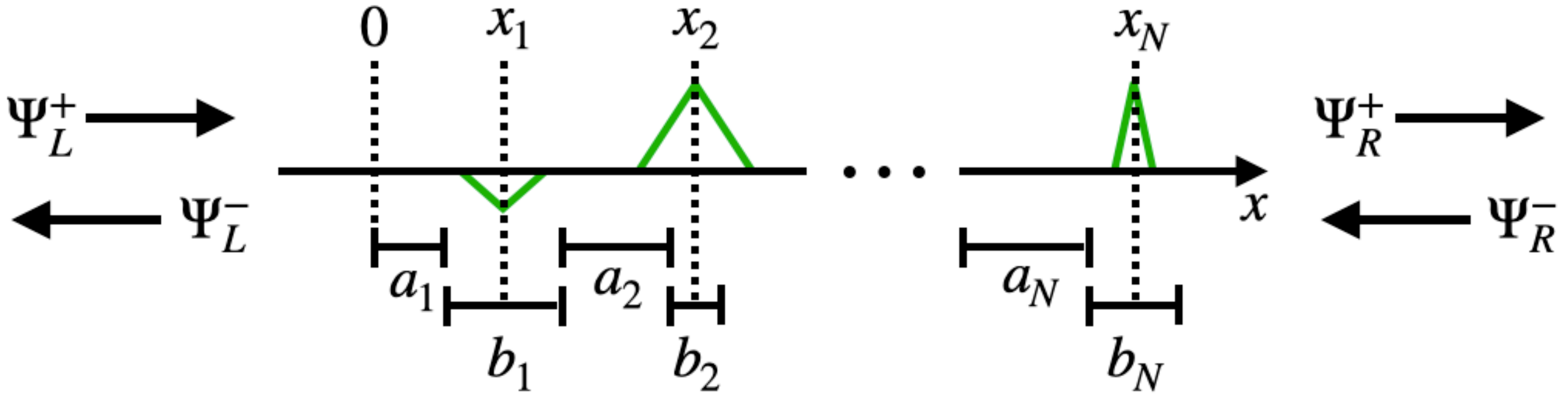}%
}\vfill
\subfloat[\label{subfig:PARS_diagram_correlated}]{%
  \includegraphics[width=\columnwidth]{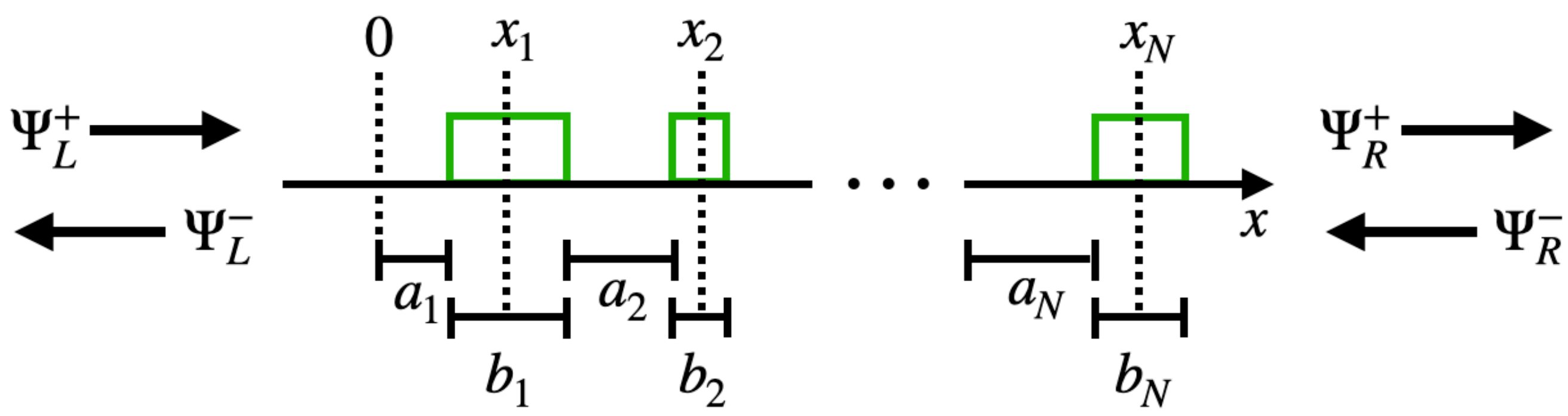}%
}
\caption{Schematic of the PARS setup defined by the Hamiltonian~\eqref{eq:H PARS}.  The scattering amplitudes $\Psi_\alpha^\pm$ ($\alpha=L,R$) refer to Eq.~\eqref{eq:scattering wavefn PARS}.  (a) The general case, in which the spacings $a_n$, the widths $b_n$, and the parameters of the potential all vary independently.  (b) A special case: square wells of equal strength.  The width $w_n$ of the $n$th well is equal to the width $b_n$.}\label{fig:PARS_diagrams}
\end{figure}

To define the problem, we consider a family of potentials $V_{\hat{\mathbf{D}}}(x)$, where $\hat{\mathbf{D}}$ is an array of parameters characterizing the potential, and where all potentials in the family have a fixed range $x_\text{max}$ independent of $\hat{\mathbf{D}}$ [that is, $V_{\hat{\mathbf{D}}}(x) = 0$ for $|x| > x_\text{max}$] \footnote{We can expect our results to hold also if the potentials are sufficiently small (e.g., exponentially decaying) for $|x|>x_\text{max}$.}.  We form a chain of $N$ such potentials, with $\hat{\mathbf{D}}$  ``promoted''  to a site-dependent array $\hat{\mathbf{D}}_n$ and with the $n$th potential centered at some position $x_n$:
\beq
    H = \frac{P^2}{2m} + \sum_{n=1}^N V_n(x - x_n),\label{eq:H PARS}
\eeq
where $V_n(x) \equiv V_{\hat{\mathbf{D}}_n}(x)$.  We set the centers to be at $x_n = \sum_{j=1}^n(a_j + b_j) - \frac{1}{2}b_n$, where $a_n$ and $b_n$ may be disordered arbitrarily as long as the potentials never overlap ($a_{n+1} + \frac{1}{2}b_{n+1}+\frac{1}{2}b_n > 2x_\text{max}$).  Thus, we have a sequence of regions of varying widths $b_n$ which are separated by spacings $a_n$ (see Fig.~\ref{fig:PARS_diagrams}).
We collect the parameters of site $n$ into a single array $\mathbf{D}_n \equiv (\hat{\mathbf{D}}_n,a_n,b_n)$ and assume that $\mathbf{D}_1, \dots, \mathbf{D}_N$ are i.i.d. (allowing correlations between the individual elements of $\mathbf{D}_n$).

In any particular disorder realization, we consider a scattering eigenstate with energy $E$, writing the wavefunction outside the sample as
\begin{widetext}
\beq
    \Psi(x) =
    \begin{cases}
        \Psi_L^+ e^{i kx} + \Psi_L^- e^{-i k x} & x < x_1 - x_\text{max},\\
        \Psi_R^+ e^{i k \left(x - x_N - \frac{1}{2}b_N \right)} + \Psi_R^- e^{-i k \left(x - x_N - \frac{1}{2}b_N \right)}
         & x_N + x_\text{max} < x,\label{eq:scattering wavefn PARS}
    \end{cases}
\eeq
\end{widetext}
where $E=k^2/(2m)$.  With this choice of phase convention, we may bring the scattering problem into the required form for applying our general results.  Indeed, we show in Appendix~\ref{sec:Relation of PARS reflection amplitudes to single-site reflection amplitudes} that if the single-site scattering problem of the potential $V_{\hat{\mathbf{D}}}(x)$ [i.e., Eqs.~\eqref{eq:H PARS} and~\eqref{eq:scattering wavefn PARS} with $N=1$, $a_1=b_1=0$, and $\hat{\mathbf{D}}_1\equiv \hat{\mathbf{D}}$] has reflection amplitudes $\hat{r}_{\hat{\mathbf{D}}}$ and $\hat{r}_{\hat{\mathbf{D}}}'$, then the scattering transfer matrix of the chain may be written as
\beq
    \mathcal{T}_{1\dots N} = \mathcal{T}_N\dots\mathcal{T}_1,
\eeq
where the local scattering transfer matrix $\mathcal{T}_n$ is equivalent to an $S$ matrix $\mathcal{S}_n$ which has reflection amplitudes $r_n$ and $r_n'$ given by 
\bseq
\begin{align}
    r_n &=e^{i k (2a_n+ b_n)}\hat{r}_{\hat{\mathbf{D}}_n},\label{eq:rn in terms of hatrn}\\
    r_n' &= e^{i k b_n}\hat{r}_{\hat{\mathbf{D}}_n}'.\label{eq:rnp in terms of hatrnp}
\end{align}
\eseq
Thus, given the reflection amplitudes of the single-site problem, we can apply our general results to the general PARS problem defined above.  We note here that throughout this section, the localization length is expressed in units of the average lattice spacing (and thus should be multiplied by $\langle a_n + b_n\rangle_n$ to restore physical dimensions).

From Eqs.~\eqref{eq:rn in terms of hatrn} and~\eqref{eq:rnp in terms of hatrnp}, we see that $a_n$ and $b_n$ may be understood as phase disorder, at least in the case when they are distributed independently of the parameters characterizing the potential.  If $a_n$ or $b_n$ is strongly disordered, the uniform phase result~\eqref{eq:lloc AndersonNew1980} is obtained; for $a_n$ we can see this from Eq.~\eqref{eq:lloc in terms of Fourier coefficients}, while for $b_n$ we can use Eq.~\eqref{eq:lloc in terms of Fourier coefficients} with the roles of $r_n$ and $r_n'$ interchanged.

\paragraph{A comparison with the literature.} We verify our fourth-order result for the inverse localization length with the literature in the following special case.  Consider the case of identical potentials with disorder only in the separations $a_n$ between the centers of the potentials; that is, set all $b_n=0$ and $V_n(x) \equiv V(x)$, so that all $R_n\equiv R$, all $r_n'\equiv r'= \sqrt{R}e^{i\hat{\phi}_{r'}}$, and all $r_n = \sqrt{R}e^{i\hat{\phi}_r} e^{2 i k a_n}$ (where we recall that the hat indicates the variable for the single-site scattering problem).  For notational convenience, define (for $j=1,2$)
\beq
    \tilde{a}_j = e^{ji (\hat{\phi}_{r} + \hat{\phi}_{r'})} \langle e^{2j i k a_n)}\rangle_n.
\eeq
Then Eq.~\eqref{eq:lloc to 4th order} yields
\begin{multline}
    \frac{2}{\lloc} =  \left( 1 - 2\text{Re}\left[\frac{\tilde{a}_1}{1+\tilde{a}_1} \right] \right)R\\
    + \left( \frac{1}{2} + \text{Re}\left[ \frac{\tilde{a}_2(1-\tilde{a}_1^2)+2\tilde{a}_1(\tilde{a}_1-\tilde{a}_2)}{(1+\tilde{a}_1)^2(\tilde{a}_2-1)}   \right] \right)R^2 \\
    + O(R^3),\label{eq:lloc PARS constant potential}
\end{multline}
in agreement with a result by Lambert and Thorpe~\cite{LambertRandom1983} (see Appendix~\ref{sec:Comparison with Lambert and Thorpe} for details).

A notable feature of Eq.~\eqref{eq:lloc PARS constant potential}, pointed out in Ref.~\cite{LambertRandom1983} for the case of a delta-function model, is that the inverse localization length can be non-monotonic in the strength of phase disorder (i.e., the disorder in $a_n$). 

\paragraph{SPS relation.}
Returning to the general case, we use the usual Born series to obtain the single-site scattering amplitudes in terms of the Fourier transform of the potential [$\widetilde{V}_n(q) = \int dx\ e^{-i q x}V_n(x)$]:
\bseq
\begin{align}
    r_n &= -\frac{im}{k}\widetilde{V}_{n}(2k)^*e^{ik(2a_n+b_n)} + O(V^2),\\
    r_n' &= -\frac{im}{k}\widetilde{V}_n(2k)e^{ikb_n} + O(V^2).
\end{align}
\eseq
Here and in the following we are really expanding in some site-independent potential strength $V$ that all $V_n$ are proportional to.  The dimensionless small parameter is of order $m k V/x_\text{max}$, but for short we continue to refer to an expansion in $V$.

Note that the reflection strength $|r_n|$ starts at linear order in $V$ but also has quadratic and higher corrections from higher terms in the Born series.  Equations~\eqref{eq:lloc to 2nd order} and~\eqref{eq:sigma to 2nd order} then show that the inverse localization length and variance each start at second order ($V^2$) and have third-order ($V^3$) corrections that must be equal, due to the absence of $|r_n|^3$ terms in~\eqref{eq:lloc to 2nd order} and~\eqref{eq:sigma to 2nd order}.  Thus, even without calculating the $V^3$ term explicitly, we have obtained
\beq
    \lim_{N\to\infty}\frac{\sigma(N)^2}{2N}= \frac{2}{\lloc}+O(V^4).\label{eq:SPS relation PARS}
\eeq
Equation~\eqref{eq:SPS relation PARS} thus demonstrates the SPS relation~\eqref{eq:SPS relation in terms of lloc} for a broad class of periodic-on-average random systems up to the first two (generally non-vanishing) orders of the potential strength, $V^2$ and $V^3$.  Let us note that Eq.~\eqref{eq:SPS relation PARS} at the leading order [i.e., with the error term as $O(V^3)$] follows straightforwardly from the result of Schrader \textit{et al}.~\cite{SchraderPerturbative2004}; our work rules out $O(|r_n|^3)$ terms in Eqs.~\eqref{eq:lloc to 2nd order} and~\eqref{eq:sigma to 2nd order}, which in turn rules out any $V^3$ term in Eq.~\eqref{eq:SPS relation PARS}.

As a special case, we may apply Eq.~\eqref{eq:SPS relation PARS} to a model used by Deych \textit{et al}. in Refs.~\cite{DeychSingle2000, DeychSingleparameter2001} to provide numerical support for their approach to the SPS relation.  The model, which is defined in more detail in Ref.~\cite{DeychStatistics1998}, has classical light passing through alternating regions with dielectric constants $\epsilon_A$ and $\epsilon_B$; the $A$ regions have fixed widths and the $B$ regions have i.i.d. disordered widths.  This problem maps exactly to a quantum problem with square-well potentials, as we now explain.  While Refs.~\cite{DeychSingle2000, DeychSingleparameter2001, DeychStatistics1998} used a position-space transfer matrix setup, we instead use a scattering setup in which the ``leads'' to the left and right of the sample are $A$ regions.    We recall that the indices of refraction in the two regions are $n_\alpha= \sqrt{\epsilon_\alpha/\epsilon_0}$ ($\alpha=A,B$), and we set $\epsilon_A=\epsilon_0$ without loss of generality.  Then this classical problem is equivalent to a quantum scattering problem of the form~\eqref{eq:H PARS} with all $a_n\equiv a$, $V_n(x) = V \Theta(\frac{1}{2}b_n-|x|)$ \footnote{This model provides an example of correlations between the parameters of $\mathbf{D}_n$, which are allowed in our setup.  In particular, the parameters $\hat{\mathbf{D}} = (V, w)$ of the square well potential $V_{\hat{\mathbf{D}}}(x) = V \Theta(\frac{1}{2}w -|x|)$ are first promoted to site-dependent parameters $\hat{\mathbf{D}}_n = (V_n,w_n)$.  We then set all $V_n\equiv V$ and $w_n=b_n$; in $\mathbf{D}_n=(V_n,w_n,a_n,b_n)$, the potential width $w_n$ is thus correlated with the width $b_n$.}, and $V= -\frac{k^2}{2m}(n_B^2 -1)$ (the mass $m$ is arbitrary).  Thus, Eq.~\eqref{eq:SPS relation PARS} yields the SPS relation in the regime of the two dielectric constants $\epsilon_A$ and $\epsilon_B$ being close to each other, with arbitrary disorder strength in the widths $b_n$.

\paragraph{Inverse localization length at leading order.} At the leading order in the potential strength, Eq.~\eqref{eq:lloc to 2nd order} yields
\begin{multline}
    \frac{2}{\lloc}= \left(\frac{m}{k}\right)^2 \biggr( \langle |\widetilde{V}_n(2k)|^2\rangle_n \\
    + 2 \text{Re}\left[\frac{\langle\widetilde{V}_n(2k)e^{ik b_n}\rangle_n \langle\widetilde{V}_n(2k)^*e^{ik(2a_n+b_n)}\rangle_n }{1-\langle e^{2ik(a_n+ b_n)} \rangle_n} \right]\biggr) + O(V^3).\label{eq:lloc PARS}
\end{multline}

In the case of equal spacings between the centers of the potentials, Eq.~\eqref{eq:lloc PARS} simplifies to yield the variance of $\frac{m}{k}\widetilde{V}_n(2k)$.  [A similar result is known for continuous random potentials; see, e.g., Eq. (2.84) of Ref.~\cite{IzrailevAnomalous2012}.]  Indeed, setting all $a_n= a$ and $b_n=0$ yields
\beq
    \frac{2}{\lloc} = \left(\frac{m}{k}\right)^2 \left(\langle |\widetilde{V}_n(2k)|^2\rangle_n - |\langle\widetilde{V}_n(2k)\rangle_n|^2 \right).\label{eq:lloc PARS equal spacings}
\eeq
The first term is the only one present when the uniform phase hypothesis holds.  Thus, we disagree with the expectation in Ref.~\cite{StonePhase1983} that the uniform phase hypothesis [and thus Eq.~\eqref{eq:lloc AndersonNew1980}] should hold for any potential that is positive as often as it is negative \footnote{For a definite counterexample, consider $V_n(x) = c_n\delta(x)$, i.e., an array of delta function potentials with random strengths.  If $\langle c_n \rangle_n\ne 0$, then Eq.~\eqref{eq:lloc AndersonNew1980} is not correct at leading order in weak disorder, even if $\langle \sgn c_n \rangle_n=0$.}.

The next three orders in $V$ in the inverse localization length may be obtained from Eq.~\eqref{eq:lloc to 4th order} and from the next order in the Born series for the single-site reflection amplitudes, but we have not done this calculation.  

\paragraph{Transparent mirror effect.}
As a particular application of our results, we consider an array of square-well potentials with equal strengths and with spacings and widths that are disordered independently (Fig.~\ref{subfig:PARS_diagram_correlated}).  This model is equivalent to a classical optics problem in which light scatters on a disordered Bragg grating (i.e., a chain of dielectric slabs); the strong reflection that occurs due to localization, even if the individual dielectrics have weak reflection coefficients, is known in that context as the transparent mirror effect~\cite{BerryTransparent1997}.  This Kronig-Penney--type model has been studied by a number of approaches (e.g., Refs.~\cite{LifshitsIntroduction1988,DeychStatistics1998,Luna-AcostaOne2009,MafiAnderson2015,UpadhyayaDisorder2018}), and the model used by Deych \textit{et al}. is a particular case.  The transparent mirror effect has recently been studied in twist-angle-disordered bilayer graphene in Ref.~\cite{JoyTransparent2020}, although it seems that our results do not apply to the model considered there because the disorder couples neighboring S matrices.

We thus consider the scattering problem for the quantum Hamiltonian~\eqref{eq:H PARS} with $V_n(x) = V \Theta(\frac{1}{2}b_n-|x|)$.  Let us recall the correspondence to the classical optics problem.  A free quantum particle with momentum $k$ has its momentum change to $k' = \sqrt{k^2 -2m V}$ in a region of constant potential $V$, while for light passing through a region with constant index of refraction $\tilde{n}$ (we use a tilde to avoid confusion with the site index $n$) we instead have the free momentum $k$ changing to $k '=\tilde{n}k$.   We thus read off the correspondence $-\frac{2m V}{k^2} = \tilde{n}^2 -1$.  We cover both cases by defining
\bseq
\begin{align}
    \delta &= \frac{mV}{k^2} = -\frac{ \tilde{n}^2 -1}{2},\label{eq:delta} \\
    k' &= k\sqrt{1-2\delta},\label{eq:kprime}
\end{align}
\eseq
and we assume $\delta<1/2$ throughout (which corresponds to real momentum $k'$).

Our results apply to the more general case in which $\delta$ (in addition to $a_n$ and $b_n$) is also disordered; the only requirement is that the disorder parameters $\mathbf{D}_n\equiv(\delta_n,a_n,b_n)$ are i.i.d.  For simplicity, we are focusing on the case of $\delta_n\equiv \delta$ and independent disorder in $a_n$ and $b_n$.

To use our general results, we first recall the reflection amplitudes for scattering on a rectangular potential of strength $V$ (or, equivalently, an index of refraction $\tilde{n}$) from $x=-\frac{1}{2}b_n$ to $\frac{1}{2}b_n$.  The reflection amplitudes and reflection coefficient are
\bseq
\begin{align}
    \hat{r}_{\hat{\mathbf{D}}_n} &= \hat{r}_{\hat{\mathbf{D}}_n}' \\
    &= \frac{-i \frac{k^2-k'^2}{2 k k'} \sin(k' b_n) e^{-ikb_n}}{\cos(k' b_n) - i\frac{k^2 + k'^2}{2k k'} \sin(k'b_n)},\label{eq:rHat rectangular well}
\end{align}
\eseq
and
\beq
    \hat{R}_{\hat{\mathbf{D}}_n} = \frac{\left(\frac{k^2-k'^2}{2 k k'}\right)^2 \sin^2(k'b_n)}{1+ \left(\frac{k^2-k'^2}{2 k k'}\right)^2\sin^2(k'b_n)}.\label{eq:RHat rectangular well}  
\eeq
When we expand these terms in small $\delta$ using Eq.~\eqref{eq:kprime}, we do not expand the term $k' b_n$ that appears in $\cos$ and $\sin$ for two reasons: (a) this makes a symmetry property (that we discuss below) more manifest, and (b) this expansion can fail to commute with the strong disorder limit in some cases (e.g., a flat disorder distribution for $b_n$).

Equations~\eqref{eq:lloc to 2nd order},~\eqref{eq:rn in terms of hatrn}, and~\eqref{eq:rnp in terms of hatrnp} then yield our main result for the transparent mirror problem:
\begin{multline}
    \frac{2}{\lloc} = \text{Re}\left[ \frac{(1-\langle e^{2ika_n}\rangle_n)(1- \langle e^{2ik'b_n}\rangle_n)}{1-\langle e^{2ika_n}\rangle_n \langle e^{2ik'b_n}\rangle_n}\right]\left(\frac{1}{2}\delta^2 + \delta^3\right)\\ + O(\delta^4).\label{eq:lloc transparent mirror}
\end{multline}
The distinctive feature of this result compared to prior work is that the spacings $a_n$ and widths $b_n$ can be disordered arbitrarily.  While the $\delta^3$ term has the same dependence on disorder as the $\delta^2$ term, the $\delta^4$ and $\delta^5$ terms [which may be obtained from the higher-order result~\eqref{eq:lloc to 4th order}] have different dependence on disorder; we omit these lengthy expressions.

Equation~\eqref{eq:lloc transparent mirror} satisfies a consistency check based on symmetry, as we now summarize (see Appendix~\ref{sec:Symmetry property} for details).  By considering an alternate scattering problem in which the leads are $b$ regions and the scatterers are $a$ regions, we show that each order in $\delta$ of $2/\lloc$ must be symmetric under the exchange
\beq
    k \leftrightarrow k',\ a_n \leftrightarrow b_n,\label{eq:exchange transparent mirror}
\eeq
which indeed is true for Eq.~\eqref{eq:lloc transparent mirror}.  We have verified that the $\delta^4$ and $\delta^5$ corrections also satisfy this symmetry.  Another consistency check is that we obtain $2\lloc= O(\delta^6)$ in the case of no disorder.

We have also compared with some analytical results from the literature, as we now summarize (see Appendix~\ref{sec:Comparison with the literature} for details).  Reference~\cite{BerryTransparent1997} obtains the uniform phase result~\eqref{eq:lloc AndersonNew1980} in the case of strong disorder in $a_n$ [c.f. Eq.~\eqref{eq:rn in terms of hatrn} and the discussion below there for how we obtain the same result].  Reference~\cite{Luna-AcostaOne2009} treats $\delta$ exactly, with weak disorder in $a_n$ and no disorder $b_n$, and we find that their result agrees with Eq.~\eqref{eq:lloc transparent mirror} in the regime of overlap.  Finally, in Ref.~\cite{LifshitsIntroduction1988}, $a_n$ and $b_n$ are considered to follow exponential distributions, with arbitrary disorder strength, and $\delta$ is treated at the leading order.  The result is of the same form as what we obtain from Eq.~\eqref{eq:lloc transparent mirror}, but seems to have different numerical factors.

Using Eq.~\eqref{eq:lloc transparent mirror}, we can show analytically that the inverse localization length can have non-monotonic dependence on disorder strength for some choices of disorder, e.g., both $a_n$ and $b_n$ uniformly distributed in an interval $[0,W]$ with varying disorder strength $W$.  The special case of all $b_n\equiv b$ can alternatively be treated using Eq.~\eqref{eq:lloc PARS constant potential} (note that one must replace $a_n\to a_n-b$ there to account for the finite width of the scatterers).  Then, a uniform distribution in $a_n$ (for example) exhibits non-monotonicity, as is also clear from the numerical results of Ref.~\cite{MafiAnderson2015}.

\subsection{Application to discrete-time quantum walks}\label{sec:Application to discrete-time quantum walks}
We consider a general two-component, single-step quantum walk in one dimension.  We explore the effect of phase disorder on the localization length using Eq.~\eqref{eq:lloc to 4th order}, showing in particular that the dependence on disorder strength can be non-monotonic.  We verify our results with the literature in the limits of weak and strong phase disorder and also with numerics.

The setup is an infinite chain with site index $n$ and an internal ``spin'' degree of freedom ($\uparrow$ or $\downarrow$).  The ``shift'' operator $\hat{S}$ moves the two spins one step in opposite directions:
\beq
    \hat{S} = \sum_n \left( \ket{n+1 ,\uparrow} \bra{n, \uparrow} + \ket{n-1,\downarrow}\bra{n,\downarrow} \right). 
\eeq
The unitary operator $\hat{U}$ that implements a single time step is $\hat{U} = \hat{S}\hat{U}_\text{coin}$, where the ``coin'' operator $\hat{U}_\text{coin}$ acts as a unitary matrix on the spin degrees of freedom at each site.  We focus on the case of a coin operator that acts on each site independently:
\beq
    \hat{U}_\text{coin} = \sum_n \ket{n}\bra{n}\otimes U_{\text{coin},n},
\eeq
where $U_{\text{coin},n}$ is a $2\times2$ unitary matrix.  Following Vakulchyk \textit{et al}.~\cite{ VakulchykAnderson2017}, we parametrize the general coin matrix as
\beq
    U_{\text{coin},n} =   e^{i\varphi_n}
    \bpmat
       e^{i\varphi_{1,n}} \cos \theta_n & e^{i\varphi_{2,n}} \sin\theta_n \\
       -e^{-i\varphi_{2,n}}\sin\theta_n & e^{-i\varphi_{1,n}}\cos\theta_n
    \epmat,
\eeq
where the site-dependent (and possibly disordered) phases $\mathbf{D}_n\equiv (\varphi_n$, $\varphi_{1,n}$, $\varphi_{2,n}, \theta_n)$ may be interpreted as potential energy, external and internal synthetic fluxes, and kinetic energy, respectively~\cite{ VakulchykAnderson2017}.

The stationary state equation $\hat{U}\ket{\Psi} = e^{-i \omega}\ket{\Psi}$ may be brought to a $2\times2$ transfer matrix form even though a $4\times4$ transfer matrix would be expected (for a bipartite lattice with nearest-neighbor coupling)~\cite{ VakulchykAnderson2017}.  Indeed, writing a general state as $\ket{\Psi} = \sum_n \left( \Psi_{\uparrow}(n)\ket{n,\uparrow} + \Psi_{\downarrow}(n) \ket{n,\downarrow} \right)$ and defining a two-component wavefunction $\Psi(n) = (\Psi_{\uparrow}(n),  \Psi_{\downarrow}(n-1))$ (note that the $\downarrow$ component is offset by $1$ unit), one finds that the stationary state equation is equivalent to
\beq
    \Psi(n+1) = 
    \mathcal{M}_n
    \Psi(n),
\eeq
where
\beq
    \mathcal{M}_n = e^{i\varphi_{1,n}}
    \bpmat
        e^{i (\omega+ \varphi_n)} \sec \theta_n & e^{i \varphi_{2,n}}\tan \theta_n\\
        e^{- i \varphi_{2,n}} \tan \theta_n & e^{-i (\omega+\varphi_n)}\sec \theta_n
    \epmat\label{eq:transfer matrix DTQW}
\eeq
is the transfer matrix~\cite{ VakulchykAnderson2017}.

To bring the problem into a scattering framework, we define a disordered ``sample'' occupying sites $n=1,\dots,N$ of the chain by taking $\mathbf{D}_1,\dots,\mathbf{D}_N$ to have i.i.d. disorder, possibly including correlations between the phases $\varphi_n$, $\varphi_{1,n}$, $\varphi_{2,n}$, and $\theta_n$ at a given site $n$.  We will see below that in order to be in the weak reflection regime captured by our general calculation, we must take $\sin\theta_n$ to be small, which corresponds to a highly biased coin.  The remaining phases $\varphi_n$, $\varphi_{1,n}$, and $\varphi_{2,n}$ are arbitrary, and in particular we can explore the crossover between weak and strong disorder in these phases.

Given the disordered sample as we have defined it above, there are many possible scattering problems corresponding to different choices for a site-independent array $\mathbf{D}_\text{leads}$ to be assigned to $\mathbf{D}_n$ in the ``leads'' (i.e., $\mathbf{D}_n = \mathbf{D}_\text{leads}$ for $n\le 0$ and for $n\ge N+1$).  We find it convenient to set $\mathbf{D}_\text{leads}=\mathbf{0}$, and we refer to this setup as ``Tarasinski leads,'' since the same is done by Tarasinski \textit{et al}. in Ref.~\cite{TarasinskiScattering2014}.  We show in Appendix~\ref{sec:Tarasinski leads and their equivalence to other leads} that for samples long enough to be in the localized regime, other choices of leads result in the same probability distribution of $-\ln T$ (in particular the localization length is independent of the choice of leads).

We proceed to set up the scattering problem with Tarasinski leads, which have a linear quasienergy spectrum.  A scattering solution to the stationary state equation may be written outside the sample as (see Appendix~\ref{sec:Tarasinski leads and their equivalence to other leads} for details)
\beq
    \Psi(n) =
    \begin{cases}
    \bpmat \Psi_L^+ e^{ik (n-1)} \\ \Psi_L^- e^{-ik (n-1)} \epmat  & n \le 1,\\
    \\
    \bpmat \Psi_R^+ e^{ik (n-1-N)} \\ \Psi_R^- e^{-ik (n-1-N)} \epmat & n \ge N+1,
    \end{cases}\label{eq:scattering wavefn DTQW Tarasinski leads}
\eeq
where $k=\omega$.  With this choice of phase convention, we have
\beq
    \bpmat
        \Psi_L^+\\
        \Psi_L^-
    \epmat
    = \Psi(1),\qquad
    \bpmat
        \Psi_R^+\\
        \Psi_R^-
    \epmat
    = \Psi(N+1),
\eeq
and so,
\beq
    \bpmat
        \Psi_R^+\\
        \Psi_R^-
    \epmat
    = \mathcal{M}_N\dots\mathcal{M}_1
    \bpmat
        \Psi_L^+\\
        \Psi_L^-
    \epmat.
\eeq
We thus obtain the scattering transfer matrix of the sample as $\mathcal{T}_{1\dots N}=\mathcal{T}_N\dots\mathcal{T}_1$ with $\mathcal{T}_n=\mathcal{M}_n$.  It may be verified that $\mathcal{M}_n$ satisfies the psuedo-unitarity condition $\mathcal{M}_n^\dagger \sigma^z \mathcal{M}_n = \sigma^z$ (which confirms that $\mathcal{M}_n$ is a valid scattering transfer matrix).  The corresponding $S$ matrix is $\mathcal{S}_n= e^{i \omega}U_{\text{coin},n}$, i.e.,
\beq
    \mathcal{S}_n = e^{i(\omega+\varphi_n)}
    \bpmat
        e^{i \varphi_{1,n}} \cos\theta_n & e^{i \varphi_{2,n}} \sin \theta_n \\
        -e^{-i\varphi_{2,n}} \sin\theta_n & e^{-i\varphi_{1,n}}\cos\theta_n
    \epmat.\label{eq:S matrix DTQW}
\eeq
We can thus apply our results to the problem of $N$ scatterers with the single-site $S$ matrix given by~\eqref{eq:S matrix DTQW}.  [Indeed, since any $2\times2$ unitary $S$ matrix can be parametrized as in Eq.~\eqref{eq:S matrix DTQW}, the DTQW we are considering in fact represents the most general problem that our results apply to.]

From Eq.~\eqref{eq:S matrix DTQW}, we read off the single-site reflection amplitudes and reflection coefficient.  The expansion parameter in our formalism is thus $|r_n| = \sin\theta_n$.  The generic condition~\eqref{eq:generic condition} reads
\beq
    \langle e^{2i\ell(\omega+\varphi_n)}\rangle_n \ne 1,\label{eq:generic condition DTQW}
\eeq
which indeed holds for all non-zero integers $\ell$ except in the special case that $\varphi_n$ is non-disordered and $\omega$ is a rational multiple of $\pi$.  Although we ignore this special case from now on, our explicit results below are still valid there unless the rational multiple is of a particular form (see Appendix~\ref{sec:Partial extension of our results to anomalies}).

Up to second order, Eqs.~\eqref{eq:lloc to 2nd order} and~\eqref{eq:sigma to 2nd order} yield the inverse localization length and variance:
\bseq
\begin{align}
    &\frac{2}{\lloc} = \langle \sin^2\theta_n\rangle_n\notag\\
    &+ 2 \text{Re}\left[e^{2i\omega}\frac{\langle e^{i(\varphi_n+\varphi_{2,n})}\sin\theta_n\rangle_n \langle e^{i(\varphi_n-\varphi_{2,n})}\sin\theta_n\rangle_n}{1 -e^{2i\omega}\langle e^{2i\varphi_n}\rangle_n} \right]\notag\\
    &\qquad \qquad +O(\sin^4\theta_n)\label{eq:lloc to 2nd order DTQW general}\\
    &\qquad =\lim_{N\to\infty}\frac{\sigma(N)^2}{2N} + O(\sin^4\theta_n).\label{eq:sigma to 2nd order DTQW general}
\end{align}
\eseq
To present the inverse localization length to fourth order in a compact form, we use the notation
\beq
    B_{a,b}^{(j)} = \langle e^{i(a\varphi_n + b \varphi_{2,n})}\sin^j\theta_n\rangle_n e^{i a\omega},\label{eq:B DTQW general}
\eeq
and we note that $\alpha_m= (1- B_{2m,0}^{(0)})^{-1}$.  Then Eq.~\eqref{eq:lloc to 4th order} yields our main result for the DTQW:
\begin{widetext}
\begin{multline}
    \frac{2}{\lloc} = \langle\sin^2\theta_n\rangle_n + 2 \text{Re}\left[\alpha_1 B_{1,-1}^{(1)}B_{1,1}^{(1)} \right]
    +\frac{1}{2}\langle\sin^4\theta_n\rangle_n\\
    -  \text{Re}\biggr\{\alpha_2 \left(B_{2,-2}^{(2)} + 2\alpha_1 B_{1,-1}^{(1)}B_{3,-1}^{(1)} \right)\left(B_{2,2}^{(2)}+2\alpha_1 B_{1,1}^{(1)}B_{3,1}^{(1)}\right) + 2\alpha_1^2 B_{1,-1}^{(1)}B_{1,1}^{(1)}B_{2,0}^{(2)} \biggr\}+ O(\sin^6\theta_n),\label{eq:lloc to 4th order DTQW general}
\end{multline}
\end{widetext}
in which the first two terms recapitulate Eq.~\eqref{eq:lloc to 2nd order DTQW general}.  Note that the external synthetic flux $\varphi_{1,n}$ does not appear (as noted by Ref.~\cite{ VakulchykAnderson2017}).

Next, we specialize Eq.~\eqref{eq:lloc to 4th order DTQW general} to various choices of disorder, checking our results with the literature and pointing out cases in which the localization length depends non-monotonically on disorder strength.  For each choice of disorder, we only need to calculate the coefficients $B_{a,b}^{(j)}$ defined in Eq.~\eqref{eq:B DTQW general}.  Throughout, we present only the localization length, bearing in mind that the variance may be obtained at the leading order from Eq.~\eqref{eq:sigma to 2nd order DTQW general}.   
\subsubsection{Disorder in individual phases.}
Here, we follow Ref.~\cite{ VakulchykAnderson2017}, introducing disorder in one phase variable at a time.

Consider first the case of disorder in $\varphi_n$ only, with $\theta_n\equiv \theta$ and $\varphi_{1,n}=\varphi_{2,n}=0$.  We obtain
\bseq
\begin{align}
    B_{a,b}^{(j)} &= \langle e^{ia \varphi_n}\rangle_ne^{ia \omega} \sin^j\theta \\
    &= \sinc(aW)e^{ia \omega}\sin^j \theta,
\end{align}
\eseq
where the second line specializes to a flat disorder distribution of width $W$ (i.e., $\varphi_n$ uniform in $[-W,W]$), and where $\sinc x = (\sin x)/x$.

We now check our result~\eqref{eq:lloc to 4th order DTQW general} with the literature in the limits of weak and strong disorder.  For weak disorder with $\langle \varphi_n\rangle_n=0$, we obtain $2/\lloc =[ \cot^2 \omega \sin^2\theta + (\csc^4\omega -1)\sin^4 \theta ]\langle \varphi_n^2\rangle_n + O(\sin^6\theta)$, which agrees with the small $\sin\theta$ expansion of a result from Ref.~\cite{ VakulchykAnderson2017} [see their Eq. (26), recall the dispersion relation $\cos \omega= \cos\theta \cos k$, and note that $\langle\varphi_n^2\rangle_n = W^2/3$ for the flat disorder distribution] except for an overall constant factor of $2$ \footnote{In the special case of weak disorder in $\varphi_{2,n}$, Ref.~\cite{ VakulchykAnderson2017} mentions a factor of $2$ discrepancy between their analytical and numerical results for $\lloc$ [see below their Eq. (29)], and it is in the right direction to agree with our result.  We believe that this factor of $2$ is also present in their analytical calculations for weak disorder in $\varphi_n$ or in $\theta_n$, as removing it restores agreement with our results in those cases as well.  The factor of $2$ that appears in our convention of presenting the localization length in the form $2/\lloc$ has been accounted for in this discussion and is not the source of the discrepancy.}.  In the strong disorder limit (i.e., the flat distribution with $W=\pi$), we get $2/\lloc = \sin^2\theta+\frac{1}{2}\sin^4\theta + O(\sin^6\theta)$, in agreement with the small $\sin\theta$ expansion of another result from Ref.~\cite{VakulchykAnderson2017}, namely [their Eq. (48)]
\beq
    \lloc = 1/|\ln\cos\theta|.\label{eq:lloc DTQW uniform phase}
\eeq
Equation~\eqref{eq:lloc DTQW uniform phase} (for any $\theta$) may also be obtained from our Eqs.~\eqref{eq:lloc in terms of Fourier coefficients} and~\eqref{eq:S matrix DTQW}; in our setup, this is a case in which the uniform phase result~\eqref{eq:lloc AndersonNew1980} holds because the local reflection phase is uniformly distributed independently of the local reflection coefficient.  In passing, we note that if there is disorder in $\theta_n$ as well (independent of the strong disorder in $\varphi_n$), then Eq.~\eqref{eq:lloc DTQW uniform phase} generalizes to $\lloc = 1/\langle |\ln\cos\theta_n|\rangle_n$ (and indeed the same result is obtained if the strong disorder is in $\varphi_{2,n}$ instead of $\varphi_n$).

At the band center ($\omega=\pi/2$) anomaly noted by Ref.~\cite{VakulchykAnderson2017}, we obtain $2/\lloc = (1/45)[\sin^2\theta + 2 \sin^4\theta + O(\sin^6\theta) ]W^4$, whereas the expansion of the result of Ref.~\cite{VakulchykAnderson2017} for small $\sin\theta$ yields the same answer with prefactor $1/40$ [or $1/20$ if the factor of $2$~\cite{Note15} is corrected].  Unlike the Anderson model anomalies mentioned above, in this case we expect our result to apply, since our assumptions are met [i.e., localization occurs and the inequality~\eqref{eq:generic condition DTQW} holds].

\begin{widetext}
\begin{figure*}[t]
\subfloat[\label{subfig:small_coin_parameter}]{%
  \includegraphics[scale=0.46]{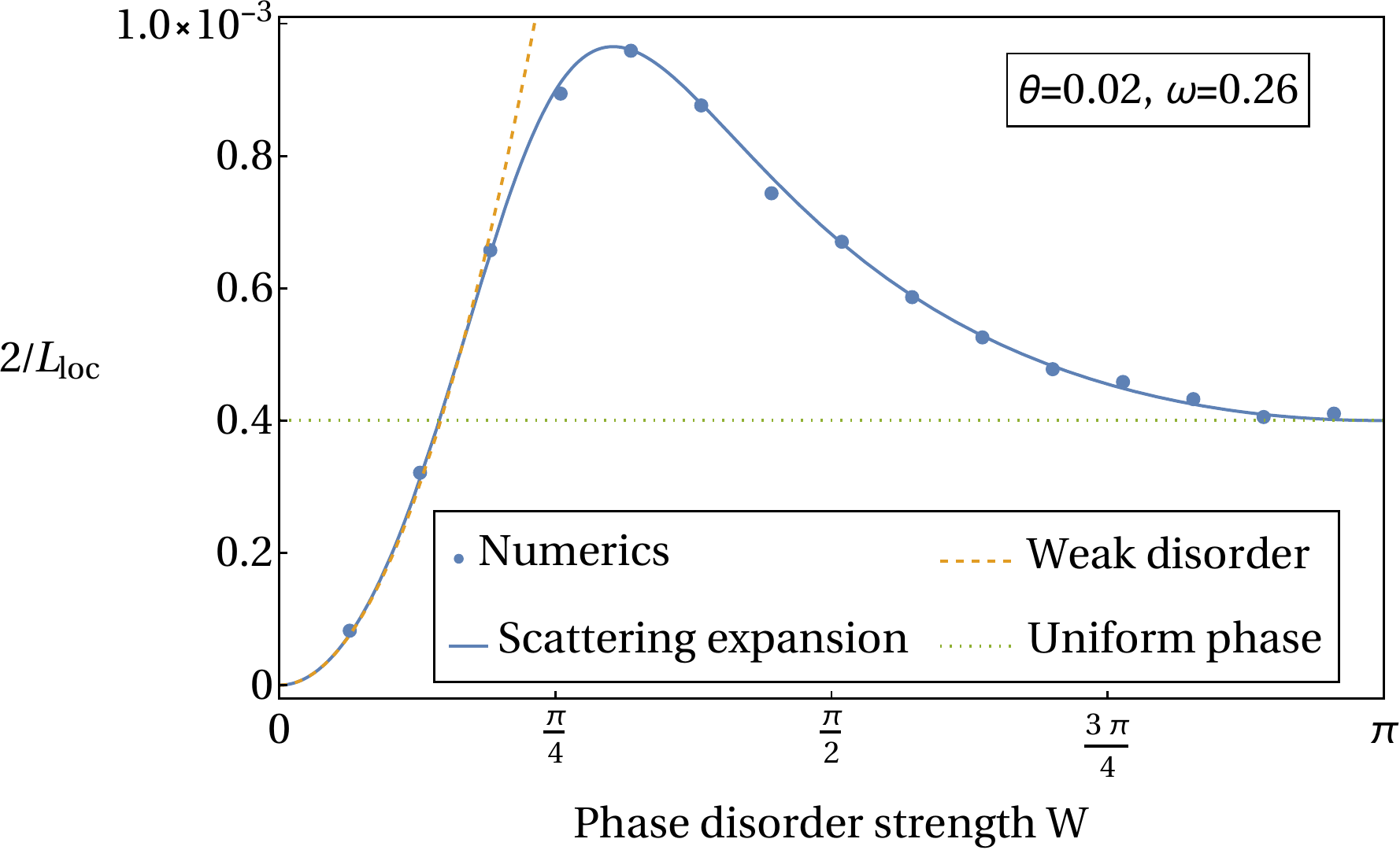}%
}
\subfloat[\label{subfig:larger_coin_parameters}]{%
  \includegraphics[scale=0.52]{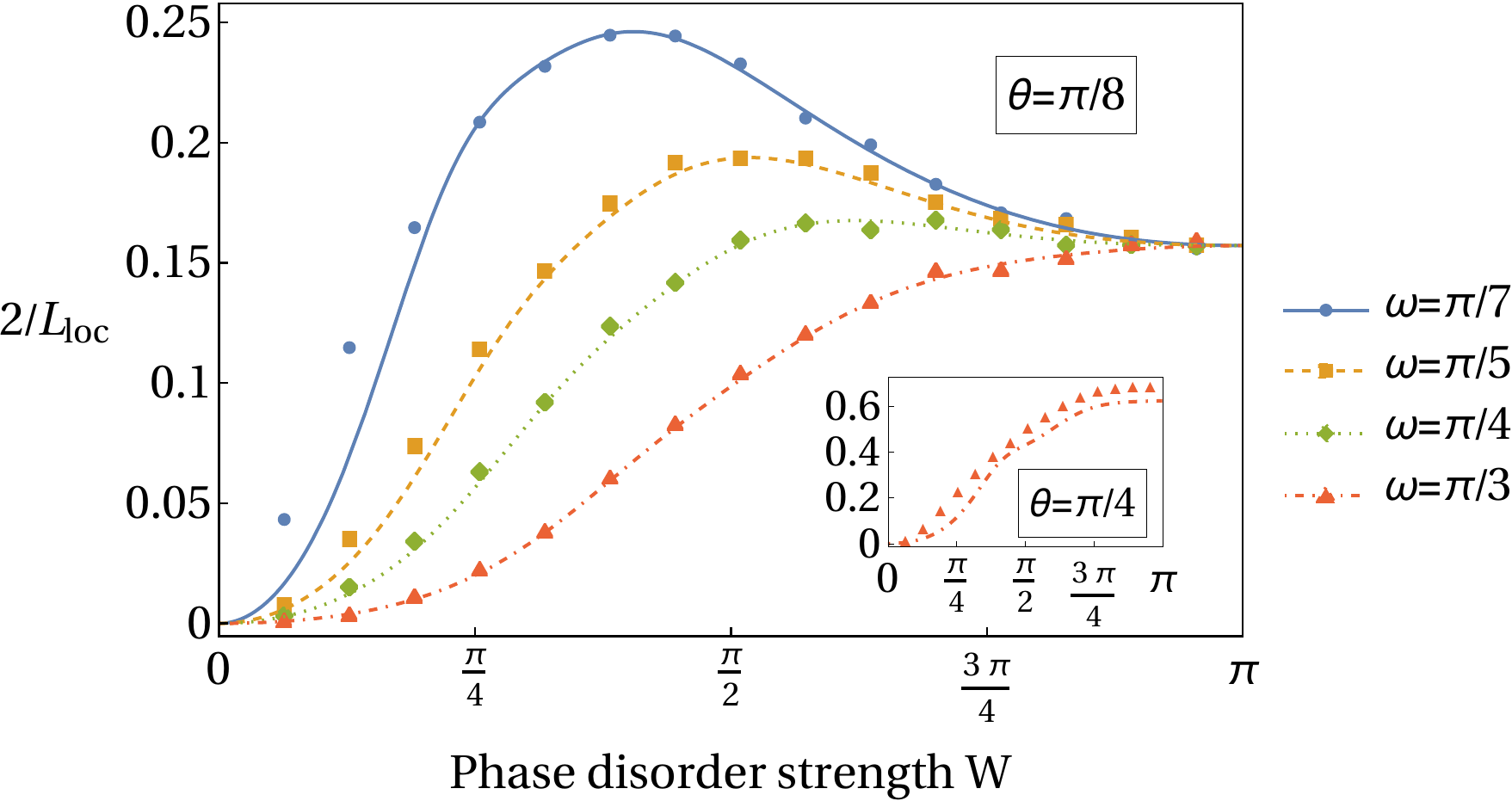}%
}
\caption{The inverse localization length ($2/\lloc$) vs. the strength of disorder ($W$) in the phase variable $\varphi_n$ in the DTQW.  All $\theta_n\equiv \theta$, $\varphi_{1,n}=0$, and $\varphi_{2,n}=0$.  Numerical points were obtained from the slope of a linear fit of $-\ln T_{1\dots N}$ vs. $N$ with $N_\text{max}/10 < N \le N_\text{max}$ in one disorder realization, where $N_\text{max}=10^7$ in (a) and $10^6$ in (b).  (a) A highly biased coin.    Solid line: theoretical leading order result~\eqref{eq:lloc varphi flat disorder} [the higher-order result Eq.~\eqref{eq:lloc to 4th order DTQW general} would be indistinguishable on the plot].  For comparison, we show the weak disorder expansion ($\frac{\cos^2\omega \tan^2\theta}{3(\cos^2\theta-\cos^2\omega)}W^2$, dashed line) from Ref.~\cite{VakulchykAnderson2017} and the prediction of the uniform phase hypothesis ($2|\ln\cos\theta|$, dot-dashed line), which holds at strong disorder ($W=\pi$) but not otherwise.  (b) Moderately biased coin parameter ($\theta = \pi/8$).  The next-to-leading order theoretical result~\eqref{eq:lloc to 4th order DTQW general} (lines) matches fairly well with numerics (points) except for small disorder near the band edge ($w\approx \theta$).  Inset to (b): A Hadamard coin ($\theta=\pi/4)$.}\label{fig:DTQW_plots}
\end{figure*}
\end{widetext}

Equation~\eqref{eq:lloc to 4th order DTQW general} thus interpolates, in the regime of strong coin bias, between the known limits of weak and strong disorder (see Fig.~\ref{subfig:small_coin_parameter}).  In certain ranges of quasienergy $\omega$, the dependence on the disorder strength $W$ is non-monotonic.  Although strictly speaking we have assumed $\sin\theta$ to be a small parameter, we find favorable agreement with numerics even for only a moderate amount of coin bias (e.g., $\theta=\pi/8$); see Fig.~\ref{subfig:larger_coin_parameters}.

For definiteness, we present the inverse localization length at second order [Eq.~\eqref{eq:lloc to 2nd order DTQW general}] with the flat distribution:
\begin{widetext}
\beq
    \frac{2}{\lloc} = \left( 1 + \frac{2\sinc^2W[\cos(2w)-\sinc(2W)]}{1-2\cos(2w)\sinc(2W)+\sinc^2(2W)}\right) \sin^2\theta + O(\sin^4\theta),\label{eq:lloc varphi flat disorder}
\eeq
\end{widetext}
which seems to qualitatively capture the non-monotonicity in $W$ [although including the $\sin^4\theta$ terms in Eq.~\eqref{eq:lloc to 4th order DTQW general} improves the agreement with numerics].

In the case of disorder in $\varphi_{2,n}$ only, with $\theta_n\equiv \theta$ and $\varphi_{n}=\varphi_{1,n}=0$, we obtain
\bseq
\begin{align}
    B_{a,b}^{(j)} &= \langle e^{i b\varphi_{2,n}} \rangle_n\sin^j\theta\\ &= \sinc(bW)\sin^j\theta,
\end{align}
\eseq
where the second line specializes to $\varphi_{2,n}$ uniform in $[-W,W]$.  In this case, the inverse localization length~\eqref{eq:lloc to 4th order DTQW general} depends on disorder monotonically, with a simple leading order expression $2/\lloc =\left( 1 - \sinc^2W\right)\sin^2 \theta + O(\sin^4\theta)$.  We again agree with Ref.~\cite{VakulchykAnderson2017} in the limits of weak and strong disorder once their results are expanded in $\sin \theta$.  Indeed, for weak disorder with $\langle \varphi_{2,n}\rangle_n=0$, we find $2/\lloc = \langle \varphi_n^2\rangle_n[\sin^2\theta + \sin^4\theta + O(\sin^6\theta)]$, in agreement with their Eq. (29)~\cite{Note15}.  For strong disorder, Ref.~\cite{VakulchykAnderson2017} again obtains Eq.~\eqref{eq:lloc DTQW uniform phase}, which we recover in the same sense as mentioned above.

Finally, we consider disorder in the coin parameter $\theta_n$, with $\varphi_n=\varphi_{1,n}=\varphi_{2,n}=0$.  In this case we can only access the weak disorder regime, since our expansion is in small $\sin \theta_n$.  We write $\theta_n = \theta_0 + \Delta\theta_n$, where $\Delta\theta_n$ is, e.g., uniformly distributed in $[-W,W]$.  We obtain $B_{a,b}^{(j)}= \langle \sin^j\theta_n\rangle_n$ and $2/\lloc = \langle(\Delta \theta_n)^2\rangle_n\left( 1 + \csc^2\omega\sin^2\theta_0\right) + O(\sin^6\theta_n)$, in agreement with Ref.~\cite{ VakulchykAnderson2017} [see their Eq. (31)~\cite{Note15}].

\subsubsection{Alternate form of phase disorder}
We now specialize to a case studied experimentally in Ref.~\cite{SchreiberDecoherence2011} and theoretically in Ref.~\cite{DerevyankoAnderson2018}.  Again we verify our results with the literature in the limits of weak and strong phase disorder, and then we explore the full range of disorder and find that the localization length in certain ranges of quasienergy is non-monotonic as a function of disorder strength.

The coin matrix in Ref.~\cite{DerevyankoAnderson2018} is, in our notation,
\beq
    U_{\text{coin},n}=
    \bpmat
        e^{i \phiup(n)} & 0 \\
        0 & e^{i \phidown(n)}
    \epmat
    \bpmat
        \cos \theta & e^{i\varphi'}\sin \theta \\
        - e^{-i \varphi'} \sin \theta & \cos \theta
    \epmat,\label{eq:coin matrix Derevyanko}
\eeq
where $\theta$, $\varphi'$ are constant phases and $\phiup(n)$, $\phidown(n)$ are disordered phases.  The coin matrix of Ref.~\cite{SchreiberDecoherence2011} is obtained by shifting $\phidown(n)\to\phidown(n)+\pi$ and $\theta\to2\theta$.

We specialize to the case~\eqref{eq:coin matrix Derevyanko} by setting $\theta_n \equiv \theta$, $\varphi_n = \frac{1}{2}[\phiup(n)+\phidown(n)]$, $\varphi_{1,n} =\frac{1}{2}[\phiup(n)-\phidown(n)]$, and $\varphi_{2,n}=\frac{1}{2}[\phiup(n)-\phidown(n)]+\varphi'$.  Then the inverse localization length up to error of sixth order in $\sin\theta$ is given by Eqs.~\eqref{eq:B DTQW general} and~\eqref{eq:lloc to 4th order DTQW general} with the above substitutions.

Our result agrees with Ref.~\cite{DerevyankoAnderson2018} in the regimes of overlap, as we now explain.  In the case of weak disorder with vanishing mean and no up-down correlation ($\langle\phiup(n)\rangle_n = \langle\phidown(n)\rangle_n = \langle\phiup(n)\phidown(n)\rangle_n=0$ and $\langle \phiup(n)^2\rangle_n=\langle \phidown(n)^2\rangle_n\equiv \langle\phi^2\rangle$), we obtain $\frac{2}{\lloc} = \frac{1}{2}\langle\phi^2\rangle\csc^2 \omega\sin^2\theta\left(1+\csc^2\omega\sin^2\theta \right) + O(\sin^6\theta)$, in agreement with the small $\sin\theta$ expansion of a result \footnote{See Eq. (11) of Ref.~\cite{DerevyankoAnderson2018}, where $\beta$ is our $\omega$ and $\lambda(\beta)$ is our $1/\lloc$.} from Ref.~\cite{DerevyankoAnderson2018}.  In the case of strong disorder, i.e., $\phi_\uparrow(n)$ and $\phi_\downarrow(n)$ independently and uniformly distributed in $(-\pi,\pi)$, we obtain $2/\lloc= \sin^2\theta+\frac{1}{2}\sin^4\theta + O(\sin^6\theta)$, in agreement with the small $\sin\theta$ expansion of another result from Ref.~\cite{DerevyankoAnderson2018}, namely Eq.~\eqref{eq:lloc DTQW uniform phase} [Eq. (10) there].  Eq.~\eqref{eq:lloc DTQW uniform phase} may also be obtained directly just as in the cases mentioned above of strong disorder in $\varphi_n$ or $\varphi_{2,n}$ (and has the same generalization to the case of independent disorder in $\theta$).

We thus interpolate, in the regime of small coin parameter, between the known limits of weak and strong phase disorder.  For definiteness, we present now the leading order result, first for general phase disorder and then for the case of $\phiup(n),\phidown(n)$ being independently and uniformly distributed in $[-W,W]$.  Equation~\eqref{eq:lloc to 2nd order DTQW general} yields
\begin{widetext}
\bseq
\begin{align}
    \frac{2}{\lloc} &= \left(1 + 2\text{Re}\left[ \frac{\langle e^{i\phiup(n)}\rangle_n\langle e^{i\phidown(n)}\rangle_n }{e^{-2i\omega} -\langle e^{i[\phiup(n)+\phidown(n)]}\rangle_n } \right] \right)\sin^2\theta + O(\sin^4\theta)\\
    &= \frac{1- \sinc^4 W}{ 1 - 2 \cos(2\omega)\sinc^2 W + \sinc^4W}\sin^2\theta + O(\sin^4\theta).\label{eq:lloc to 2nd order DTQW Derevyanko flat}
\end{align}
\eseq
\end{widetext}
Note that in the regime of $0 < \omega < \pi/4$ or $3\pi/4 < \omega < \pi$, there is a value of phase disorder strength $W = W_0$ beyond which any further increase in disorder strength (up to the maximum $W=\pi$) \emph{increases} the localization length.

\section{Joint probability distribution}\label{sec:Joint probability distribution}

We proceed to apply the same scattering expansion approach to the joint probability distribution $P_{N}(s,\phi_{r'})$.  We present and discuss the results first, then the analytical calculation and numerical checks.

\subsection{Overview of results}
Our main result is the following general form for sufficiently large sample length $N$: there is a constant $c$ and two functions $\hat{s}(\phi_{r'})$, $\eta(\phi_{r'})$ for which we have
\begin{multline}
    P_{N}(s,\phi_{r'}) =\\
    \frac{\exp\left\{-\frac{1}{2} \left[s- \frac{2}{\lloc}N - \hat{s}(\phi_{r'}) \right]^2/\sigma(N,\phi_{r'})^2\right\}}{\sqrt{2\pi \sigma(N,\phi_{r'})^2}}\\
    \times p_{\infty}(\phi_{r'}),\label{eq:joint prob dist final}
\end{multline}
where the phase-dependent variance $\sigma(N,\phi_{r'})^2$ grows linearly with $N$ and has a sub-leading, phase-dependent correction:
\beq
    \sigma(N,\phi_{r'})^2 = 2 N c + \eta(\phi_{r'}).
\eeq
For large $N$, the marginal distribution $P_{N}(s)$ determined by Eq.~\eqref{eq:joint prob dist final} takes the known Gaussian form:
\beq
    P_{N}(s) = \frac{\exp\left\{-\frac{1}{2} \left[s-\frac{2}{\lloc} N + O(N^0) \right]^2/\sigma(N)^2 \right\}}{\sqrt{2\pi \sigma(N)^2}} ,\label{eq:marginal dist of s}
\eeq
where the slope of the variance is determined by the same constant $c$:
\beq
    \sigma(N)^2 = 2 c N  + O(N^0).\label{eq:sigma in terms of b}
\eeq

Equation~\eqref{eq:joint prob dist final} may be understood as the statement that the conditional distribution of $s$ given $\phi_{r'}$ is Gaussian for large $N$.  Furthermore, the leading behavior of the mean and variance is linear in $N$ with slope independent of the phase.

We obtain Eq.~\eqref{eq:joint prob dist final} to all orders in the scattering expansion by a calculation described in the following section.  This calculation, which relies on some assumptions (that we discuss below), provides a procedure for calculating all parameters and functions that appear in Eq.~\eqref{eq:joint prob dist final} order by order in the scattering expansion (with coefficients only involving local averages), except that the functions $\hat{s}(\phi_{r'})$ and $\eta(\phi_{r'})$ each have a $\phi_{r'}$-independent additive constant that is not determined.   [These constants appear in the $O(N^0)$ error terms in Eqs.~\eqref{eq:marginal dist of s} and~\eqref{eq:sigma in terms of b}.]  The expansions for $2/\lloc$ and $p_{\infty}(\phi_{r'})$ have already been presented in Sec.~\ref{sec:Scattering expansion of the localization length}; here we develop similar recursive expansions for $c$, $\hat{s}(\phi_{r'})$, and $\eta(\phi_{r'})$.

A notable feature that we demonstrate for the function $\hat{s}(\phi_{r'})$ is that it coincides, at leading order and up to a factor of $2\pi$, with the first order correction to uniformity in the phase distribution $p_{\infty}(\phi_r')$.  In particular, we recall from~\eqref{eq:p to 3rd order} that the phase distribution is given at first order by $p_{\infty}(\phi_{r'}) = 1/(2\pi) + p_{\infty}^{(1)}(\phi_{r'})$, where
\beq
    2\pi p_{\infty}^{(1)}(\phi_{r'}) = 2\text{Re}\left[\frac{ \langle r_n'\rangle_n e^{-i\phi_{r'}} }{1+ \langle r_n r_n'/R_n\rangle_n} \right].\label{eq:p1 solution}
\eeq
This same function describes the leading-order correlations between the transmission coefficient and reflection phase, that is,
\beq
    \hat{s}^{(1)}(\phi_{r'}) = 2\pi p_{\infty}^{(1)}(\phi_{r'}) + \text{const.},\label{eq:s1 in terms of p1}
\eeq
where the constant is independent of $\phi_{r'}$.

In the framework of the scaling theory mentioned in the Introduction, our result for the joint distribution may be regarded as demonstrating three-parameter scaling in the limit of weak local scattering.  To see this, consider Eq.~\eqref{eq:joint prob dist final} with $2/\lloc$, $c$, $\hat{s}(\phi_{r'})$, and $p_{\infty}(\phi_{r'})$ each expanded to the first non-vanishing order and with the contribution from the function $\eta(\phi_{r'})$ dropped.  [Note that $\eta(\phi_{r'})$ is a $1/N$ correction relative to $\sigma(N)^2$, which itself is already a $1/N$ correction relative to the mean.  We find below that we must take the function $\eta(\phi_{r'})$ into account in the derivation of Eq.~\eqref{eq:joint prob dist final}, but it seems to be highly suppressed for large $N$ in the final answer.]  The joint probability distribution is then entirely determined by three real parameters: the mean $(2/\lloc)N$ [which determines $\sigma(N)^2 = 2 cN $ since the SPS relation holds at leading order] and the real and imaginary parts of the complex parameter $z \equiv \langle r_n'\rangle_n/(1+ \langle r_n r_n'/R_n\rangle_n)$ that determines the function $p_{\infty}^{(1)}(\phi_{r'})$.  These last two parameters may alternatively be taken to be the mean and variance of $\phi_{r'}$, since they are given at leading order by $\overline{\phi_{r'}} = 2 \Im(z)$ and $\overline{\phi_{r'}^2} - \overline{\phi_{r'}}^2 = \pi^2/3 - 4 \Re(z)$, and thus they determine $z$.

The correlation between $s$ and $\phi_{r'}$ in~\eqref{eq:joint prob dist final} is a finite-size effect, as we now explain.  We write the average of $s$ as $\langle s \rangle = 2N/\lloc + O(N^0)$, and we consider how accurate $\langle s\rangle$ is as an estimate of the conditional average of $s$ with fixed $\phi_{r'}$ in~\eqref{eq:joint prob dist final}.  The phase-dependent variation of the mean introduces a relative error of order $\hat{s}(\phi_{r'})/\langle s\rangle \sim1/N$, while the finite standard deviation introduces a relative error of $\sigma(N,\phi_{r'})/\langle s\rangle = c\lloc/\sqrt{N} + O(N^{-3/2})$, where the $N^{-3/2}$ term contains the contribution of the function $\eta(\phi_{r'})$.  Prior work has found the joint probability distribution to factorize into a transmission coefficient part times a phase part~\cite{RobertsJoint1992, PendrySymmetry1994}, in apparent contradiction to our Eq.~\eqref{eq:joint prob dist final}; this suggests that the prior work only accounted for the $1/\sqrt{N}$ term in the above discussion and neglected the $1/N$ and $N^{-3/2}$ terms that contain the  correlations between $s$ and $\phi_{r'}$.

\subsection{Analytical calculation}\label{sec:Calculation}
\subsubsection{Setup}
We arrive at our result~\eqref{eq:joint prob dist final} by verifying that it satisfies the recursion relation for the joint probability distribution (taking $N$ large and neglecting a small error term).  We do not address the question of the uniqueness of the solution. Although our approach is partially heuristic, we note that (a) the results can be checked numerically (see Fig.~\ref{fig:numerics_for_joint_dist}), (b) the result we get for the constant $c$ yields Eq.~\eqref{eq:sigma to 2nd order}, and (c) we obtain the correct probability distribution when we apply the same approach to a soluble toy model (Appendix~\ref{sec:Toy recursion relation}). 

Our task is to determine the joint probability distribution as defined by Eq.~\eqref{eq:joint prob dist def}.  From this definition (with $N$ and with $N+1$) and the recursion relations~\eqref{eq:s recursion Rto1} and~\eqref{eq:phirp recursion Rto1}, we obtain the following recursion in the localized regime:
\beq
    P_{N+1}(s,\phi_{r'}) = \mathcal{F}[s,\phi_{r'};\{ P_N\}],\label{eq:P recursion} 
\eeq
where $\mathcal{F}$ is a linear functional in its last argument and is defined by (here we replace the disorder average over site $N+1$ by any site $n$)
\begin{multline}
    \mathcal{F}[s,\phi_{r'};\{ P_N\}] = \int ds'\int_{-\pi}^\pi d\phi\ P_N\left(s', \phi\right)\\
    \times 
    \langle \delta\bpl s' + g_n(s')  -s \bpr\delta\bpl\phi + h_n(\phi) - \phi_{r'}\bpr\rangle_n.\label{eq:mathcalF def}
\end{multline}

Before going into further details, we first give an overview of what our calculation will show and what assumptions we make.  We start by emphasizing that Eq.~\eqref{eq:P recursion} holds in the localized regime, i.e., it holds for $N\ge N_0(\param)$, where $N_0(\param)$ is of the order of $\lloc$ [recall the discussion below Eq.~\eqref{eq:hn}].  There is also an exact recursion relation that holds for all $N$ and that coincides with Eq.~\eqref{eq:P recursion} for $N\ge N_0(\param)$.  This exact relation can be obtained in a similar way from the exact recursion relations~\eqref{eq:transmission coeff recursion} and~\eqref{eq:reflection phase recursion}.

The given onsite disorder distribution (of the matrix elements of $\mathcal{S}_n$) determines both the initial condition (the function $P_{N=1}$) and the exact recursion relation.  Iterating the exact recursion relation, one obtains $P_{N_0(\param)}$, which can then be regarded as the initial condition for~\eqref{eq:P recursion}.  However, due to the complexity of the exact recursion relation, we do not have any precise characterization of $P_{N_0(\param)}$.  Conceptually, we regard the problem as consisting of the recursion relation~\eqref{eq:P recursion} with some unknown initial condition $P_{N_0(\param)}$.

Let us call any family of functions $\{ P_N \}$, parametrized by $N=1,2,\dots$, a ``trajectory.'' Our approach is to define a family of trajectories that satisfy Eq.~\eqref{eq:P recursion} when $N$ is large (up to an error term that we expect to be negligible), in the hope that a ``generic'' initial condition $P_{N_0(\param)}$ will ``flow'' to being arbitrarily close to one of these trajectories after sufficiently many iterations of~\eqref{eq:P recursion}.

Below, we make an ansatz for this family of trajectories.  In particular, we define a function $P_N^{(\text{ansatz})}(s,\phi_{r'})$ which is proportional to $1/\sqrt{N}$ and normalized to $1$.  The ansatz is parametrized by two undetermined constants.  [These constants are the $\phi_{r'}$-independent additive constants referred to below Eq.~\eqref{eq:sigma in terms of b}, and they have no effect on the large-$N$ behavior.]  The main content of our calculation is the demonstration that, for large $N$ and to all orders in the scattering expansion,
\beq
    P_{ N+1}^{(\text{ansatz})}(s,\phi_{r'}) = \mathcal{F}[s,\phi_{r'}; \{ P_N^{(\text{ansatz})}\}] + O(1/N^2).\label{eq:joint dist recursion general main claim}
\eeq

We then make two assumptions: (1) that Eq.~\eqref{eq:joint dist recursion general main claim} implies that $P_N^{(\text{ansatz})}$ gets arbitrarily close, as $N\to\infty$, to a trajectory that satisfies the recursion relation exactly; and (2) that all trajectories that satisfy the recursion relation exactly (or at least, all such trajectories that start from a ``generic'' initial condition) can be approximated in this way.  We then conclude that there are some values for the two undetermined constants for which $P_N^{(\text{ansatz})}$ is a good approximation of the true solution when $N$ is large.  

Although we do not have direct evidence for these assumptions, we can clarify (1) as follows.  The basic point is that the $O(1/N^2)$ error terms in Eq.~\eqref{eq:joint dist recursion general main claim} must not accumulate to any significant amount over many iterations.  Let us write Eq.~\eqref{eq:joint dist recursion general main claim} to one more order in $1/\sqrt{N}$: $ P_{N+1}^{(\text{ansatz})}(s,\phi_{r'}) = \mathcal{F}[s,\phi_{r'}; \{ P_N^{(\text{ansatz})}\}] + f_1(s,\phi_{r'})/N^2 + O(N^{-5/2})$ for some function $f_1(s,\phi_{r'})$.  To cancel $f_1$ up to $O(N^{-5/2})$ error, we take the joint distribution to be $P_N(s,\phi_{r'}) = P_N^{(\text{ansatz})}(s,\phi_{r'})+P_N^{(\text{ansatz}),1}(s,\phi_{r'})$, where $P_N^{(\text{ansatz}),1}(s,\phi_{r'})$ satisfies
\begin{multline}
    P_{N+1}^{(\text{ansatz}),1}(s,\phi_{r'}) = \mathcal{F}[s,\phi_{r'}; \{ P_N^{(\text{ansatz}),1}\}] \\
    - f_1(s,\phi_{r'})/N^2,\label{eq:inhomogeneous equation}
\end{multline}
which we regard as an ``inhomogeneous'' version of the ``homogeneous'' equation~\eqref{eq:P recursion}.  Our assumption is that Eq.~\eqref{eq:inhomogeneous equation} has a particular solution that is of order of the integral in $N$ of the forcing term (i.e., of order $1/N$) or smaller.  We set $P_{N+1}^{(\text{ansatz}),1}$ to be this particular solution.  Higher orders proceed similarly, so that, granting the assumption just mentioned, Eq.~\eqref{eq:joint dist recursion general main claim} does indeed show that $P_N^{(\text{ansatz})}(s,\phi_{r'})$ is the leading term in $1/\sqrt{N}$ of the solution to the homogeneous equation~\eqref{eq:P recursion}.  We do not study the inhomogeneous equation~\eqref{eq:inhomogeneous equation} further.

We turn next to the demonstration of Eq.~\eqref{eq:joint dist recursion general main claim}.

\subsubsection{Ansatz}
It is convenient to work in a set of variables in which the joint probability distribution turns out to be separable.  Thus, we change variables from $(s,\phi_{r'})$ to $(\tilde{s},\phi_{r'})$, where $\tilde{s}$ is defined below.  The tilde will be used throughout to indicate quantities defined with reference to $\tilde{s}$ rather than $s$.

We define a shifted and rescaled variable $\tilde{s}$ as
\beq
    \tilde{s}_{1\dots N} = \frac{ s_{1\dots N} - \frac{2}{\lloc}N- \hat{s}(\phi_{r_{1\dots N}'}) }{1 + \frac{\eta(\phi_{r_{1\dots N}'})}{2cN}},\label{eq:stilde}
\eeq
where $\hat{s}(\phi)$ and $\eta(\phi)$ are functions to be determined later.  To reduce clutter in our calculation below, we temporarily re-define $\eta(\phi)/(2c)\to\eta(\phi)$, restoring the factor of $2c$ at the end.  From Eqs.~\eqref{eq:s recursion Rto1} and~\eqref{eq:phirp recursion Rto1}, we read off the recursion relation for $\tilde{s}$ in the localized regime:
\beq
    \tilde{s}_{1\dots N+1} = \frac{\left[ 1 + \eta(\phi_{r_{1\dots N}'})/N \right] \tilde{s}_{1\dots N} + \tilde{g}_{N+1}(\phi_{r_{1\dots N}'}) }{ 1 + \eta\bpl\phi_{r_{1\dots N}'} + h_{N+1}(\phi_{r_{1\dots N}'}) \bpr/(N+1)},\label{eq:stilde recursion} 
\eeq
where we have defined
\beq
    \tilde{g}_n(\phi) = g_n(\phi) - 2/\lloc + \hat{s}(\phi) - \hat{s}\bpl\phi+h_n(\phi)\bpr.\label{eq:gtilde}
\eeq
Naively, it would seem that for large $N$ we could drop the $\eta(\phi)$ terms in Eq.~\eqref{eq:stilde recursion}; however, we will see that they sometimes contribute in the recursion relation for the probability distribution due to being multiplied by a factor of $N$.

It is convenient for calculations below that the variable $\tilde{s}$ should have, on average and for large $N$, no change as $N$ is increased.  (In Appendix~\ref{sec:Toy recursion relation}, we provide a simple example to show that using a variable without this property can lead to an incorrect answer for the variance unless the approximation of the discrete variable $N$ as a continuous variable is treated with particular care.)  To verify this property, we first note that by taking $N\to\infty$ in the recursion relation~\eqref{eq:p recursion finite N} for the phase distribution and integrating, we obtain the identity
\beq
    \int_{-\pi}^\pi d\phi\ p_{\infty}(\phi) [f(\phi) - \langle f\bpl\phi + h_n(\phi)\bpr\rangle_n]=0\label{eq:phi integration identity}
\eeq
for any function $f(\phi)$.  Then from Eqs.~\eqref{eq:stilde recursion} and~\eqref{eq:gtilde} and the normalization of the phase distribution, we find that the average increase of $\tilde{s}$ in one step is (for large $N$)
\bseq
\begin{align}
    &\langle\tilde{s}_{1\dots +1}\rangle_{1\dots N+1}-\langle\tilde{s}_{1\dots N}\rangle_{1\dots N}= \notag\\
    &\qquad \int_{-\pi}^\pi d\phi\ p_{\infty}(\phi) \langle \tilde{g}_n(\phi)\rangle_n+\dots\\
    &= \int_{-\pi}^\pi d\phi\ p_{\infty}(\phi) \langle g_n(\phi)\rangle_n - 2/\lloc + \dots,\label{eq:stilde average increase}
\end{align}
\eseq
where the $\eta(\phi)$ function only contributes a negligible $O(1/N)$ error term.  Thus, the requirement that $\tilde{s}$ does not increase on average with $N$ becomes $2/\lloc = \int_{-\pi}^\pi d\phi\  p_{\infty}(\phi)\langle g_n(\phi)\rangle_n$, which is indeed the same equation as found earlier for the localization length [Eq.~\eqref{eq:lloc as integral}].

The joint probability distribution of $\tilde{s}$ and $\phi_{r'}$ is, by definition,
\beq
    \tilde{P}_N(\tilde{s},\phi_{r'}) = \langle \delta(\tilde{s}_{1\dots N} - \tilde{s} )\delta(\phi_{r_{1\dots N}'} - \phi_{r'})\rangle_{1\dots N},\label{eq:Ptilde def}
\eeq
and is related to $P_N$ by a change of variables.  The recursion relation in the localized regime is readily found to be
\beq
    \tilde{P}_{N+1}(\tilde{s},\phi_{r'}) = \tilde{\mathcal{F}}[N,\tilde{s},\phi_{r'};\{ \tilde{P}_N\}],\label{eq:Ptilde recursion} 
\eeq
where $\tilde{\mathcal{F}}$ is a linear functional in its last argument and is given by
\begin{multline}
    \tilde{\mathcal{F}}[s,\phi_{r'};\{ P_N\}]  = \int d\tilde{s}'\int_{-\pi}^\pi d\phi\ \tilde{P}_{1\dots N}\left(\tilde{s}', \phi\right)\\
    \times 
    \langle \delta\left( \frac{\left[ 1 + \eta(\phi)/N \right] \tilde{s}' + \tilde{g}_n(\phi) }{ 1 + \eta\bpl\phi + h_n(\phi) \bpr/(N+1)} -\tilde{s} \right)\\
    \times\delta\bpl\phi + h_n(\phi) - \phi_{r'}\bpr\rangle_n.\label{eq:mathcalFtilde def}
\end{multline}
These relations are just Eqs.~\eqref{eq:P recursion} and~\eqref{eq:mathcalF def} in the new variables.  Our goal is to make an ansatz $\tilde{P}_N^{(\text{ansatz})}$ that satisfies Eq.~\eqref{eq:joint dist recursion general main claim} in the new variables, i.e.,
\beq
    \tilde{P}_{ N+1}^{(\text{ansatz})}(\tilde{s},\phi_{r'}) = \tilde{\mathcal{F}}[N,\tilde{s},\phi_{r'}; \{ \tilde{P}_N^{(\text{ansatz})}\}] + O(1/N^2).\label{eq:joint dist recursion general main claim tilde}
\eeq

Our ansatz is
\beq
    \tilde{P}_N^{(\text{ansatz})}(\tilde{s},\phi_{r'}) = \tilde{F}_N(\tilde{s}) p_{\infty}(\phi_{r'}),\label{eq:Ptilde ansatz}
\eeq
where the phase distribution $p_{\infty}(\phi_{r'})$ is as calculated in Sec.~\ref{sec:Scattering expansion of the localization length} (alternatively, we may leave it as an undetermined function for now and recover the same series by our calculation below) and where $\tilde{F}_N(\tilde{s})$ is a Gaussian with variance $2cN$ (with a constant $c$ to be determined later):
\beq
    \tilde{F}_N(\tilde{s})=\frac{1}{\sqrt{4\pi c N }}e^{-\tilde{s}^2/(4 c N)}.\label{eq:Ftilde def}
\eeq

Changing variables from $(\tilde{s},\phi_{r'})$ back to $(s,\phi_{r'})$ yields
\begin{multline}
    P_N^{(\text{ansatz})}(s,\phi_{r'}) = \left[1+ \frac{\eta(\phi_{r'})}{2c N}\right]^{-1}\\
    \times \tilde{P}_N^{(\text{ansatz})}\left(\frac{s- \frac{2}{\lloc}N - \hat{s}(\phi_{r'})}{1+ \frac{\eta(\phi_{r'})}{2c N}}, \phi_{r'} \right),
\end{multline}
where we have taken $N$ large enough that the Jacobian determinant is positive.  This indeed produces the claimed general form of Eq.~\eqref{eq:joint prob dist final}.

We show below that the constant $c$ and the functions $\hat{s}(\phi_{r'})$ and $\eta(\phi_{r'})$ are uniquely determined (up to $\phi_{r'}$-independent additive constants in the two functions), order by order in the scattering expansion, by the requirement that the ansatz must satisfy Eq.~\eqref{eq:joint dist recursion general main claim tilde}.  Thus, each choice of the two additive constants determines a trajectory $\{P_N^{(\text{ansatz})} \}$ that satisfies Eq.~\eqref{eq:joint dist recursion general main claim}.

Let us note that Eq.~\eqref{eq:joint dist recursion general main claim tilde} can also be obtained starting from a more general ansatz.  In particular, one can replace $\tilde{s}\to (1+ \beta_2/N)(\tilde{s}-\beta_1)$ in Eq.~\eqref{eq:Ftilde def} (for some constants $\beta_1,\beta_2)$.  However, one finds that for large $N$, the constants $\beta_1$ and $\beta_2$ only affect $P_N^{(\text{ansatz)}}$ by additive shifts of $\hat{s}(\phi)$ and $\eta(\phi)$, respectively, and these have already been accounted for.  Thus, this particular modification does not introduce any additional generality (although we cannot rule out the possibility of other, non-Gaussian solutions).

We proceed to show that Eq.~\eqref{eq:joint dist recursion general main claim} reduces to certain integral equations constraining $\hat{s}(\phi_{r'})$ and $\eta(\phi_{r'})$.  Then, we show that these integral equations can be solved order by order in the scattering expansion, which in turn provides a scattering expansion also of the constant $c$.

\subsubsection{Reduction to integral equations}\label{eq:Reduction to integral equations}
It is convenient to Fourier transform in $\tilde{s}$; we write the transformed function with the same symbol [e.g., $\tilde{F}_N(q) = \int d\tilde{s}\ e^{-i q\tilde{s}}\tilde{F}_N(\tilde{s})$].  Note in particular that
\beq
    \tilde{F}_N(q) = e^{- c N q^2}.\label{eq:Ftilde Gaussian FT}
\eeq
We proceed to evaluate $\mathcal{F}[N,q,\phi_{r'};\{\tilde{P}_N^{(\text{ansatz})} \}] \equiv \int d\tilde{s}\ e^{-i q\tilde{s}}\mathcal{F}[N,\tilde{s},\phi_{r'};\{\tilde{P}_N^{(\text{ansatz})}\}]$ with the goal of imposing Eq.~\eqref{eq:joint dist recursion general main claim} in $q$ space.  From Eq.~\eqref{eq:mathcalFtilde def}, we obtain the functional $\tilde{\mathcal{F}}$ in Fourier space:
\begin{widetext}
\begin{multline}
    \tilde{\mathcal{F}}[N,q,\phi_{r'};\{\tilde{P}_N\}] = \int_{-\pi}^\pi d\phi\ \langle \exp\left[-i q \frac{\tilde{g}_n(\phi)}{1+ \eta\bpl\phi+ h_n(\phi)\bpr/(N+1)}\right] \\
    \times \tilde{P}_N\left(\frac{1 + \eta(\phi)/N}{1+ \eta\bpl\phi+ h_n(\phi)\bpr/(N+1)}  q, \phi\right) \delta\bpl\phi + h_n(\phi) - \phi_{r'}\bpr\rangle_n.\label{eq:mathcalF FT}
\end{multline}
Evaluating this functional on an ansatz of the form~\eqref{eq:Ptilde ansatz}, we find
\begin{multline}
    \tilde{\mathcal{F}}[N,q,\phi_{r'}; \{\tilde{P}_N^{(\text{ansatz})}\}] = \int_{-\pi}^\pi d\phi\ p_{\infty}(\phi) \langle\tilde{F}_N\left( \frac{1 + \eta(\phi)/N}{1+ \eta\bpl\phi+ h_n(\phi)\bpr/(N+1)} q\right)\\
    \times \exp\left[-i q \frac{\tilde{g}_n(\phi)}{1+ \eta\bpl\phi+ h_n(\phi)\bpr/(N+1)}\right]\delta\bpl\phi + h_n(\phi)- \phi_{r'}\bpr \rangle_n.\label{eq:Ptilde recursion Ftilde integral form}
\end{multline}
\end{widetext}
Next, we expand in small $q$, using the specific form~\eqref{eq:Ftilde Gaussian FT} to write Eq.~\eqref{eq:Ptilde recursion Ftilde integral form} as
\beq
    \tilde{\mathcal{F}}[N,q,\phi_{r'}; \{\tilde{P}_N^{(\text{ansatz})}\}] = C(N,q,\phi_{r'}) \tilde{F}_N(q),\label{eq:Ptilde recursion FT in terms of C}
\eeq
where $C(N,q,\phi_{r'})$ is a power series in $q$:
\beq
    C(N,q,\phi_{r'})\equiv \sum_{j=0}^\infty  C_j(N,\phi_{r'})q^j,\label{eq:C series}
\eeq
with some coefficients $C_j(N,\phi_{r'})$.

The essential point that simplifies the calculation is that the order in $1/\sqrt{N}$ of a given term differs between $\tilde{s}$ space and $q$ space.  We have
\beq
    \int\frac{dq}{2\pi}\ e^{i q \tilde{s}} q^j \tilde{F}_N(q) = O(N^{-(j+1)/2}) \qquad (j\ge0).\label{eq:FT error estimate}
\eeq
In particular: for the purpose of imposing the condition~\eqref{eq:joint dist recursion general main claim}, any term $q^j\tilde{F}_N(q)$ with $j\ge 3$ is negligible because its contribution in $\tilde{s}$ space is $O(1/N^2)$.  Thus, provided that we establish that each $C_j(N,\phi_{r'})$ is finite as $N\to\infty$, it follows that we can drop all but the terms $j=0,1,2$ in Eq.~\eqref{eq:C series}.

To show that each $C_j(N,\phi_{r'})$ is finite as $N\to\infty$, we inspect the two terms in the integrand in~\eqref{eq:Ptilde recursion Ftilde integral form} that need to be expanded in $q$: the $\tilde{F}_N$ term and the exponential.  The latter clearly only produces terms that are finite as $N\to\infty$.  The $\tilde{F}_N$ term is of the form $\tilde{F}_N([1+O(1/N)]q)$, so it suffices to note that $[\tilde{F}_N(q)]^{-1}\pd^j \tilde{F}_N(q)/\pd q^j= O(N^j)$.

Next, we need the specific form of the $j=0$ term.  From Eq.~\eqref{eq:Ptilde recursion Ftilde integral form}, we read off
\bseq
\begin{align}
    C_0(N,\phi_{r'}) &= \int_{-\pi}^\pi d\phi\ p_{\infty}(\phi) \langle \delta\bpl\phi+h_n(\phi)-\phi_{r'}\bpr\rangle_n\\
    &= p_\infty(\phi_{r'}),\label{eq:C0}
\end{align}
\eseq
where we have noted that the first line is just the right-hand side of the recursion relation~\eqref{eq:p recursion finite N} (with $N\to\infty$) for the probability distribution of the reflection phase.  We also note
\beq
    \tilde{F}_{N+1}(q) = \left[ 1- c q^2 + O(q^4)\right]\tilde{F}_N(q),
\eeq
which may be obtained by treating the discrete variable $N$ in Eq.~\eqref{eq:Ftilde Gaussian FT} as continuous.  From Eqs.~\eqref{eq:Ptilde ansatz},~\eqref{eq:Ptilde recursion FT in terms of C}, and~\eqref{eq:C series}, we then obtain
\begin{multline}
    \tilde{P}_{N+1}^{(\text{ansatz})}(q,\phi_{r'}) - \mathcal{F}[N,q,\phi_{r'};\{\tilde{P}_N^{(\text{ansatz})}] = \\-C_1(N,\phi_{r'})q\tilde{F}_N(q) - \left[ C_2(N,\phi_{r'})+ c p_\infty(\phi_{r'})\right]q^2 \tilde{F}_N(q)\\
    + \dots,
\end{multline}
where the omitted terms are $O(1/N^2)$ in $\tilde{s}$ space due to~\eqref{eq:FT error estimate}.

Again using~\eqref{eq:FT error estimate}, we then see that imposing~\eqref{eq:joint dist recursion general main claim} yields the following two requirements:
\bseq
\begin{align}
    C_1(N,\phi_{r'}) &= O(1/N),\label{eq:C1 condition}\\
    C_2(N,\phi_{r'}) + c p_{\infty}(\phi_{r'}) &= O(1/\sqrt{N}).\label{eq:C2 condition}
\end{align}
\eseq

To put these requirements into a more explicit form, we calculate $C_1(N,\phi_{r'})$ and $C_2(N,\phi_{r'})$ for large $N$ as follows: first, if we temporarily set the function $\eta(\phi)$ to be identically zero, then we may read off $C(N,q,\phi_{r'}) = \int_{-\pi}^\pi d\phi\ p_{\infty}(\phi) \langle e^{-i q \tilde{g}_n(\phi)}\delta\bpl\phi+h_n(\phi)-\phi_{r'}\bpr\rangle_n$.  To then include the function $\eta(\phi)$, we expand $\tilde{F}_N$ in Eq.~\eqref{eq:Ptilde recursion Ftilde integral form} about the point $q$ and use Eq.~\eqref{eq:Ftilde Gaussian FT} to write $\frac{\pd^j}{\pd q^j}\tilde{F}_N(q)$ in terms of $\tilde{F}_N(q)$.  The only contribution from the function $\eta(\phi)$ that survives for large $N$ comes from the following expansion:
\bseq
\begin{align}
    &\tilde{F}_N\left( \frac{1 + \eta(\phi)/N}{1+ \eta\bpl\phi+ h_n(\phi)\bpr/(N+1)} q\right) = \tilde{F}_N(q) \notag\\  + &\left(\frac{\eta(\phi)}{N} - \frac{\eta\bpl\phi+h_n(\phi)\bpr}{N+1} \right)q \frac{\pd}{\pd q} \tilde{F}_N(q) + \dots\\
    &= \{ 1 - 2 c [\eta(\phi)- \eta\bpl\phi+h_n(\phi)\bpr + O(1/N)] q^2 \notag\\
    &\qquad\qquad\qquad\qquad\qquad+ O(q^4)\}\tilde{F}_N(q),
\end{align}
\eseq
where we have used $\frac{\pd}{\pd q}\tilde{F}_N(q) = - 2 c N q \tilde{F}_N(q)$. We thus obtain
\begin{widetext}
\bseq
\begin{align}
    C_1(N,\phi_{r'}) &= -i \int_{-\pi}^\pi d\phi\ p_{\infty}(\phi) \langle \tilde{g}_n(\phi) \delta\bpl\phi+h_n(\phi)-\phi_{r'}\bpr\rangle_n +O(1/N),\label{eq:C1}\\
    C_2(N,\phi_{r'}) &= - \int_{-\pi}^\pi d\phi\ p_{\infty}(\phi) \langle \left\{\frac{1}{2}\tilde{g}_n(\phi)^2 + 2 c \left[ \eta(\phi)- \eta\bpl\phi+h_n(\phi)\bpr \right] \right\} \delta\bpl\phi+h_n(\phi)-\phi_{r'}\bpr\rangle_n + O(1/N).\label{eq:C2}
\end{align}    
\eseq
\end{widetext}

Thus, the two requirements~\eqref{eq:C1 condition} and~\eqref{eq:C2 condition} yield two integral equations involving the constant $c$ and the functions $\hat{s}(\phi)$ and $\eta(\phi)$.  We show in the next section that these integral equations uniquely determine $c$, $\hat{s}(\phi)$, and $\eta(\phi)$ (except for the addition of $\phi$-independent constants to the functions) order by order in the scattering expansion.

\subsubsection{Series solution of integral equations}\label{sec:Series solution of integral equations}
It is convenient to take the Fourier transform in $\phi_{r'}$.  We write $C_{j,\ell} \equiv \lim_{N\to\infty} \int_{-\pi}^\pi\frac{d\phi}{2\pi}\ e^{-i\ell\phi}C_j(N,\phi)$, where $j=1,2$.  Taking the Fourier transform of Eq.~\eqref{eq:C1} yields
\begin{multline}
    C_{1,\ell} = -i (-1)^\ell \int_{-\pi}^\pi \frac{d\phi}{2\pi}\ e^{-i\ell\phi} p_{\infty}(\phi)\\
    \times\langle v_n^{-\ell} A_{n,\ell}(\phi)\tilde{g}_n(\phi)\rangle_n,
\end{multline}
where we have recalled Eqs.~\eqref{eq:hn} and~\eqref{eq:A}.  Thus, recalling Eq.~\eqref{eq:gtilde}, we see that the requirement~\eqref{eq:C1 condition} is equivalent to the following equation holding for all integers $\ell$:
\begin{multline}
    \int_{-\pi}^\pi \frac{d\phi}{2\pi}\ e^{-i\ell\phi} p_{\infty}(\phi)\langle v_n^{-\ell} A_{n,\ell}(\phi)\\
    \times\left[g_n(\phi)- 2/\lloc + \hat{s}(\phi)-\hat{s}\bpl\phi+h_n(\phi)\bpr\right]\rangle_n = 0.\label{eq:C1 integral equation}
\end{multline}

We now show that Eq.~\eqref{eq:C1 integral equation} determines, to all orders in the scattering expansion, all Fourier coefficients $\hat{s}_\ell\equiv\int_{-\pi}^\pi\frac{d\phi}{2\pi}\ e^{-i\ell\phi}\hat{s}(\phi)$ except for $\hat{s}_{\ell=0}$ (which is one of the undetermined constants we have discussed above).  The fact that $\hat{s}_{\ell=0}$ is unconstrained follows immediately from the invariance of the equation under shifting $\hat{s}(\phi)$ by an additive constant.  To determine the $\ell\ne0$ coefficients, we start by noting that since $g_n(\phi)$ and $2/\lloc$ start at first and second order (respectively) in the scattering expansion, $\hat{s}(\phi)$ must in general start at first order.  Noting that $\int_{-\pi}^\pi \frac{d\phi}{2\pi}\  e^{-i\ell\phi}\hat{s}\bpl\phi + h_n(\phi)\bpr = \int_{-\pi}^\pi \frac{d\phi}{2\pi}\  e^{-i\ell(\phi-h_n^{(0)})}\hat{s}^{(1)}(\phi) + O(\param^2) = (-v_n)^\ell \hat{s}_\ell^{(1)}+ O(\param^2) $, we re-arrange the first-order part of Eq.~\eqref{eq:C1 integral equation} to yield
\beq
    \hat{s}_\ell^{(1)} = \frac{\langle v_n^{-\ell} g_{n,\ell}^{(1)}\rangle_n}{(-1)^\ell - \langle v_n^{-\ell} \rangle_n } = \gamma^{(1)}\delta_{\ell,-1} + \gamma^{(1)*}\delta_{\ell,1},
\eeq
which demonstrates Eq.~\eqref{eq:s1 in terms of p1}.

Consider next the order $j$ part of Eq.~\eqref{eq:C1 integral equation} for some $j\ge 2$.  By a similar calculation as above, the quantity $\hat{s}_\ell^{(j)}$ only appears in the combination $[\langle v_n^{-\ell}\rangle_n-(-1)^\ell]s_\ell^{(j)}$, which means that we can solve for $\hat{s}_\ell^{(j)}$ in terms of $\hat{s}_{\ell'}^{(j')}$ with $1\le j' < j$.  This completes the demonstration that Eq.~\eqref{eq:C1 integral equation} determines the function $\hat{s}(\phi)$ to all orders in the scattering expansion.

Although we do not present the explicit expressions in this case, it may be shown that $\hat{s}_\ell^{(j)}$ is non-vanishing for only finitely many $\ell$ and may be expressed as a finite sum over Fourier coefficients of the other quantities appearing in Eq.~\eqref{eq:C1 integral equation}.  We have calculated the second-order correction as well, with the result (in $\phi$ space)
\begin{multline}
    \hat{s}(\phi) = 2\text{Re}\left\{ \gamma^{(1)}e^{-i\phi} + \left[\frac{3}{2}\gamma^{(2)} - \left(\gamma^{(1)}\right)^2\right]e^{-2i\phi} \right\}\\ + O(\param^3) +\text{const.}\label{eq:sHat to 2nd order}
\end{multline}
Thus, we see that the correspondence with the non-uniform part of the marginal distribution of the reflection phase [Eq.~\eqref{eq:s1 in terms of p1}] does not generally extend beyond the first order.

We proceed to the second constraint,~\eqref{eq:C2 condition}.  Equation~\eqref{eq:C2} yields [rescaling $2c\eta(\phi)\to\eta(\phi)$ to return to our original definition of $\eta(\phi)$]
\begin{multline}
    C_{2,\ell}= -(-1)^\ell \int_{-\pi}^\pi d\phi\ p_{\infty}(\phi)e^{-i\ell\phi} \langle v_n^{-\ell}A_{n,\ell}(\phi)\\
    \times \left[\frac{1}{2}\tilde{g}_n(\phi)^2 + \eta(\phi) - \eta\bpl\phi+h_n(\phi)\bpr \right] \rangle_n,
\end{multline}
so that~\eqref{eq:C2 condition} is equivalent to the following equation holding for all integers $\ell$:
\begin{multline}
    cp_{\infty,\ell} = (-1)^\ell \int_{-\pi}^\pi d\phi\ p_{\infty}(\phi)e^{-i\ell\phi} \langle v_n^{-\ell}A_{n,\ell}(\phi)\\
    \times\left[\frac{1}{2}\tilde{g}_n(\phi)^2 + \eta(\phi) - \eta\bpl\phi+h_n(\phi)\bpr \right] \rangle_n.\label{eq:C2 integral equation}
\end{multline}
We now show that the $\ell=0$ component of Eq.~\eqref{eq:C2 integral equation} determines the constant $c$ to all orders, while the $\ell\ne 0$ coefficients determine the function $\eta(\phi)$ to all orders aside from a $\phi$-independent additive constant.  Using Eqs.~\eqref{eq:lloc as integral} and~\eqref{eq:phi integration identity}, we simplify the $\ell=0$ component of Eq.~\eqref{eq:C2 integral equation} to
\begin{multline}
    c= \frac{1}{2}\int_{-\pi}^\pi d\phi\ p_{\infty}(\phi)\langle \frac{1}{2}g_n(\phi)^2 +2 \hat{s}(\phi)g_n(\phi)\\
    + 2 \hat{s}\bpl\phi+h_n(\phi)\bpr \left[\hat{s}\bpl\phi+h_n(\phi)\bpr - \hat{s}(\phi) - g_n(\phi) \right]\rangle_n\\ - \frac{1}{2}\left(\frac{2}{\lloc}\right)^2,
\end{multline}
which thus yields a series expansion $c= c^{(2)}+c^{(3)} + \dots$, since we have already established series expansions of $p_{\infty}(\phi)$, $\hat{s}(\phi)$, and $2/\lloc$.  We readily obtain $c^{(2)}= \gamma^{(1)}\langle r_n\rangle_n$ and $c^{(3)}=0$ [all odd orders indeed vanish by the symmetry argument presented below Eq.~\eqref{eq:sigma to 2nd order}], which thus yields the SPS relation up to errors of the fourth order [Eq.~\eqref{eq:sigma to 2nd order}].  We have not calculated the next coefficient ($c^{(4)}$) explicitly, but it could be done in our formalism.  As commented earlier, Schrader \textit{et al}. have already shown in a specific model~\cite{SchraderPerturbative2004} that $c^{(4)}$ need not coincide with the fourth-order part of $2/\lloc$, as would be expected if the SPS relation were exactly true.

Note that Eq.~\eqref{eq:C2 integral equation} is invariant under an additive shift of the function $\eta(\phi)$ by any $\phi$-independent constant, i.e., the component $\eta_{\ell=0}$ is unconstrained.  However, all $\ell\ne0$ coefficients are fixed by Eq.~\eqref{eq:C2 integral equation}, and we can then calculate $\eta(\phi)$ order by order in the scattering expansion [since all other quantities in Eq.~\eqref{eq:C2 integral equation} can be calculated order by order].  A straightforward calculation shows that the series starts at second order with
\beq
    \eta(\phi) = \text{Re}\left\{ \left[\gamma^{(2)} - (\gamma^{(1)})^2 \right] e^{-2i\phi} \right\} +O(\param^3) + \text{const.}\label{eq:eta to leading order}
\eeq
In our discussion above of three-parameter scaling, we ignored the function $\eta(\phi)$ for reasons discussed there.  If $\eta(\phi)$ is included at the leading order, then an additional two real parameters (the real and imaginary parts of $\gamma^{(2)}$) must be taken into account for a total of five parameters to determine the joint probability distribution in the regime of weak local reflection.

We check Eqs.~\eqref{eq:sHat to 2nd order} and~\eqref{eq:eta to leading order} numerically in the following way.  For a fixed and large system size $N$ (in a given model), we calculate the tuple $(s_{1\dots N},\phi_{r_{1\dots N}'})$ in many disorder realizations.  We then bin the tuples based on the value of $\phi_{r_{1\dots N}'}$ into bins of some width $\phi_{\text{width}}\ll 2\pi$.  Within each bin we do a Gaussian fit to the PDF of the values of $s_{1\dots N}$ in the bin.  We thus obtain an estimate of the mean and variance of $s_{1\dots N}$ conditional on a certain value of $\phi_{r_{1\dots N}'}$. According to our result~\eqref{eq:joint prob dist final}, the conditional mean and variance should be given by $2 N/ \lloc + \hat{s}(\phi_{r_{1\dots N}'})$ and $2cN + 2\eta(\phi_{r_{1\dots N}'})$, respectively, up to the addition of constants that are independent of $N$ and of $\phi_{r_{1\dots N}'}$.

\begin{widetext}
\begin{figure*}[htb]
\subfloat[\label{subfig:reflection_phase_dist_DTQW}]{%
  \includegraphics[width=\columnwidth]{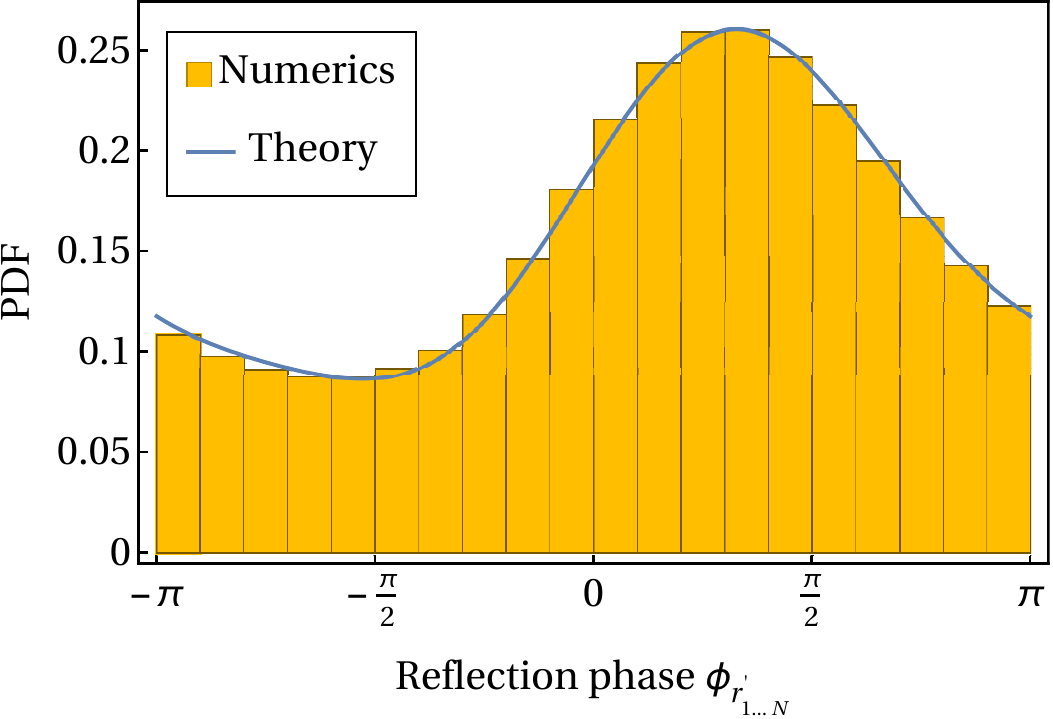}%
}\hfill
\subfloat[\label{subfig:phase-dependent_s_dist_DTQW}]{%
  \includegraphics[width=\columnwidth]{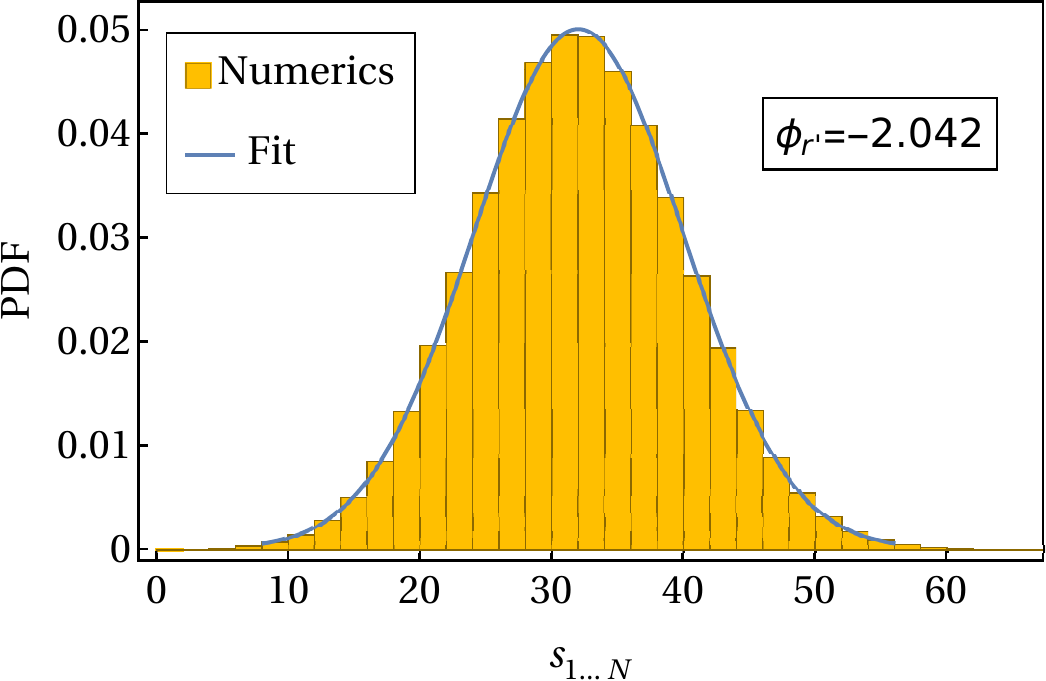}%
}
\vfill
\subfloat[\label{subfig:phase-dependent_mean_DTQW}]{%
  \includegraphics[width=\columnwidth]{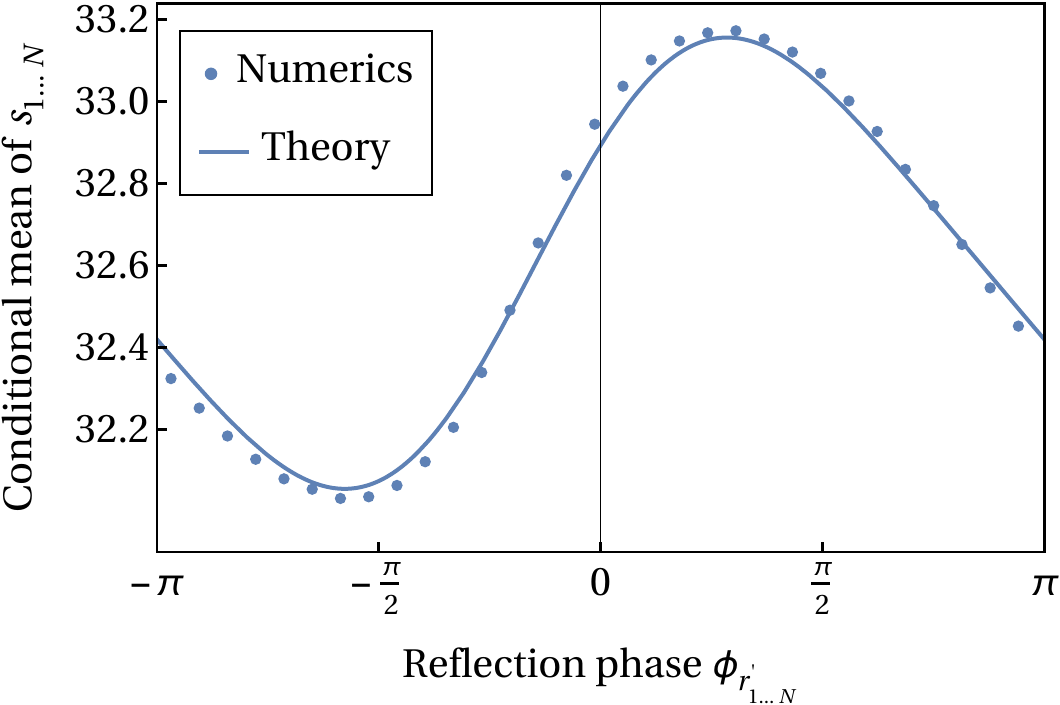}%
}
\hfill
\subfloat[\label{subfig:phase-dependent_variance_DTQW}]{%
  \includegraphics[width=\columnwidth]{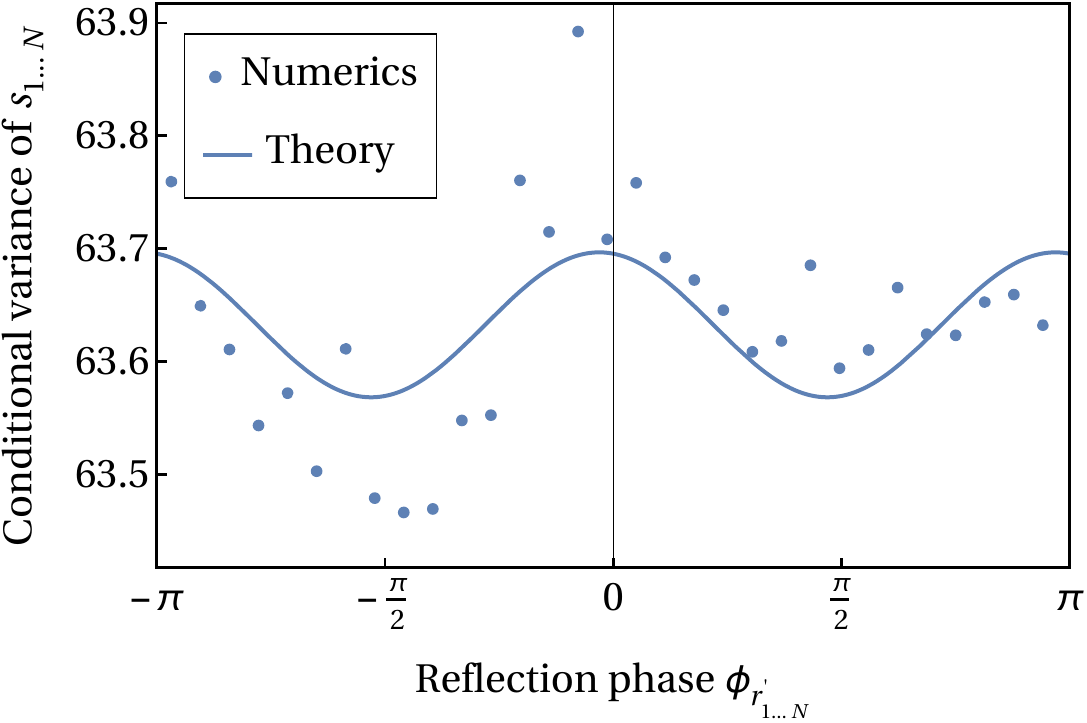}%
}
\caption{Numerical checks of our result~\eqref{eq:joint prob dist final} for the joint probability distribution of $s_{1\dots N}$ and $\phi_{r_{1\dots N}'}$.  In the DTQW with Tarasinski leads (see Sec.~\ref{sec:Application to discrete-time quantum walks}), $1.1\times10^8$ disorder realizations were generated with the following parameters: $N=850$ sites, $\omega= 0.25$, all $\theta_n\equiv\theta = 0.14$, all $\varphi_{1,n}\equiv\varphi_{2,n}=0$, and $\varphi_n$ uniformly distributed in $[W,W]$ with $W=0.6$.  (a) The marginal distribution of the reflection phase.  The theoretical result is calculated up to an error of $O(\sin^4\theta)$ [see Eqs.~\eqref{eq:gamma1}--\eqref{eq:gamma33},~\eqref{eq:p to 3rd order}, and~\eqref{eq:S matrix DTQW}].  For (b), (c), and (d), the data was binned with $\phi_\text{width}=0.2$ (see main text).  (b) The conditional distribution of $s_{1\dots N}$ given the value $\phi_{r_{1\dots N}'} =-2.042$ (which is the center of the bin for the numerics).  The fit is to a Gaussian, and the resulting mean and variance yield the sixth data point from the right in (c) and (d).  (c) The mean of $s_{1\dots N}$ conditional on the value of $\phi_{r_{1\dots N}'}$.  The theory curve is given by Eq.~\eqref{eq:sHat to 2nd order} with the phase-independent, additive constant as a fit parameter.  The similarity between the plots (a) and (c) is due to Eq.~\eqref{eq:s1 in terms of p1}.  (d) The variance of $s_{1\dots N}$ conditional on the value of $\phi_{r'}$.  The theory curve is $2\eta(\phi)$, where $\eta(\phi)$ is given by Eq.~\eqref{eq:eta to leading order} with the phase-independent constant as a fit parameter.}\label{fig:numerics_for_joint_dist}
\end{figure*}
\end{widetext}

Figure~\ref{fig:numerics_for_joint_dist} shows the results for the DTQW considered in Sec.~\ref{sec:Application to discrete-time quantum walks}, and also illustrates the leading-order correspondence~\eqref{eq:s1 in terms of p1} between the marginal distribution of the reflection phase and the phase-dependent mean.
Equation~\eqref{eq:sHat to 2nd order} for the phase-dependent mean agrees well with numerics (Fig.~\ref{subfig:phase-dependent_mean_DTQW}).  Equation~\eqref{eq:eta to leading order} for the phase-dependent variance is in rough agreement (Fig.~\ref{subfig:phase-dependent_variance_DTQW}); we believe that the discrepancies could be due to a combination of (a) Eq.~\eqref{eq:eta to leading order} being a leading order expression only (with the expansion parameter being $\sin\theta\approx 0.14$), and (b) statistical error due to having insufficiently many data points.  We note that the explicit expression~\eqref{eq:eta to leading order} for the phase-dependent variance is not a central part of our work, and that the numerical check of the phase-dependent mean suffices to confirm that the joint probability distribution is in general non-separable.

\section{Conclusion}\label{sec:Conclusion}

In this paper, we developed a scattering expansion for the general problem of single-channel scattering through a disordered region with i.i.d. disorder.  We found a systematic expansion of the inverse localization length and reflection phase distribution in terms of averages over the local reflection amplitudes, with the local reflection strength being the small parameter.  The leading order of this expansion was shown to be equivalent to an earlier result by Schrader \textit{et al}.~\cite{SchraderPerturbative2004}, and the next order was checked in special cases from the literature.  Using these first two orders, we explored the effect of phase disorder on DTQWs, interpolating between known results in the limits of weak and strong disorder and showing analytically that the localization length can vary non-monotonically in between.

In the scattering framework, we further developed the general understanding of SPS without the uniform phase hypothesis, as demonstrated by Schrader \textit{et al}.~\cite{SchraderPerturbative2004}.  In particular, we showed that their small parameter can be interpreted as the local reflection strength, and we showed that the SPS relation holds to another order beyond what they showed.  We also explicitly showed that the SPS relation holds for weak potential strength in a broad class of PARS.  The picture of SPS that emerges, primarily from Ref.~\cite{SchraderPerturbative2004} but extended by our work, is that instead of there being a length scale for complete phase randomization, there is instead a phase \emph{convergence} length at which the probability distribution of the reflection phase reaches its limiting form (which in general is non-uniform).  When the local reflection strength is weak and the scattering region is larger than both the localization length and the phase convergence length, SPS holds. 

We proceeded to apply the scattering expansion to the joint probability distribution of the minus logarithm of the transmission coefficient ($-\ln T$) and the reflection phase ($\phi_{r'}$).  In the limit of large system size, we obtained a general form for this distribution to all orders in the scattering expansion.  This calculation relied on an ansatz and some plausible, though unproven, assumptions (and we checked our final result numerically).  We showed that all quantities that appear in this general form may be expressed order by order in terms of local averages, and we explicitly evaluated some of the leading-order terms.  We found that at the leading order in the local reflection strength, the same function describes both the deviations from uniformity in the marginal distribution of $\phi_{r'}$ and the correlations between $-\ln T$ and $\phi_{r'}$.  This showed that the joint distribution satisfies, at the leading order, a three-parameter scaling theory, with one parameter being the usual one for the SPS description of the marginal distribution of $-\ln T$ and the remaining two parameters describing both the marginal distribution of the reflection phase and the phase-transmission correlations.

It would be interesting to explore implications that our scattering-based approach might have for the more usual DTQW setup, in which a walker starts in a spatially confined initial state and evolves in time.  Ballistic spread (i.e., variance increasing quadratically with time) is an important property of DTQWs and is known to be suppressed by localization.  However, if the localization length is sufficiently large, then this suppression would be unimportant, since the walker can be expected to travel ballistically until reaching a distance of order $\lloc$.  (It has indeed been found in a particular model that the maximum distance reached by the walker has the same scaling with disorder strength as the localization length~\cite{KeatingLocalization2007}.)  Our scattering-based results for $\lloc$ might yield a lower bound (after appropriate minimization over quasienergy) on the $\lloc$ that appears in the time-dependent problem.  Also, our technique for calculating the reflection phase distribution might extend to the distribution of the Wigner delay time ($d\phi_{r'}/d\omega$), which would characterize the time that a walker spends in being reflected from a disordered region in an otherwise non-disordered environment.

Another direction to explore would be applications of our approach to other problems involving products of random matrices, even outside the setting of scattering theory.  For instance, in the study of randomly driven conformal field theories, Ref.~\cite{WenPeriodically2022} encounters a problem that seems to fit our framework [a product of random SU$(1,1)$ matrices]; each matrix represents a time step, and the Lyapunov exponent (inverse localization length) is shown to be the rate of entanglement entropy growth (and to be a lower bound on the heating rate). 

Finally, this work could be a step towards an analytical treatment of the quasi-one-dimensional case (i.e., many scattering channels rather than one).  This would be significant because the quasi-one-dimensional case can be used to study delocalization transitions in dimensions higher than one; in particular, one studies (usually numerically) the scaling, as the number of transverse modes goes to infinity, of the largest localization length~\cite{RomerNumerical2022}.  If we can carry out the approach with multiple scattering channels, then the possibility could arise of taking this limit analytically.  We note that Ref.~\cite{ChalkerScattering1993} takes a similar approach in a particular model and finds that departure from the isotropy assumption (c.f. Sec.~\ref{sec:Setup and review of prior work}) is necessary in order to obtain a metal-insulator transition.  Ultimately, we would like to use the scattering expansion as an analytical handle on critical exponents in higher-dimensional localization-delocalization transitions, such as the plateau transition in the integer quantum Hall effect~\cite{ChalkerPercolation1988}.

\acknowledgments{We thank Albert Brown and Fenner Harper for collaboration on related work, and we thank Victor Gurarie, Daniel Lidar, and Abhinav Prem for discussion.  A.B.C., P.S., and R.R. acknowledge financial support from the University of California Laboratory Fees Research Program funded by the UC Office of the President (UCOP), grant number LFR-20-653926.  A.B.C acknowledges financial support from the Joseph P. Rudnick Prize Postdoctoral Fellowship (UCLA).  P.S. acknowledges financial support from the Center for Quantum Science and Engineering Fellowship (UCLA) and the Bhaumik Graduate Fellowship (UCLA).  This work used computational and storage services associated with the Hoffman2 Shared Cluster provided by UCLA Institute for Digital Research and Education’s Research Technology Group.}

\begin{widetext}
\appendix

\section{S matrices and scattering transfer matrices}\label{sec:S matrices and scattering transfer matrices}
We recall some basic facts about single-channel scattering in one dimension.  A generic $2\times2$ $S$ matrix $\mathcal{S}$ relates incoming to outgoing scattering amplitudes by
\beq
    \bpmat
        \Psi_R^+\\
        \Psi_L^-
    \epmat
    =\mathcal{S}
    \bpmat
        \Psi_L^+\\
        \Psi_R^-
    \epmat,
\eeq
and may be parametrized as
\beq
    \mathcal{S} =
    \bpmat
        t & r'\\
        r & t'
    \epmat,\label{eq:S general parametrization amplitudes}
\eeq
where $t$ and $t'$ ($r$ and $r'$) are the transmission (reflection) amplitudes.  It is straightforward to show that the unitarity condition ($\mathcal{S}^\dagger \mathcal{S} = 1$) yields $|t|^2=|t'|^2\equiv T$, $|r|^2=|r'|^2\equiv R$, $R+T=1$, and $r' t^* + r^* t'= r t^* + r'^* t' = 0$ .  We can thus write
\beq
    \mathcal{S} = \bpmat
        \sqrt{T} e^{i \phi_t} & \sqrt{1-T} e^{i \phi_{r'}}\\
        \sqrt{1-T} e^{i \phi_r} & \sqrt{T} e^{i \phi_{t'} }
    \epmat,
\eeq
with $t= \sqrt{T} e^{i\phi_t}$, $t' = \sqrt{T} e^{i \phi_{t'}}$, $r = \sqrt{R}e^{i\phi_r}$, $r' = \sqrt{R}e^{i \phi_{r'}}$, and (unless $T=0$ or $1$) $\phi_t + \phi_{t'} -  \phi_r - \phi_{r'} = \pi$ (modulo $2\pi$).

Provided that the transmission coefficient $T$ is non-zero, one can relate left to right amplitudes using a scattering transfer matrix $\mathcal{T}$ as follows:
\beq
    \bpmat
        \Psi_R^+\\
        \Psi_R^-
    \epmat
    =\mathcal{T}
    \bpmat
        \Psi_L^+\\
        \Psi_L^-
    \epmat,
\eeq
where
\beq 
    \mathcal{T} = 
    \bpmat
        1/t^* & r' / t' \\
        -r/t' & 1/t'
    \epmat.\label{eq:general scattering transfer matrix}
\eeq
Two scatterers in sequence ($\mathcal{S}_1$ to the left of $\mathcal{S}_2$) are described by an $S$ matrix denoted $\mathcal{S}_{12}$.  Let us parametrize $\mathcal{S}_1$, $\mathcal{S}_2$, $\mathcal{S}_{12}$ as in Eq.~\eqref{eq:S general parametrization amplitudes}.  $\mathcal{S}_{12}$ may be obtained from its corresponding scattering transfer matrix $\mathcal{T}_{12}=\mathcal{T}_2\mathcal{T}_1$.  Using unitarity, it is straightforward to obtain from this matrix equation the following exact relations:
\bseq
\begin{align}
    t_{12}&= \frac{t_1 t_2}{1-r_1' r_2},\label{eq:t12}\\
    t_{12}'&= \frac{t_1' t_2'}{1-r_1' r_2},\label{eq:t12p}\\
    r_{12}&= r_1\frac{1 -r_2/r_1'^*}{1-r_1' r_2},\label{eq:r12}\\
    r_{12}'&= r_2'\frac{1-r_1'/r_2^* }{1-r_1' r_2}.\label{eq:r12p}
\end{align}
\eseq
We can now derive the recursion relations for the transmission coefficient and reflection phase.  Identifying $\mathcal{S}_{1\dots N}\equiv\mathcal{S}_1$, $\mathcal{S}_{N+1}\equiv\mathcal{S}_{2}$, and $\mathcal{S}_{1\dots N+1}\equiv\mathcal{S}_{12}$, we obtain Eq.~\eqref{eq:transmission coeff recursion} from the main text by taking the modulus-squared of either Eq.~\eqref{eq:t12} or Eq.~\eqref{eq:t12p}.  Also, we can write Eq.~\eqref{eq:r12p} as
\beq
    r_{12}' = -\frac{r_1'r_2'}{r_2^*}\frac{1-r_2^*/r_1'}{1-r_1'r_2},
\eeq    
and taking the argument of both sides yields Eq.~\eqref{eq:reflection phase recursion} from the main text.

\section{The Schrader \textit{et al}. formula}\label{sec:Schrader et al.'s formula}

First we state the main results of Schrader \textit{et al}. (Ref.~\cite{SchraderPerturbative2004}) with some adjustments of notation to match our conventions.  Then we show that Eqs.~\eqref{eq:lloc to 2nd order} and~\eqref{eq:sigma to 2nd order} in the main text follow straightforwardly from their result.

\subsection{Statement of their result}
Schrader \textit{et al}. consider a product of $2\times2$ matrices $K_{1\dots N}= K_N\dots K_1$, where, as in the main text, we are using the convention that the subscript $n$ stands for dependence on some disorder variables that are i.i.d.  Each $K_n$ also depends on a global (i.e., site-independent and disorder-independent) real parameter $\lambda$.  By assumption, each $K_n$ satisfies three conditions.  The first is a pseudo-unitarity condition:
\beq
    K_n^\dagger \sigma^y K_n = \sigma^y.\label{eq:pseudo-unitarity condition with sigmay}
\eeq
The second condition is that all $K_n$ commute at $\lambda = 0$:
\beq
    [K_n, K_{n'}] \rvert_{\lambda=0} =0,\label{eq:commutativity}
\eeq
and the third is a trace inequality at $\lambda=0$:
\beq
    | \text{Tr } K_n\rvert_{\lambda=0} | < 2.\label{eq:trace condition}
\eeq
Note that the pseudo-unitarity condition~\eqref{eq:pseudo-unitarity condition with sigmay} with $\sigma^z$ instead of $\sigma^y$ is the defining condition for a scattering transfer matrix $\mathcal{T}_n$ (as it is equivalent to the unitarity of the corresponding $S$ matrix).  Schrader \textit{et al}. point out that the two types of matrices are isomorphic, being related by a unitary transformation:
\beq
    C \equiv \frac{1}{\sqrt{2}}
    \bpmat
    i & i \\
    1 & -1
    \epmat
    ,\qquad 
    K_n \equiv C \mathcal{T}_n C^{-1}.\label{eq:isomorphism}
\eeq
They write the matrix that rotates by an angle $\eta$ as
\beq
    R_\eta = 
    \bpmat
        \cos\eta & -\sin\eta\\
        \sin\eta & \cos\eta
    \epmat.
\eeq    
They show that any $K_n$ may be written as
\beq
    M K_n M^{-1} = e^{i\xi_n} R_{\eta_n} \exp[\lambda P_n + \lambda^2 Q_n + O(\lambda^3)],\label{eq:decomposition}
\eeq
where $M \in \text{SL}(2,\mathbb{R})$, $P_n$ and $Q_n$ are real, traceless $2\times2$ matrices, and $M$, $\eta_n$, $P_n$, and $Q_n$ are all independent of $\lambda$ (while $\xi_n$ may depend on $\lambda$).  The phases $\xi_n$ and $\eta_n$ are explicitly found as follows.  From $\text{det }K_n^\dagger K_n=1$ [which follows from Eq.~\eqref{eq:pseudo-unitarity condition with sigmay}] we have $\text{det }K_n = e^{2i\xi_n}$ for some $\xi_n \in [0,\pi)$.  The phase $\eta_n$ is defined by $\cos \eta_n = \frac{1}{2}(e^{-i\xi_n} \text{Tr } K_n)\rvert_{\lambda=0}$.

They further define a $\lambda$-independent constant $\beta_n$ in terms of the matrix elements of $P_n$:
\beq
    \beta_n = (P_n)_{1,1} - \frac{1}{2}i \left[ (P_n)_{1,2} + (P_n)_{2,1} \right].\label{eq:beta definition}
\eeq
Their results for the localization length and variance [see their Eqs. (6), (7), and (22)] are given in our conventions by
\beq
    \frac{2}{\lloc} = \left\{ \langle |\beta_n|^2 \rangle_n +2 \text{Re}\left[\frac{ \langle\beta_n\rangle_n \langle \beta_n^* e^{i 2\eta_n} \rangle_n}{1- \langle e^{2i\eta_n} \rangle_n} \right] \right\}\lambda^2 + O(\lambda^3)\label{eq:lloc Schrader et al.}
\eeq
and
\beq
\lim_{N\to\infty}\frac{\sigma(N)^2}{2N} = \frac{2}{\lloc} + O(\lambda^3).\label{eq:sigma Schrader et al.}
\eeq

\subsection{Corollary to the Schrader \textit{et al}. result}
We now use the isomorphism~\eqref{eq:isomorphism} to apply Eqs.~\eqref{eq:lloc Schrader et al.} and~\eqref{eq:sigma Schrader et al.} to a scattering transfer matrix $\mathcal{T}_n$, which in general may be parametrized as
\beq
    \mathcal{T}_n = 
    \bpmat
        1/t_n^* & r_n'/t_n'\\
        -r_n/t_n' & 1/t_n'
    \epmat.
\eeq
Note that by unitarity (or, equivalently, the pseudo-unitarity condition $\mathcal{T}_n^\dagger\sigma^z\mathcal{T}_n=\sigma^z$), we have $|r_n|=|r_n'| \equiv \sqrt{R_n}$, $|t_n|=|t_n'|\equiv\sqrt{T_n}$, and $R_n+T_n =1$.  We write $t_n = \sqrt{T_n}e^{i\phi_{t_n}}$ and $t_n' = \sqrt{T_n}e^{i\phi_{t_n'}}$.  We assume that the $S$ matrix elements depend on the global parameter $\lambda$, and in particular that $r_n$ and $r_{n'}$ are each proportional to $\lambda$.  Then by unitarity, $T_n = 1 + O(\lambda^2)$, and so
\beq
    t_n = e^{i\phi_{t_n}} + O(\lambda),\ t_n' = e^{i\phi_{t_n'}} + O(\lambda).
\eeq
We expand the transmission phases as $\phi_{t_n}=\phi_{t_n}^{(0)} + \phi_{t_n}^{(1)} + O(\lambda^2)$ and $\phi_{t_n'}=\phi_{t_n'}^{(0)} + \phi_{t_n'}^{(1)} + O(\lambda^2)$, where the superscript indicates the order in $\lambda$.  The scattering transfer matrix at $\lambda=0$ simplifies to
\beq
    \mathcal{T}_n\rvert_{\lambda=0} =
    \bpmat
        e^{i\phi_{t_n}^{(0)}} & 0 \\
        0 & e^{-i\phi_{t_n'}^{(0)}}.
    \epmat
\eeq
It is then clear that all $\mathcal{T}_n$ commute at $\lambda=0$.  Furthermore, $\Tr \mathcal{T}_n\rvert_{\lambda=0} = e^{i\phi_{t_n}^{(0)}}+e^{-i\phi_{t_n'}^{(0)}}$, which has absolute value that we can assume to be less than $2$ in almost all disorder realizations.  In this last step we are appealing to ``typical'' disorder, since even if $\phi_{t_n}^{(0)}=\phi_{t_n'}^{(0)}$ is enforced by symmetry, the point $\phi_{t_n}^{(0)}=0$ is a set of measure zero.  The isomorphism~\eqref{eq:isomorphism} then shows that Eq.~\eqref{eq:commutativity} and inequality~\eqref{eq:trace condition} hold, while Eq.~\eqref{eq:pseudo-unitarity condition with sigmay} also follows from the isomorphism.  Thus, the conditions for the Schrader et al. formula are met.

We proceed to identify the various terms in the decomposition~\eqref{eq:decomposition}.  From Eq.~\eqref{eq:isomorphism}, we readily obtain
\beq
    \text{Det }K_n = e^{i(\phi_{t_n} - \phi_{t_n'})},
\eeq
and thus
\beq
    \xi_n = \frac{1}{2}(\phi_{t_n} - \phi_{t_n'}).\label{eq:xi}
\eeq
We then obtain $\eta_n =\frac{1}{2}(\phi_{t_n}^{(0)} + \phi_{t_n'}^{(0)})$.

We then define a real, traceless, $\lambda$-independent matrix $P_n$ by
\beq
    \lambda P_n =\frac{1}{2}
    \bpmat
        -(r_n + r_n^*) & -i (r_n - r_n^*) -\phi_{t_n}^{(1)} - \phi_{t_n'}^{(1)} \\
        -i (r_n - r_n^*) +\phi_{t_n}^{(1)} + \phi_{t_n'}^{(1)} & r_n + r_n^*
    \epmat.
\eeq
Then it is straightforward to verify
\beq
    K_n = e^{i\xi_n} R_{\eta_n}\left(1+\lambda P_n\right) + O(\lambda)^2,
\eeq
which confirms the decomposition~\eqref{eq:decomposition} with $M=1$ (and $Q$ will not be needed).

From Eq.~\eqref{eq:beta definition}, we then read off the constant $\beta_n$:
\beq
    \lambda \beta_n = - r_n.\label{eq:lambda beta}
\eeq
Also note
\bseq
\begin{align}
    e^{2i\eta_n} &= t_n'/t_n^* + O(\lambda)\\
    &= - \frac{r_n r_n'}{R_n} + O(\lambda),\label{eq:2eta average}
\end{align}
\eseq
where we used the unitarity constraint $r_n' t_n^* + r_n^* t_n'=0$.  Substituting into their results~\eqref{eq:lloc Schrader et al.} and~\eqref{eq:sigma Schrader et al.} then yields Eqs.~\eqref{eq:lloc to 2nd order} and~\eqref{eq:sigma to 2nd order} from the main text with error terms $O(\lambda^3)$ instead of $O(\lambda^4)$.  The vanishing of the third-order contributions is discussed in the main text.

\subsection{Correspondence of the phase variables}
As we have mentioned in the main text, Schrader \textit{et al}.~\cite{SchraderPerturbative2004} emphasize the importance of a non-uniform phase distribution in obtaining the correct answer for the inverse localization length and variance at leading order.  To further clarify the connection between their work and ours, we show in this section that the phase variable they consider (in a calculation that does not explicitly refer to scattering) is, in the localized regime, related by a factor of $1/2$ to the reflection phase in our setup.  We then confirm that our calculation yields the same result for the first Fourier coefficient of the phase distribution ($p_{\infty,\ell=-1}$) as their calculation.

Their phase variable is $\theta_n$ (not to be confused with the coin parameter in the DTQW in our main text).  Note from the paragraph above their Sec. 3 that they set $\xi_\sigma(\lambda)=0$ ($\xi_n=0$ in our notation).     The recursion relation they obtain is [their Eqs. (12) and (13)]
\beq
    e^{2i\theta_{n+1}}= \frac{ \bra{v} \mathcal{K}_n \ket{e_\theta}}{\bra{\bar{v}} \mathcal{K}_n\ket{e_\theta}},\label{eq:theta recursion}
\eeq
where [see their Eqs. (9) and (10)]
\bseq
\begin{align}
    \ket{v} &= \frac{1}{\sqrt{2}}
    \bpmat
        1\\
        -i
    \epmat,\\
    e_\theta &= 
    \bpmat
        \cos\theta\\
        \sin\theta
    \epmat.
\end{align}
\eseq
From Eq.~\eqref{eq:xi} and $\xi_n=0$ we see that $\phi_{t_n}=\phi_{t_n}'$, i.e., $t_n=t_n'$.  Using Eq.~\eqref{eq:isomorphism} and unitarity, we then find that their recursion relation~\eqref{eq:theta recursion} becomes
\beq
    e^{2i\theta_{n+1}}= e^{i (\phi_{r_n}+\phi_{r_n'} + \pi)} \frac{1- r_n^* e^{-2i\theta_n}}{1-r_n e^{2i \theta_n}}.
\eeq
Identifying $2\theta_n = \phi_{r_{1\dots n}'}$ and taking the logarithm of both sides yields exactly the recursion relation~\eqref{eq:phirp recursion Rto1} for the reflection phase in the localized regime (which we recall consists of $n$ large enough that $R_{1\dots n}\approx 1$).  This suffices to show that the probability distribution $p_{\infty}(\phi_{r'})$ coincides with the probability distribution of $2\theta_n$ for large $n$; while $2\theta_n$ and $\phi_{r_{1\dots n}'}$ need not coincide for small $n$, this has no effect in the limit of many sites.  Note in particular that the ``invariant measure'' discussed by Schrader \textit{et al}. (and also in the prior mathematics literature~\cite{FurstenbergNoncommuting1963,BougerolProducts1985,JitomirskayaDelocalization2003}) is the same, up to trivial rescaling, as $p_\infty(\phi_{r'})$.

Taking $N\to\infty$ in their Lemma 1 should thus yield, in our notation, the first Fourier component $p_{\infty,\ell=-1}$ explicitly up to error $O(|r_n|^2)$ and also the error estimate $p_{\infty,\ell=-2}= O(|r_n|)$ for the second Fourier component.  Recall that we obtained $2\pi p_{\infty,\ell=-1}= \gamma^{(1)} + O(|r_n|^3)$ and $2\pi p_{\infty,\ell=-2}= \gamma^{(2)}+ O(|r_n|^4)$ [see Eq.~\eqref{eq:p to 3rd order}], where the constants $\gamma^{(1)}$ and $\gamma^{(2)}$ are given by Eqs.~\eqref{eq:gamma1} and~\eqref{eq:gamma2}.  Recalling Eqs.~\eqref{eq:lambda beta} and~\eqref{eq:2eta average}, we see that $\lim_{N\to\infty}I_1(N)$ in the unnumbered equation above their Eq. (20) agrees with what we found for the first Fourier component (though we find a stronger error estimate).  Furthermore, our explicit result for the second Fourier component is consistent with (and stronger than) their error estimate because $\gamma^{(2)} = O(|r_n|^2)$.

\section{Comparison with the weak scattering approximation}\label{sec:Comparison with the weak scattering approximation}
Here we expand on our comments in footnotes~\cite{Note2,Note6}, providing a detailed comparison of our approach with the WSA developed in Refs.~\cite{AsatryanSuppression2007,AsatryanAnderson2010} and reviewed in Ref.~\cite{GredeskulAnderson2012}.

We note first that Refs.~\cite{AsatryanSuppression2007,AsatryanAnderson2010,GredeskulAnderson2012} use the convention that new scatterers are added to the left edge of the sample, rather than our convention of adding them to the right edge.  The two conventions are related by exchanging $\mathcal{S}_n\leftrightarrow\mathcal{S}_n^\dagger$ and $\mathcal{S}_{1\dots N}\leftrightarrow\mathcal{S}_{1\dots N}^\dagger$, i.e.,
\begin{align}
    &t_n \leftrightarrow t_n^*,\  t_n' \leftrightarrow t_n'^*,\  r_n \leftrightarrow r_n'^*,\notag\\
    &t_{1\dots N} \leftrightarrow t_{1\dots N}^*,\  t_{1\dots N}' \leftrightarrow t_{1\dots N}'^*,\  r_{1\dots N} \leftrightarrow r_{1\dots N}'^*.\label{eq:convention switch}
\end{align}
We also note that Ref.~\cite{GredeskulAnderson2012} considers the special case $t_n=t_n'$ and $r_n=r_n'$ (see below).  [Note that in this case, Eqs.~\eqref{eq:t12}--\eqref{eq:r12p} show that $t_{1\dots N}'=t_{1\dots N}$, but $r_{1\dots N}'$ may differ from $r_{1\dots N}$.]

With $\mathcal{S}_{1\dots N}\equiv\mathcal{S}_1$, $\mathcal{S}_{N+1}\equiv\mathcal{S}_{2}$, and $\mathcal{S}_{1\dots N+1}\equiv\mathcal{S}_{12}$, the exact recursion relations~\eqref{eq:t12}and~\eqref{eq:r12p} become
\bseq
\begin{align}
    t_{1\dots N+1}&= \frac{t_{1\dots N} t_{N+1}}{1-r_{1\dots N}' r_{N+1}},\label{eq:t recursion}\\
    r_{1\dots N+1}' &= r_{N+1}' + \frac{r_{1\dots N}' t_{N+1}t_{N+1}'}{1-r_{1\dots N}'r_{N+1}}\label{eq:rp recursion for later WSA},
\end{align}
\eseq
where we have used a relation that follows from the unitarity constraints $r_n' t_n'^*+ r_n^* t_n =0$ and $R_n+T_n=1$:
\beq
    r_n'/r_n^* = - t_n t_n' + r_n r_n'.\label{eq:ttp relation} 
\eeq
Setting $t_n=t_n'$ and $r_n=r_n'$ and making the change of conventions given by~\eqref{eq:convention switch}, we see that Eqs.~\eqref{eq:t recursion} and~\eqref{eq:rp recursion for later WSA} recover the complex conjugates of the exact recursion relations as presented in Ref.~\cite{GredeskulAnderson2012} [Eqs. (2.16) and (2.17) there], in a slightly different notation.

We proceed next to review the WSA calculation of the inverse localization length~\cite{ AsatryanAnderson2010}.  We generalize the calculation slightly by not assuming $t_n=t_n'$ or $r_n=r_n'$.  We confirm that the WSA result agrees with Eq.~\eqref{eq:lloc to 4th order} at the leading order ($|r_n|^2$) but disagrees at the next non-vanishing order ($|r_n|^4$).  Throughout, we follow the same convention as used in our main text of adding new sites to the right.   

The WSA makes the following expansions of Eqs.~\eqref{eq:t recursion} and~\eqref{eq:rp recursion for later WSA}~\cite{GredeskulAnderson2012}:
\bseq
\begin{align}
   \ln t_{1\dots N+1} &= \ln t_{1\dots N} + \ln t_{N+1} + r_{1\dots N}'r_{N+1} + \dots,\label{eq:ln t recursion WSA}\\
   r_{1\dots N+1}' &= r_{N+1} + r_{1\dots N}' t_{N+1}t_{N+1}' + \dots,\label{eq:rp recursion WSA} 
\end{align}
\eseq
[Recalling~\eqref{eq:convention switch}, we see that Eqs.~\eqref{eq:ln t recursion WSA} and~\eqref{eq:rp recursion WSA} are equivalent to Eqs. (2.16) and (2.17) of Ref.~\cite{GredeskulAnderson2012} once we specialize to $t_n=t_n'$ and $r_n=r_n'$.].  Equation~\eqref{eq:ln t recursion WSA} is the expansion of Eq.~\eqref{eq:t recursion} to first order (in $|r_{N+1}|$).  As noted below Eq. 2.19 in Ref.~\cite{GredeskulAnderson2012}, Eq.~\eqref{eq:rp recursion WSA} is an uncontrolled approximation to Eq.~\eqref{eq:rp recursion for later WSA} because the first-order term $r_{1\dots N}'^2 t_{N+1}t_{N+1}'r_{N+1}$ is omitted (while $r_{N+1}$ is kept).

In the WSA approach, the inverse localization length is calculated by taking Eqs.~\eqref{eq:ln t recursion WSA} and~\eqref{eq:rp recursion WSA} to be the exact recursion relations.  In this way it is straightforward to obtain, by induction,
\beq
    r_{1\dots N}' = \sum_{j=1}^N r_j'\sum_{m=j+1}^N t_m t_m'
\eeq
and
\beq
    \ln t_{1\dots N} = \sum_{j=1}^N \ln t_j + \sum_{m=2}^N\sum_{j=m}^N r_{j-m+1}'r_j \prod_{p=j-m+2}^{j-1}t_pt_p',
\eeq
the latter of which recovers Eq. (13) of Ref.~\cite{ AsatryanAnderson2010} if we set $t_n=t_n'$ and $r_n=r_n'$ [also recall~\eqref{eq:convention switch}].  Then, relabeling some summation variables and using the assumption that the disorder is i.i.d., we obtain
\beq
    \langle \ln t_{1\dots N} \rangle_{1\dots N} = N \langle \ln t_n\rangle_n + \langle r_n\rangle_n \langle r_n'\rangle_n \sum_{j=0}^{N-2}\sum_{m=0}^j \langle t_n t_n'\rangle_n^m.
\eeq
The geometric sums may be carried out explicitly.  Recalling that
\beq
    2/\lloc = \lim_{N\to\infty}\langle -\ln T_{1\dots N}\rangle_{1\dots N}/N
\eeq
and $\ln T_{1\dots N} = 2 \Re[\ln t_{1\dots N}]$, we thus obtain
\beq
    \left(\frac{2}{\lloc}\right)_{\text{WSA}} = \langle -\ln T_n\rangle_n - 2 \text{Re}\left[ \frac{\langle r_n\rangle_n \langle r_n'\rangle_n }{1-\langle t_n t_n'\rangle_n}\right],\label{eq:lloc WSA}
\eeq
which indeed recovers Eq. (43) of Ref.~\cite{AsatryanAnderson2010} in the case $t_n=t_n'$ and $r_n=r_n'$ [note that the change of conventions~\eqref{eq:convention switch} has no effect here].  Noting that $r_n'/r_n^* = r_n r_n'/R_n$ and recalling Eq.~\eqref{eq:ttp relation}, we then expand Eq.~\eqref{eq:lloc WSA} to obtain
\beq
    \left(\frac{2}{\lloc}\right)_{\text{WSA}} = \langle R_n\rangle_n -2 \text{Re}\left[ \alpha_1\langle r_n\rangle_n \langle r_n'\rangle_n\right]
    + \frac{1}{2}\langle R_n^2\rangle_n  -2 \text{Re}\left[ \alpha_1^2\langle r_n\rangle_n \langle r_n'\rangle_n \langle r_n r_n'\rangle_n \right] + O(|r_n|^6).
\eeq
Comparing to Eq.~\eqref{eq:lloc to 4th order} from the main text, we see that the WSA calculation yields the correct answer for $2/\lloc$ at the leading order, but misses several terms that appear at the next non-vanishing order.

To compare the WSA with our work in further detail, we examine the WSA recursion relations~\eqref{eq:ln t recursion WSA} and~\eqref{eq:rp recursion WSA} in the localized regime.  We show that if these relations are treated consistently as first-order approximations (rather than following the WSA approach of treating them as exact), then a factor of $2$ is missing from one of the terms in the leading-order expression for the inverse localization length.

In the localized regime, $r_{1\dots N}$ becomes a pure phase ($r_{1\dots N}\to e^{i\phi_{r_{1\dots N}}}$), and Eqs.~\eqref{eq:ln t recursion WSA} and~\eqref{eq:rp recursion WSA} yield the following recursion relations.  For the minus logarithm of the transmission coefficient, the WSA yields
\beq
    s_{1\dots N+1} = s_{1\dots N} + s_{N+1} - 2 \text{Re}[ r_{N+1}e^{i\phi_{r_{1\dots N}'}}] + \dots,
\eeq
which is indeed the leading-order expansion of Eq.~\eqref{eq:s recursion Rto1} from the main text.  For the reflection phase, the WSA yields
\beq
    \phi_{r_{1\dots N+1}'}= \phi_{r_{1\dots N}'}+\pi + \phi_{r_{N+1}} +\phi_{r_{N+1}'}+ \beta\text{Im}[ r_{N+1}e^{i\phi_{r_{1\dots N}'}}] + \dots,\label{eq:phirp recursion with beta}
\eeq
with $\beta = 1$; this agrees with Eq.~\eqref{eq:phirp recursion Rto1} at the zeroth order but disagrees at the first order by a constant factor ($\beta=2$ is obtained if all first order terms are accounted for).

It is straightforward to repeat the calculation of Sec.~\ref{sec:General calculation} to obtain the limiting distribution of the reflection phase at first order under the assumption of the recursion relation~\eqref{eq:phirp recursion with beta} (with $\beta$ as a free parameter).  The result is
\beq
    2\pi p_{\infty}(\phi_{r'}) = 1 + \beta \text{Re}\left[\alpha_1 \langle r_n' \rangle_ne^{-i\phi_{r'}} \right] + O(|r_n|^2).
\eeq
From Eq.~\eqref{eq:lloc in terms of Fourier coefficients}, we then read off the inverse localization length at leading order:
\beq
    \frac{2}{\lloc} = \langle R_n \rangle_n - \beta \text{Re}\left[ \alpha_1 \langle r_n\rangle_n\langle r_n'\rangle_n \right] + O(|r_n|^4),\label{eq:lloc with beta}
\eeq
which indeed agrees with Eq.~\eqref{eq:lloc to 2nd order} for $\beta=2$ (but not for $\beta=1$).

\section{Existence of the limiting distribution of the reflection phase}\label{sec:Existence of the limiting distribution of the reflection phase}
Given the assumption that localization occurs, we show that the probability distribution of the reflection phase, $p_N(\phi)$, has a limit as $N\to\infty$.  The calculation here is a slight generalization of that of Lambert and Thorpe~\cite{LambertPhase1982} with some further comments on its implications.  In view of the phase correspondence discussed in Appendix~\ref{sec:Schrader et al.'s formula}, the existence of $p_\infty(\phi)$ may also be viewed as a consequence of theorems for the invariant measure~\cite{FurstenbergNoncommuting1963,BougerolProducts1985}.

In the main text, we subdivided an $(N+1)$-site sample into an $N$-site sample plus an additional site to the right.  This yielded a relation between $p_{N+1}(\phi)$ and $p_N(\phi)$ for large $N$ [Eq.~\eqref{eq:p recursion finite N}].  We now show that subdividing the $(N+1)$-site sample in another way (an $N$-site sample plus an additional site to the left) yields another relation between $p_{N+1}(\phi)$ and $p_N(\phi)$ for large $N$, namely, the simple relation that the probability distribution stops changing as $N$ increases:
\beq
    p_{N+1}(\phi)=p_N(\phi).\label{eq:p recursion 2nd equation}
\eeq
Equation~\eqref{eq:p recursion 2nd equation} shows that $p_\infty(\phi)$ exists, which in turn allows us to solve Eq. ~\eqref{eq:p recursion finite N} (for large $N$) order by order in the scattering expansion, as we have done in the main text.

To obtain Eq.~\eqref{eq:p recursion 2nd equation} we start by noting that, since the disorder is i.i.d., we are free to label the $N$ sites as $n=-N,\dots,-1$; in particular, we have \footnote{One can confirm that Eq.~\eqref{eq:p alternate definition} agrees with the expression in the main text by writing the integrals over the disorder variables $\mathbf{D}_n$ explicitly and then re-labeling the variables; see footnote~\cite{Note4}.}
\beq
    p_N(\phi)= \langle \delta(\phi- \phi_{r_{-N \dots -1}'} )\rangle_{-N \dots -1}.\label{eq:p alternate definition} 
\eeq
With $\mathcal{S}_{-(N+1)}\equiv\mathcal{S}_1$, $\mathcal{S}_{-N\dots -1}\equiv\mathcal{S}_{2}$, and $\mathcal{S}_{-(N+1)\dots -1}\equiv\mathcal{S}_{12}$, Eq.~\eqref{eq:r12p} becomes
\beq
    r_{-(N+1)\dots -1}'= r_{-N\dots -1}'\frac{1-r_{-(N+1)}'/r_{-N\dots -1}^* }{1-r_{-(N+1)}' r_{-N\dots -1}}\label{eq:rp recursion}.
\eeq
Assuming localization occurs, $r_{1\dots N}$ and $r_{1\dots N}'$ become pure phases for large $N$: $r_{-N\dots -1}\to e^{i\phi_{r_{-N\dots -1}}}$, $r_{-N\dots -1}'\to e^{i\phi_{r_{-N\dots -1}'}}$.  Then Eq.~\eqref{eq:rp recursion} simplifies to
\beq
    e^{i(\phi_{r_{-(N+1)\dots -1}} - \phi_{r_{-1\dots -N}'})} = 1,\label{eq:left-to-right reflection phase constant}
\eeq
which implies that the recursion relation for the reflection phase simplifies to $\phi_{r_{-(N+1)\dots -1}'}=\phi_{r_{-N \dots -1}'}$ for large $N$.  From Eq.~\eqref{eq:p alternate definition} we then read off Eq.~\eqref{eq:p recursion 2nd equation}.

Next, we show the connection between the calculation above and that of Lambert and Thorpe ~\cite{LambertPhase1982}.  To make the precise comparison, we repeat the calculation considering the left-to-right reflection phase ($\phi_r$) instead of right-to-left reflection phase ($\phi_{r'}$).  We write the probability distribution of $\phi_r$ as $p_N^{[\phi_r]}(\phi)$, where the superscript is only a label, and we subdivide an $(N+1)$-site sample in the same way as in the main text (i.e., the ``additional'' site is on the right).  From Eq.~\eqref{eq:r12} with $\mathcal{S}_{1\dots N}\equiv\mathcal{S}_1$, $\mathcal{S}_{N+1}\equiv\mathcal{S}_{2}$, and $\mathcal{S}_{1\dots N+1}\equiv\mathcal{S}_{12}$, we then obtain the analogs of Eqs.~\eqref{eq:rp recursion} and~\eqref{eq:left-to-right reflection phase constant}:
\bseq
\begin{align}
    r_{1\dots N+1}&= r_{1\dots N}\frac{1 -r_{N+1}/r_{1\dots N}'^*}{1-r_{1\dots N}' r_{N+1}},\label{eq:r recursion}\\
    e^{i(\phi_{r_{1\dots N+1}} - \phi_{r_{1\dots N}})} &= 1,\label{eq:reflection phase constant}
\end{align}
\eseq
where the second equation takes $N$ to be large and uses the assumption of localization ($r_{1\dots N}\to e^{i\phi_{r_{1\dots N}}}$, $r_{1\dots N}'\to e^{i\phi_{r_{1\dots N}'}}$).  From here we can demonstrate $p_{N+1}^{[\phi_r]}(\phi) = p_N^{[\phi_r]}(\phi)$ in the same way as before [and then Eq.~\eqref{eq:p recursion 2nd equation} follows by symmetry].

In a slightly more restricted setting, Eqs.~\eqref{eq:r recursion} and~\eqref{eq:reflection phase constant} have been obtained previously in an equivalent form by Lambert and Thorpe~\cite{LambertPhase1982}, as we now explain.  The restriction is the assumption that the transmission amplitudes of the local $S$ matrix are equal: $t_n=t_n'$.  [Note that Eqs.~\eqref{eq:t12} and~\eqref{eq:t12p} then imply that the sample $S$ matrix has the same property, i.e., $t_{1\dots N}=t_{1\dots N}'$.]  Equation~\eqref{eq:r recursion} is then equivalent to Eq. (5c) of Ref.~\cite{LambertPhase1982}.  (To see this, we note the following correspondence between their notation and ours: $\theta_n = -\phi_{t_n}$, $\phi_n = \phi_{r_n} - \phi_{t_n}+ \pi$, $\alpha_n=-\phi_{t_{1\dots n}}$, $\beta_n=\phi_{r_{1\dots n}} - \phi_{t_{1\dots n}}+ \pi$, $\rho_n = R_n/T_n$, $s_n = \sqrt{R_n}$, $z_n = R_{1\dots n}/T_{1\dots n}$, $r_n = \sqrt{R_{1\dots n}}$, and $\epsilon_n = \phi_{r_{1\dots n-1}}-2\phi_{t_{1\dots n-1}}- \phi_{r_n} = \pi - \phi_{r_{1\dots n-1}'}-\phi_{r_n}$, where their quantities appear on the left-hand side and ours on the right.)  Also, Eq.~\eqref{eq:reflection phase constant} is stated in an equivalent form (regarding the quantity $\alpha_n-\beta_n$) in the paragraph below Eq. (6) of Ref.~\cite{LambertPhase1982}.

The above calculation has a consequence for numerics, and possibly experiment, which we now explain.  Sampling the reflection phase distribution could be done by finding the reflection phase in a number of different disorder realizations of a chain of fixed length; however, it may be advantageous to instead find the reflection phase for chains over a range of lengths, with no change in the disorder realization of the existing sites as new sites are added.  According to the above calculation, the latter procedure will not work for sampling $\phi_r$ if the chain is increased to the right, since only a constant value of $\phi_r$ will be obtained (although our numerics indicate that $\phi_{r'}$ \emph{can} be sampled this way).  Similarly, $\phi_{r'}$ cannot be sampled by increasing the chain to the left (but we expect that $\phi_r$ can).

\section{Evaluation of a Fourier integral}\label{sec:Evaluation of a Fourier integral}
Our series for the inverse localization length is based on Eq.~\eqref{eq:lloc in terms of Fourier coefficients}, which we arrive at by writing the integral in Eq.~\eqref{eq:lloc as integral explicit} as a sum over Fourier coefficients.  Here we present this calculation explicitly.

Define $G_n(\phi)= \ln(1- r_n e^{i\phi} - r_n^* e^{-i \phi} +R_n)$.  Since $G_n(\phi)$ and $p_{\infty}(\phi)$ are both real, Eq.~\eqref{eq:lloc as integral explicit} may be written in terms of Fourier coefficients [$G_{n,\ell} \equiv \int_{-\pi}^\pi \frac{d\phi}{2\pi}\ e^{-i \ell \phi} G_n(\phi)$] as
\beq
    \frac{2}{\lloc}= \langle -\ln T_n\rangle_n + 2\pi p_{\infty,0}\langle G_{n,0}\rangle_n
    + 4\pi\text{Re}\left[ \sum_{\ell=1}^\infty p_{\infty,-\ell}\langle G_{n,\ell}\rangle_n\right].
\eeq
Thus, Eq.~\eqref{eq:lloc in terms of Fourier coefficients} follows from $G_{n,\ell}= 0$ for $\ell=0$ and $-r_n^\ell/\ell$ for $\ell>0$, which we now show. The same Fourier integral has been done in Ref.~\cite{RauhAnalytical2009} [Eq. (52) there)].

Writing $r_n= \sqrt{R_n}e^{i\phi_{r_n}}$, we shift $\phi+\phi_{r_n}\to \phi$ to obtain
\beq
    G_{n,\ell} = e^{i \ell \phi_{r_n}} \int_{-\pi}^\pi\frac{d\phi}{2\pi}\ e^{-i\ell\phi} \ln\left[ 1- \sqrt{R_n} e^{i\phi} -\sqrt{R_n}e^{-i\phi} + R_n\right].
\eeq
To show $G_{n,\ell=0}=0$, we note that the integrand may be written as $\ln|1- \sqrt{R_n}e^{i\phi}|^2$, so the integral vanishes.  For $\ell>0$, we define $z=e^{i\phi}$ and $\alpha=\sqrt{R_n}/(1+R_n)$ (the disorder parameters represented by the site index $n$ are spectators); then we get
\beq
    G_{n,\ell} = e^{i \ell \phi_{r_n}} \int_{-\pi}^\pi\frac{d\phi}{2\pi}\ e^{-i\ell\phi} \ln\left[ 1- \alpha(z+1/z) \right].
\eeq
Next, we expand the logarithm to all orders in $\alpha$ to calculate the coefficient of $z^\ell$, which is what is the integral over $\phi$ yields.  We get
\beq
    \ln \left[1 - \alpha(z+ 1/z) \right] = -\sum_{j=1}^\infty\frac{\alpha^j}{j} \sum_{m=0}^j\binom{j}{m}z^{2m-j},
\eeq
from which we calculate the coefficient of $z^\ell$ to be
\bseq
\begin{align}
    (\text{coeff. of }z^\ell) &= -\sum_{j=1}^\infty\frac{\alpha^j}{j} \sum_{m=0}^j\binom{j}{m}\delta_{2m-j,\ell}\\
    &= - \sum_{m=0}^\infty \sum_{j=\text{max}\{1,m\}} \frac{\alpha^j}{j}\binom{j}{m}\delta_{2m-j,\ell}\\
    &= -\sum_{m=\ell}^\infty \frac{\alpha^{2m-\ell}}{2m-\ell}\binom{2m-\ell}{m}.
\end{align}
\eseq
This sum evaluates to  $-\left(\frac{1-\sqrt{1-4\alpha^2}}{2\alpha}\right)^\ell/\ell = - (\sqrt{R_n})^\ell/\ell$, which completes the calculation.

\section{Partial extension of our results to anomalies}\label{sec:Partial extension of our results to anomalies}
We consider the case that the inequality~\eqref{eq:generic condition} is only assumed for $0<|\ell| \le \ell_\text{max}$ with some finite value $\ell_\text{max} \ge 1$, thus generalizing the calculation of the main text (in which $\ell_\text{max}\to\infty$).  The main conclusion of this calculation is that if $\ell_\text{max}$ is finite, then the scattering expansion that we find for $2/\lloc$ in the main text is valid up to an error term $O(|r_n|^{\ell_\text{max}+1})$.  (We do not know if the odd powers are guaranteed to vanish in the case that $\ell_\text{max}$ is finite.)  For instance, the leading order expression~\eqref{eq:lloc to 2nd order} is valid provided that $\ell_\text{max}\ge3$, and the next-to-leading order expression~\eqref{eq:lloc to 4th order} is valid provided that $\ell_\text{max}\ge 5$.  We also discuss how our results for the reflection phase distribution are modified, and we present more detail for the calculation of the first few Fourier coefficients.

One motivation for doing this calculation is simply to confirm that having $\ell_\text{max}$ be large but finite, as opposed to infinite, makes no practical difference to our final results (as only very high orders are affected).  A second motivation is that the case of finite $\ell_\text{max}$ arises naturally in the Anderson model for anomalous values of the momenta, as we discuss in Sec.~\ref{sec:Anderson model with diagonal disorder}.  (More generally, we refer to any cases in which $\ell_\text{max}$ is finite \footnote{To avoid cluttering the notation, we sometimes use the same symbol $\ell_\text{max}$ to refer to the \emph{maximum} value of $\ell_\text{max}$ for which~\eqref{eq:generic condition} holds for all $|\ell|\le\ell_\text{max}$.} as ``anomalies.'')

For any $q\ge 0$, we write the order-$q$ part of Eq.~\eqref{eq:p first eqn with A} in Fourier space:
\beq
    \left[ 1-(-1)^\ell \langle v_n^{-\ell}\rangle_n\right] p_{\infty,\ell}^{(q)} = (-1)^\ell \langle v_n^{-\ell} \sum_{j=0}^{q-1} \sum_{\ell'=-\infty}^\infty p_{\infty,-\ell'}^{(j)} A_{n,\ell;\ell+\ell'}^{(q-j)}\rangle_n.\label{eq:p recursion sum over all ellprime general form Appendix} 
\eeq
Note that Eq.~\eqref{eq:p0 equation} is just Eq.~\eqref{eq:p recursion sum over all ellprime general form Appendix} for $q=0$.  If $0< |\ell| \le \ell_\text{max}$ then $1-(-1)^\ell \langle v_n^{-\ell}\rangle_n\ne 0$ by assumption, so we can divide this quantity on both sides of Eq.~\eqref{eq:p recursion sum over all ellprime general form Appendix} [yielding Eq.~\eqref{eq:p recursion sum over all ellprime}].

We first prove the following statement: For any $q\ge 0$ and any $\ell_\text{max}\ge q$, the Fourier coefficients $p_{\infty,\ell}^{(q)}$ for $|\ell| \le \ell_\text{max}-q$ are exactly the same as they are in the $\ell_\text{max}\to\infty$ case (which is treated in the main text).  Roughly speaking, this means that for large but finite $\ell_\text{max}$, corrections to the calculation of the main text can only occur either in large frequencies or in large orders of the expansion.  For example, if $\ell_\text{max} = 6$, then Eq.~\eqref{eq:p to 3rd order} is valid up to the addition of $j$th order terms of the form $e^{\pm i |\ell-j|\phi}$ (where $\ell\ge 7$ and $j=0,1,2,3$).

The proof by induction is as follows.  For $q=0$, Eq.~\eqref{eq:p recursion sum over all ellprime general form Appendix} shows that $p_{\infty,\ell}^{(0)}=0$ for $0< |\ell|\le \ell_\text{max}$, and also we still have $p_{\infty,0}^{(0)}=1/(2\pi)$ by normalization; thus, $p_{\infty,\ell}^{(0)}$ agrees with the $\ell_\text{max}\to\infty$ calculation for $|\ell|\le \ell_\text{max}$ [both yielding $p_{\infty,\ell}^{(0)} = \delta_{\ell,0}/(2\pi)$].  We then assume that for some $q\ge1$, the Fourier coefficients $p_{\infty,\ell}^{(j)}$ are the same as in the $\ell_\text{max}\to\infty$ case when $0\le j \le q-1$, $j\le \ell_\text{max}$, and $|\ell| \le \ell_\text{max} - j$.

We consider $q\le \ell_\text{max}$.  By normalization, $p_{\infty,0}^{(q)}=0$.  Thus, it is enough to consider $\ell$ with $0< |\ell|\le \ell_\text{max}-q$.  For these $\ell$, we can divide the quantity $1-(-1)^\ell \langle v_n^{-\ell}\rangle_n$ on both sides of Eq.~\eqref{eq:p recursion sum over all ellprime general form Appendix}.  Taking $\ell < 0$ without loss of generality, we then use the properties of the function $A$ that are mentioned below Eq.~\eqref{eq:p recursion} to truncate the $\ell'$ sum, yielding:
\begin{equation}
    p_{\infty,\ell}^{(q)} = \frac{(-1)^\ell}{1-(-1)^\ell \langle v_n^{-\ell}\rangle_n} \langle v_n^{-\ell} \sum_{j=0}^{q-1} \sum_{\ell'=0}^{|\ell|+q-j} p_{\infty,-\ell'}^{(j)}A_{n,\ell;\ell+\ell'}^{(q-j)}\rangle_n.
\end{equation}
In the sum on the right-hand side, we have $0\le j \le q-1$, $j\le \ell_\text{max}$, and $\ell' \le |\ell| + q -j \le \ell_\text{max}-j$, so each $p_{\infty,-\ell'}^{(j)}$ is as in the $\ell_\text{max}\to\infty$ case.  We can then use Eq.~\eqref{eq:vanishing of Fourier coefficients} to replace the upper limit of the $\ell'$ sum by $\text{min}\{|\ell|+q-j,j\}$.  We have thus arrived at an equation equivalent to Eq.~\eqref{eq:p recursion} (since the $\ell'$ sum there can be truncated to the same form using a property of the function $A$).  Thus, we have shown that $p_{\infty,\ell}^{(q)}$ is as in the $\ell_\text{max}\to\infty$ case, which completes the proof.

We proceed to consider the order-$q$ part (for some $q\ge 1$) of the formula~\eqref{eq:lloc in terms of Fourier coefficients} for the inverse localization length.  We have
\beq
    \left(\frac{2}{\lloc}\right)^{(q)} = \langle -(\ln T_n)^{(q)}\rangle_n - 4\pi \text{Re}\left[ \sum_{\ell=1}^{q} \frac{1}{\ell}p_{\infty,-\ell}^{(q-\ell)} \langle r_n^\ell \rangle_n\right].
\eeq
We assume $q\le \ell_\text{max}$.  Then $\ell \le \ell_\text{max}-(q-\ell)$ in the sum on the right-hand side, which implies that each $p_{\infty,-\ell}^{(q-\ell)}$ is as in the $\ell_\text{max}\to\infty$ case.  Thus, the inverse localization length is given by the calculation of the main text up to and including order $q$.

We conclude this appendix by providing some further detail on the calculation of the first few Fourier coefficients $p_{\infty,\ell}^{(q)}$ [resulting in Eqs.~\eqref{eq:alpha}--\eqref{eq:gamma33} and~\eqref{eq:p to 3rd order}].  Throughout, we use the general property $A_{n,\ell;-\ell'}= A_{n,-\ell;\ell'}^*$ (which in particular holds at any given order) and the entries in Table~\ref{tab:Table of Fourier coefficients}.
\begin{table}[htp]
    \caption{\label{tab:Table of Fourier coefficients}
    Table of Fourier coefficients $A_{n,\ell;\ell'}^{(j)}$
    }
    \begin{ruledtabular}
    \begin{tabular}{ l C C C C }
    \diagbox[dir=NW]{$j$}{$\ell'$} &
    0 & 1 & 2 & 3\\
    \colrule
    0 & 1 & 0 &0 &0\\
    1 & 0 & -\ell r_n &0 &0\\
    2 & -\ell^2 R_n &0 &\frac{1}{2}\ell(\ell-1)r_n^2 &0\\
    3 & 0 & \frac{1}{2}\ell^2(\ell-1)R_n r_n &0 &-\frac{1}{6}\ell(\ell-1)(\ell-2)r_n^3\\
    \end{tabular}
    \end{ruledtabular}
\end{table}

Setting $q=1$ in Eq.~\eqref{eq:p recursion} yields $p_{\infty,-1}^{(1)}= -\alpha_1 \langle v_n p_{\infty,0}^{(0)}A_{n,-1;-1}^{(1)}\rangle_n = \gamma^{(1)}/(2\pi)$.  Then, setting $q=2$ yields $p_{\infty,\ell}^{(2)}= (-1)^\ell \alpha_\ell \langle v_n^{-\ell} ( p_{\infty,0}^{(0)}A_{n,\ell;\ell}^{(2)}+ p_{\infty,-1}^{(1)}A_{n,\ell;\ell+1}^{(1)})\rangle_n$, hence $p_{\infty,-1}^{(2)}=0$ and $p_{\infty,-2}^{(2)}=\gamma^{(2)}/(2\pi)$.  Finally, setting $q=3$ yields $p_{\infty,\ell}^{(3)}= (-1)^\ell \alpha_\ell \langle v_n^{-\ell}( p_{\infty,0}^{(0)}A_{n,\ell;\ell}^{(3)} +p_{\infty,-1}^{(1)}A_{n,\ell;\ell+1}^{(2)} + p_{\infty,-2}^{(2)}A_{n,\ell;\ell+2}^{(1)}) \rangle_n$, hence $p_{\infty,-1}^{(3)}=\gamma_1^{(3)}/(2\pi)$, $p_{\infty,-2}^{(3)}=0$, and $p_{\infty,-3}^{(3)} = \gamma_3^{(3)}/(2\pi)$.

\section{Further details for PARS}\label{sec:Further details for PARS}

\subsection{Relation of PARS reflection amplitudes to single-site reflection amplitudes}\label{sec:Relation of PARS reflection amplitudes to single-site reflection amplitudes}

Here we demonstrate that the scattering transfer matrix for the PARS potential~\eqref{eq:H PARS} can be written as $\mathcal{T}_{1\dots N}=\mathcal{T}_N\dots\mathcal{T}_1$, where $\mathcal{T}_n$ is simply related to the scattering transfer matrix for a single potential at the origin [with the reflection amplitudes in particular satisfying Eqs.~\eqref{eq:rn in terms of hatrn} and~\eqref{eq:rnp in terms of hatrnp}].  

We start by considering the scattering problem of a single potential at the origin.  A scattering eigenstate wavefunction $\hat{\Psi}(x)$ for the Hamiltonian $\hat{H}= \frac{P^2}{2m} + V_{\hat{\mathbf{D}}}(x)$ may be written as
\beq
    \hat{\Psi}(x) =
    \begin{cases}
        \hat{\Psi}_L^+ e^{i k x} + \hat{\Psi}_L^- e^{-i k x} & x < -x_\text{max},\\
        \hat{\Psi}_R^+ e^{i k x} + \hat{\Psi}_R^- e^{-i k x} & x > x_\text{max},
    \end{cases}\label{eq:single-site scattering wavefn general}
\eeq
where the energy is $E=k^2/(2m)$ and where the four scattering amplitudes are related by
\beq
    \bpmat
        \hat{\Psi}_R^+\\
        \hat{\Psi}_R^-
    \epmat
    = \hat{\mathcal{T}}_{\hat{\mathbf{D}}}
    \bpmat
        \hat{\Psi}_L^+\\
        \hat{\Psi}_L^-
    \epmat,\label{eq:single-site scattering transfer matrix general}
\eeq
with some scattering transfer matrix $\hat{\mathcal{T}}_{\hat{\mathbf{D}}}$ that is assumed to be known.

Consider next the scattering problem for the chain defined by Eq.~\eqref{eq:H PARS}.  In regions where the potential vanishes, the wavefunction is a linear combination of $e^{\pm i kx}$.  We choose the following phase convention for these regions: for $n=1,\dots,N$, we write
\beq
\Psi(x) =\begin{cases}
    \Psi_{L,n}^+ e^{i k(x-x_n)}+\Psi_{L,n}^- e^{-ik(x-x_n)} & x_{n-1}+x_\text{max} < x < x_n - x_\text{max},\\
    \Psi_{R,n}^+ e^{i k(x-x_n)}+\Psi_{R,n}^- e^{-ik(x-x_n)} & x_n+x_\text{max} < x < x_{n+1} - x_\text{max},
\end{cases}\label{eq:scattering wavefn inside sample PARS}
\eeq
where $x_0 = -\infty$ and $x_{N+1}=\infty$.  By shifting $x-x_n \to x$, we can then refer to the single-site problem to obtain
\beq
    \bpmat
        \Psi_{R,n}^+\\
        \Psi_{R,n}^-
    \epmat
    =
    \hat{\mathcal{T}}_{\hat{\mathbf{D}}_n}
    \bpmat
        \Psi_{L,n}^+\\
        \Psi_{L,n}^-
    \epmat.
\eeq
Equation~\eqref{eq:scattering wavefn inside sample PARS} also implies, for $n=1,\dots,N-1$,
\bseq
\begin{align}
    \bpmat
        \Psi_{L,n+1}^+\\
        \Psi_{L,n+1}^-
    \epmat
    &=
    \bpmat
        e^{ik(x_{n+1}-x_n)} & 0\\
        0 & e^{-ik(x_{n+1}-x_n)}
    \epmat
    \bpmat
        \Psi_{R,n}^+\\
        \Psi_{R,n}^-
    \epmat\\
    &=\bpmat
        e^{ik(a_{n+1}+\frac{1}{2}b_{n+1}+\frac{1}{2}b_n)} & 0\\
        0 & e^{-ik(a_{n+1}+\frac{1}{2}b_{n+1}+\frac{1}{2}b_n)}
    \epmat
    \bpmat
        \Psi_{R,n}^+\\
        \Psi_{R,n}^-
    \epmat.
\end{align}
\eseq
Comparing Eq.~\eqref{eq:scattering wavefn inside sample PARS} for $n=1$ and $N$ to Eq.~\eqref{eq:scattering wavefn PARS}, we also obtain
\beq
    \bpmat
        \Psi_{L,1}^+\\
        \Psi_{L,1}^-
    \epmat
    =
    \bpmat
        e^{i k x_1} & 0\\
        0 & e^{-i k x_1}
    \epmat
    \bpmat
        \Psi_L^+\\
        \Psi_L^-
    \epmat
    = \bpmat
        e^{i k (a_1+\frac{1}{2}b_1)} & 0\\
        0 & e^{-i k (a_1+\frac{1}{2}b_1)}
    \epmat
    \bpmat
        \Psi_L^+\\
        \Psi_L^-
    \epmat
\eeq
and
\beq
    \bpmat
        \Psi_R^+\\
        \Psi_R^-
    \epmat
    =
    \bpmat
        e^{\frac{1}{2}ik b_N} & 0\\
        0 & e^{-\frac{1}{2}ikb_N}
    \epmat
    \bpmat
        \Psi_{R,N}^+\\
        \Psi_{R,N}^-
    \epmat.
\eeq
We define
\beq
    \mathcal{T}_n =
    \bpmat
        e^{\frac{1}{2}ikb_n} & 0\\
        0 & e^{-\frac{1}{2}ikb_n}
    \epmat
    \hat{\mathcal{T}}_{\hat{\mathbf{D}}_n}
    \bpmat
        e^{ik(a_n+\frac{1}{2}b_n)} & 0\\
        0 & e^{-ik(a_n+\frac{1}{2}b_n)}
    \epmat.\label{eq:matTn in terms of hatmatTn}
\eeq
Then, using the last few equations, we obtain
\bseq
\begin{align}
    \bpmat
        \Psi_R^+\\
        \Psi_R^-
    \epmat
    = \mathcal{T}_N
    \bpmat
        e^{\frac{1}{2}i k b_{N-1} }& 0\\
        0 & e^{-\frac{1}{2}ikb_{N-1}}
    \epmat
    \bpmat
        \Psi_{R,N-1}^+\\
        \Psi_{R,N-1}^-
    \epmat
    &= \mathcal{T}_N\dots \mathcal{T}_2 \bpmat
        e^{\frac{1}{2}i k b_1} & 0\\
        0 & e^{-\frac{1}{2}i kb_1}
    \epmat
    \bpmat
        \Psi_{R,1}^+\\
        \Psi_{R,1}^-
    \epmat\\
    &= \mathcal{T}_N \dots \mathcal{T}_1 
    \bpmat
        \Psi_L^+\\
        \Psi_L^-
    \epmat,
\end{align}
\eseq
which confirms that the scattering transfer matrix for the sample is $\mathcal{T}_{1\dots N}= \mathcal{T}_N\dots\mathcal{T}_1$.

The scattering transfer matrix for the single-site problem may be parametrized as
\beq
    \hat{\mathcal{T}}_{\hat{\mathbf{D}}_n} =
    \bpmat
        1/\hat{t}_{\hat{\mathbf{D}}_n}^* & \hat{r}_{\hat{\mathbf{D}}_n}'/\hat{t}_{\hat{\mathbf{D}}_n}' \\
        -\hat{r}_{\hat{\mathbf{D}}_n} / \hat{t}_{\hat{\mathbf{D}}_n}' & 1/\hat{t}_{\hat{\mathbf{D}}_n}'
    \epmat,\label{eq:mathcalTHat parametrization}
\eeq
and $\mathcal{T}_n$ may be parametrized similarly.  (In fact, $\hat{t}_{\hat{\mathbf{D}}_n}'=\hat{t}_{\hat{\mathbf{D}}_n}$ by time-reversal symmetry.  However, $\hat{r}_{\hat{\mathbf{D}}_n}'$ and $\hat{r}_{\hat{\mathbf{D}}_n}$ are generally unequal unless the potential is parity symmetric.)  From Eq.~\eqref{eq:matTn in terms of hatmatTn} we then read off $t_n = e^{ik(a_n+b_n)}\hat{t}_{\hat{\mathbf{D}}_n}$, $t_n'=e^{ik(a_n+b_n)}\hat{t}_{\hat{\mathbf{D}}_n}'$, and Eqs.~\eqref{eq:rn in terms of hatrn} and~\eqref{eq:rnp in terms of hatrnp} from the main text.

\subsection{Comparison with Lambert and Thorpe}\label{sec:Comparison with Lambert and Thorpe}
In Ref.~\cite{LambertRandom1983}, Lambert and Thorpe derive a formula for the inverse localization length to the first two non-vanishing orders in the reflection coefficient $R$ in the case of a random chain which satisfies the following properties: each scatterer has the same reflection coefficient $R$ and the phase $\delta_n\equiv -(\phi_{r_n}-\phi_{r_{n-1}}) -2 \phi_{t_{n-1}}$ is i.i.d.  They present the case of randomly-spaced delta function potentials [in our notation, all $V_n(x)= c\delta(x)$ and all $b_n=0$] as an example which has these properties.  However, as they comment and as we now show explicitly, their result applies more generally.  Here, we show that their setup applies whenever all sites have the same potential $V_n(x)\equiv V(x)$ with strictly finite range $x_{\text{max}}$ and with all $b_n=0$ and all $a_n < 2 x_\text{max}$.  We verify that Eq.~\eqref{eq:lloc PARS constant potential} from the main text agrees with their formula.

To use their result, we set up the scattering problem with different phase conventions than we used in the main text.  We write the wavefunction outside the sample as
\beq
    \Psi(x) =
    \begin{cases}
        \Psi_L^+ e^{i kx} + \Psi_L^- e^{-i k x} & x < x_1 - x_\text{max},\\
        \Psi_R^+ e^{i k x } + \Psi_R^- e^{-i k x } & x_N + x_\text{max} < x, \label{eq:scattering wavefn PARS LT}
    \end{cases}
\eeq
while for $n=1,\dots,N$, we write
\beq
\Psi(x) =\begin{cases}
    \Psi_{L,n}^+ e^{i k x}+\Psi_{L,n}^- e^{-ik x} & x_{n-1}+x_\text{max} < x < x_n - x_\text{max},\\
    \Psi_{R,n}^+ e^{i kx}+\Psi_{R,n}^- e^{-ikx} & x_n+x_\text{max} < x < x_{n+1} - x_\text{max},
\end{cases}\label{eq:scattering wavefn inside sample PARS LT}
\eeq
where $x_0 = -\infty$ and $x_{N+1}=\infty$.  By shifting $x-x_n \to x$, we can then refer to the single-site problem [Eqs.~\eqref{eq:single-site scattering wavefn general} and~\eqref{eq:single-site scattering transfer matrix general}] to obtain
\beq
    \bpmat
        \Psi_{R,n}^+ e^{i k x_n}\\
        \Psi_{R,n}^- e^{-ikx_n}
    \epmat
    =
    \hat{\mathcal{T}}_{\hat{\mathbf{D}}_n}
    \bpmat
        \Psi_{L,n}^+e^{ikx_n}\\
        \Psi_{L,n}^-e^{-ikx_n}
    \epmat.
\eeq
Equation~\eqref{eq:scattering wavefn inside sample PARS LT} also implies, for $n=1,\dots,N-1$,
\beq
    \bpmat
        \Psi_{L,n+1}^+\\
        \Psi_{L,n+1}^-
    \epmat
    =
    \bpmat
        \Psi_{R,n}^+\\
        \Psi_{R,n}^-
    \epmat.
\eeq
Equation~\eqref{eq:scattering wavefn PARS LT} also implies
\beq
    \bpmat
        \Psi_{L,1}^+\\
        \Psi_{L,1}^-
    \epmat
    =
    \bpmat
        \Psi_L^+\\
        \Psi_L^-
    \epmat
\eeq
and
\beq
    \bpmat
        \Psi_R^+\\
        \Psi_R^-
    \epmat
    =
    \bpmat
        \Psi_{R,N}^+\\
        \Psi_{R,N}^-
    \epmat.
\eeq
Collecting the past several equations, we obtain the scattering transfer matrix $\mathcal{T}_{1\dots N}$ of the sample as a product of local scattering transfer matrices:
\beq
    \bpmat
        \Psi_R^+\\
        \Psi_R^-
    \epmat
    = \mathcal{T}_{1\dots N}
    \bpmat
        \Psi_L^+\\
        \Psi_L^-
    \epmat,
\eeq
where $\mathcal{T}_{1\dots N} = \mathcal{T}_N\dots \mathcal{T}_1$ and
\beq
    \mathcal{T}_n = 
    \bpmat
        e^{-i k x_n} & 0\\
        0 & e^{ikx_n}
    \epmat
    \hat{\mathcal{T}}_{\hat{\mathbf{D}}_n}
    \bpmat
        e^{i k x_n} & 0\\
        0 & e^{-ikx_n}
    \epmat.
\eeq
From Eq.~\eqref{eq:mathcalTHat parametrization} we then read off $t_n = \hat{t}_{\hat{\mathbf{D}}_n}$, $t_n' = \hat{t}_{\hat{\mathbf{D}}_n}'$, $r_n= e^{2i k x_n}\hat{r}_{\hat{\mathbf{D}}_n}$, and $r_n'=e^{-2i k x_n}\hat{r}_{\hat{\mathbf{D}}_n}'$.

The phase convention used above is not suitable for the setup we use in the main text because the disorder does not separate; the parameter $x_n$ that $r_n$ and $r_n'$ depend on is not i.i.d.  However, the phase convention is suitable for the setup of Lambert and Thorpe, since if we set all $b_n=0$ and take the potentials to be constant [all $V_n(x)\equiv V(x)$, hence all $\hat{r}_{\hat{\mathbf{D}}_n} \equiv \sqrt{R}e^{i\hat{\phi}_{r}}$, all $\hat{r}_{\hat{\mathbf{D}}_n}' \equiv \sqrt{R}e^{i\hat{\phi}_{r}'}$, and all $\hat{t}_{\hat{\mathbf{D}}_n} = \hat{t}_{\hat{\mathbf{D}}_n}' \equiv \sqrt{T}e^{i\hat{\phi}_t}$], then $\delta_n = -2 k(x_n-x_{n-1}) -2 \hat{\phi}_t = -2 k a_n -  \hat{\phi}_r - \hat{\phi}_{r'}-\pi$ is indeed i.i.d. (here we have noted $2\hat{\phi}_t= \hat{\phi}_r+\hat{\phi}_{r'} + \pi$ by unitarity).  Eqs. (14)--(16), (46), and (49)--(51) of Ref.~\cite{LambertRandom1983} then yield Eq.~\eqref{eq:lloc PARS constant potential} from the main text \footnote{In their Eq. (51), $\alpha_2$ should be $a_2$.}; note that our $\tilde{a}_j$ is related to their $a_j$ by $\tilde{a}_j = (-1)^j a_j^*$ [and see below Eq.~\eqref{eq:reflection phase constant} for further details of the correspondence between our notation and theirs]. 

\subsection{Further details for transparent mirror effect}\label{sec:Further details for transparent mirror effect}
\subsubsection{Symmetry property}\label{sec:Symmetry property}
The scattering transfer matrix $\mathcal{T}_n$ that we have used describes the scattering of a wave with momentum $k$ though a single barrier ($b_n$ region).  This scattering event can be decomposed into two events in sequence (as done in, e.g., Ref.~\cite{BerryTransparent1997}): first the wave passes from the $a_n$ region to the $b_n$ region with a scattering transfer matrix $\mathcal{T}_n^{(A)}$, then from the $b_n$ region to the $a_{n+1}$ region with scattering transfer matrix $\mathcal{T}_n^{(B)}$.  Indeed, one obtains
\beq
    \mathcal{T}_n= \mathcal{T}_n^{(B)}\mathcal{T}_n^{(A)},\label{eq:T equals TBTA}
\eeq
where
\bseq
\begin{align}
   \mathcal{T}_n^{(A)} &= 
    \frac{1}{2\sqrt{k k'}}
   \bpmat
        (k+k')e^{ik a_n} & (k'-k)e^{-i k a_n} \\
        (k'- k )e^{i k a_n} & (k+k')e^{-ik a_n}
   \epmat,\label{eq:TA}\\
   \mathcal{T}_n^{(B)} &= 
    \frac{1}{2\sqrt{k k'}}
   \bpmat
        (k+k')e^{ik' b_n} & (k-k')e^{-i k' b_n} \\
        (k- k' )e^{i k'b_n} & (k+k')e^{-ik' b_n}
   \epmat.\label{eq:TB}
\end{align}
\eseq
For the purpose of calculating the localization length, we can just as well treat the $b$ regions at the ends ($b_1$ and $b_N$) as the leads (with incoming and outgoing momentum $k'$) and the $a_2,\dots, a_N$ regions as scatterers.  Indeed, the scattering transfer matrix for this alternate problem may be written as
\beq
    \widetilde{\mathcal{T}}_{1\dots N-1} = \widetilde{\mathcal{T}}_{N-1}\dots\widetilde{\mathcal{T}}_1,
\eeq
where
\beq
    \widetilde{\mathcal{T}}_n = \mathcal{T}_{n+1}^{(A)} \mathcal{T}_n^{(B)}.\label{eq:tildeT equals TATB}
\eeq
We assume that the disorder distributions of $a_n$ and $b_n$ are uncorrelated.  Then, this alternate problem also fits into the framework of the scattering expansion: we can relabel $a_{n+1}\to a_n$ (for $n=1,\dots, N-1$) to show explicitly that $\widetilde{\mathcal{T}}_n$ depends only on the disorder variables of site $n$.  Comparing the last several equations, we see that the scattering transfer matrix $\widetilde{\mathcal{T}}_n$ for the alternate scattering problem may be obtained by exchanging $k\leftrightarrow k'$ and $a_n\leftrightarrow b_n$ in $\mathcal{T}_n$.

Let us now show that the alternate problem has the same localization length as the original problem.  The key point is that the scattering transfer matrices for the two samples differ only by boundary terms that do not matter for large $N$.  In particular, we have
\beq
    \widetilde{\mathcal{T}}_{1\dots N-1} = \mathcal{T}_N^{(A)} \mathcal{T}_{2\dots N-1}\mathcal{T}_1^{(B)}.
\eeq
Note from Eq.~\eqref{eq:T equals TBTA} that $\mathcal{T}_{2\dots N-1}$ does not depend on the disorder variables that appear in the boundary terms (i.e., $a_N$ and $b_1$).  Thus, by a calculation similar to that in Appendix~\ref{sec:Tarasinski leads and their equivalence to other leads} below, we may take $N\to\infty$ to show that the sample defined by a product of many $\widetilde{\mathcal{T}}_n$ has the same localization length as the sample defined by a product of many $\mathcal{T}_n$. 

Finally, let us see what consequences this symmetry has on the explicit formula for $2/\lloc$ as an expansion in $\delta$.  From the above considerations, we must get the same expansion for $2/\lloc$ if we exchange $k\leftrightarrow k'$ and $a_n\leftrightarrow b_n$ (note that this includes exchanging the disorder distributions of the $a$ and $b$ variables) in Eqs.~\eqref{eq:rHat rectangular well} and~\eqref{eq:RHat rectangular well}.  As discussed in the main text, we leave momenta $k$ and $k'$ as they are [i.e., not expanding in $\delta$ using Eq.~\eqref{eq:kprime}] when they appear in trigonometric functions.  This leaves the terms $\frac{k^2-k'^2}{2 k k'}$ and $\frac{k^2+k'^2}{2 k k'}$ as the only ones that will be expanded in $\delta$.  The latter of these is invariant under $k\leftrightarrow k'$.  The former is odd, but this change of sign always cancels because $r_n$ and $r_n'$ must appear in the same number of powers in each term of the expansion of $2/\lloc$ (as we recall from the symmetry argument in Sec.~\ref{sec:Overview}).  Thus, the expansion of $2/\lloc$ in powers of $\delta$ must be symmetric under the exchange of $k\leftrightarrow k'$ and $a_n\leftrightarrow b_n$ at each order (with the understanding that $\delta$ is unaffected by the exchange).

\subsubsection{Comparison with the literature}\label{sec:Comparison with the literature}

We compare our results with Refs.~\cite{BerryTransparent1997,Luna-AcostaOne2009,LifshitsIntroduction1988}, starting with Ref.~\cite{BerryTransparent1997}.  Although this reference focuses on the case that disorder is strong in both $a_n$ and $b_n$, an intermediate result is presented there that is valid for strong disorder in $a_n$ and arbitrary disorder in $b_n$.  In our notation, the unnumbered equation above their Eq. (32) is equivalent to
\bseq
\begin{align}
    \frac{2}{\lloc}&= \langle- \ln \frac{16 \tilde{n}^2}{|(\tilde{n}+1)^2 - (\tilde{n}-1)^2e^{2i\tilde{n} k b_n} |^2}\rangle_n\\
    &= \langle -\ln\left( \frac{1}{1+\frac{\delta^2}{1-2\delta}\sin^2(k' b_n)}\right) \rangle_n,
\end{align}
\eseq
where $\tilde{n}$ is the index of refraction and where we have used Eqs.~\eqref{eq:delta} and~\eqref{eq:kprime}.  From~\eqref{eq:RHat rectangular well}, we see that the above equation is another way of writing the uniform phase result, i.e., $2/\lloc = \langle - \ln (1-\hat{R}_{\hat{\mathbf{D}}_n})\rangle_n$.

Next, we verify that Eq.~\eqref{eq:lloc transparent mirror} from the main text agrees with Eq. (5.14) from Ref.~\cite{Luna-AcostaOne2009} in the regime in which the two calculations overlap.  Ref.~\cite{Luna-AcostaOne2009} considers the case of weak disorder in $a_n$ and no disorder in $b_n$, with $\delta$ treated without approximation.  To convert their notation to ours and expand their result in $\delta$, we set $k_a=k,k_b=k',d=\langle a_n\rangle_n + b, \alpha_+ = 1 + O(\delta^2), \alpha_- = \delta+\delta^2 +O(\delta^3)$, and $\sin^2(\kappa d)= \sin^2(k\langle a\rangle_n + k' b) + O(\delta^2)$, where their quantities appear on the left and ours on the right.  Thus, their Eq. (5.14) has the following expansion in $\delta$:
\beq
    \frac{2}{\lloc}= (k\sigma)^2 \frac{\sin^2(k'b)}{(\langle a_n\rangle_n +b)\sin^2(k\langle a_n\rangle_n + k'b)} \left(\delta^2 + 2\delta^3\right) + O(\delta^4),
\eeq
where $\sigma^2$ is the variance of $a_n$ about its average.  We indeed recover this answer from Eq.~\eqref{eq:lloc transparent mirror} by expanding in weak disorder in $a_n$ [c.f. also the comment below Eq.~\eqref{eq:lloc PARS} about restoring dimensions].

Finally, we compare Eq.~\eqref{eq:lloc transparent mirror} to a result from~\cite{LifshitsIntroduction1988}, apparently finding some differences in numerical factors.  This reference considers $a_n$ and $b_n$ to follow exponential distributions:
\beq
    P(a_n) = \frac{1}{A}e^{-a_n/A},\ P(b_n) =\frac{1}{B}e^{-b_n/B}, 
\eeq
where $A$ and $B$ here are site-independent constants.  With these disorder distributions, Eq.~\eqref{eq:lloc transparent mirror} yields
\beq
    \frac{2}{\lloc} = \frac{2 A^2 B^2 k^2}{(A+B)[(A+B)^2 + 4 A^2B^2 k^2] }\delta^2 + O(\delta^3),\label{eq:lloc exponential dists}
\eeq
in which we have expanded $k'$ in $\delta$ and restored physical dimensions.  To compare with the last unnumbered equation on page 400 of Ref.~\cite{LifshitsIntroduction1988}, we use the following conversion from their notation to ours.  They present the logarithm of the average of the transmission, rather than the average of the logarithm; thus, their result is $\gamma_T$ [defined in their Eq. (29.19)].  By their Eqs. (30.34)--(30.36), our $2/\lloc$ is the same quantity as their $2\gamma=\overline{\gamma}=4\gamma_T$.  We have $a_0=A,a_1=B,\kappa^2 =k^2/(2m)$, and  $U_0 = k^2 \delta/m$ [c.f. their page 76 and Eqs. (29.1) and (29.21); note in particular that they set the mass $m=1/2$], where their quantities appear on the left and ours on the right.  All together, their result seems to yield Eq.~\eqref{eq:lloc exponential dists} with the factors of $2$ and $4$ on the right-hand side both absent.

We note that Eq.~\eqref{eq:lloc exponential dists} (regardless of whether or not the $2$ and $4$ are included) is monotonic in disorder strength (i.e., either $A$ or $B$ is varied) if the localization length is expressed in units of the average lattice spacing, that is, if the factor $1/(A+B)$ is removed.  If instead this factor is included, then non-monotonicity is obtained.

\section{Tarasinski leads and their equivalence to other leads}\label{sec:Tarasinski leads and their equivalence to other leads}

Let us first fill in the details for the calculation of the transfer matrix~\cite{VakulchykAnderson2017}.  Writing a general state as $\ket{\Psi}=\sum_n (\Psi_{\uparrow}(n)\ket{n,\uparrow} + \Psi_{\downarrow}(n)\ket{n,\downarrow})$ one finds that the stationary state equation $\hat{U}\ket{\Psi}= e^{-i\omega}\ket{\Psi}$ is equivalent to the following pair of equations:
\bseq
\begin{align}
    e^{i\varphi_n}\left[ e^{i\varphi_{1,n}} \cos\theta_n\Psi_{\uparrow}(n) + e^{i \varphi_{2,n}}\sin\theta_n \Psi_{\downarrow}(n)\right]=e^{-i\omega}\Psi_{\uparrow}(n+1),\label{eq:Schrod1 DTQW}\\
    e^{i\varphi_n}\left[-e^{-i\varphi_{2,n}}\sin\theta_n \Psi_{\uparrow}(n) +e^{-i\varphi_{1,n}}\cos\theta_n \Psi_{\downarrow}(n) \right]=e^{-i\omega}\Psi_{\downarrow}(n-1),\label{eq:Schrod2 DTQW}
\end{align}
\eseq
which are Eqs. (11) and (12) of Ref.~\cite{VakulchykAnderson2017} with $n\to n\pm 1$ \footnote{We correct sign errors in Eqs. (11) and (12) of Ref.~\cite{ VakulchykAnderson2017} (the sign of the second term on the right-hand side of each equation)}.  Rearranging these yields the transfer matrix [Eq.~\eqref{eq:transfer matrix DTQW} from the main text].

We proceed to set up the scattering problem with Tarasinski leads, which we recall are defined by setting $\varphi_n=\varphi_{1,n}=\varphi_{2,n}=\theta_n= 0$ for $n\le 0 $ and for $n\ge N+1$.  Note that for these $n$, the transfer matrix simplifies to
\beq
    \mathcal{M}_n = 
    \bpmat
        e^{i\omega} & 0\\
        0 & e^{-i\omega}
    \epmat.
\eeq
The solutions in the leads are therefore linear combinations of $(e^{i k n}, 0)$ and $(0, e^{-i k n})$, with quasienergy $\omega=k$.  A scattering solution may then be parametrized outside the sample as
\beq
    \Psi(n) = 
    \begin{cases}
        \Psi_L^+e^{ik(n-1)} \bpmat 1 \\ 0 \epmat  +\Psi_L^-e^{-ik(n-1)} \bpmat 0 \\ 1 \epmat  & n \le 1,\\
        \\
        \Psi_R^+e^{ik(n-N-1)} \bpmat 1 \\ 0 \epmat  +\Psi_R^-e^{-ik(n-N-1)} \bpmat 0 \\ 1 \epmat  & n \ge N+1,
    \end{cases}\label{eq:scattering wavefn DTQW Tarasinski leads expanded form}
\eeq
which is Eq.~\eqref{eq:scattering wavefn DTQW Tarasinski leads} from the main text (written in a form that we generalize below).

To set up the scattering problem with other choices of leads, we next recall the solution in the clean case, i.e., $\varphi_n\equiv\varphi$, $\varphi_{1,n}\equiv \varphi_1$, $\varphi_{2,n}\equiv \varphi_2$, and $\theta_n\equiv \theta$.  Here we follow Ref.~\cite{ VakulchykAnderson2017}, correcting some sign errors there.  Writing a Bloch wave solution as $\Psi(n) = e^{i k n}u_k$ for some two-component Bloch ``spinor'' $u_k \equiv (u_{k,\uparrow}, u_{k,\downarrow})$, we find that Eqs.~\eqref{eq:Schrod1 DTQW} and~\eqref{eq:Schrod2 DTQW} become
\beq
    \bpmat
        e^{i(\varphi_1 - k)}\cos\theta & e^{i\varphi_2}\sin\theta \\
        -e^{-i\varphi_2}\sin\theta & e^{-i(\varphi_1 - k)}\cos\theta 
    \epmat
    u_k = e^{-i(\omega+\varphi)}u_k.
\eeq
Thus, the clean quasienergy spectrum is determined by \footnote{This corrects a sign error in Eq. (6) of Ref.~\cite{ VakulchykAnderson2017} ($\varphi\to-\varphi$).}
\beq
    \cos(\omega +\varphi)=\cos\theta \cos(k-\varphi_1),\label{eq:spectrum DTQW}
\eeq
and the Bloch spinors are determined up to overall phase and normalization by \footnote{This corrects a sign error in Eq. (7) of Ref.~\cite{ VakulchykAnderson2017} ($\omega\to -\omega$).  Note that Eq. (7) there is derived using the convention $\Psi_n = (\Psi_{\uparrow}(n)\ \Psi_{\downarrow}(n))$, which accounts for the extra factor of $e^{ik}$ in our equation ($e^{ik}u_{k,\downarrow}\to u_{k,\downarrow}$).}
\beq
    u_{k,\uparrow} = e^{i(\varphi_2- \varphi_1+k)} \frac{\cos \theta - e^{i(\varphi_1 - \omega - k - \varphi)}}{\sin\theta}u_{k,\downarrow}.
\eeq
See Fig. 2 of Ref.~\cite{ VakulchykAnderson2017} for the quasienergy spectrum at various values of $\theta$.

We can now consider a scattering state with quasienergy $\omega$ in the problem with $(\varphi_n,\varphi_{1,n},\varphi_{2,n},\theta_n)= (\varphi_\text{leads},\varphi_{1,\text{leads}},\varphi_{2,\text{leads}},\theta_\text{leads})$ for $n\le 0$ and for $n\ge N+1$.  At any $\omega$ (aside from exceptional points) there are two corresponding momenta $k$ in the first Brillouin zone [c.f. Eq.~\eqref{eq:spectrum DTQW}].  One momentum is right moving and the other left moving; we label them as $k_+$ and $k_-$, respectively.  (Explicitly, the group velocity $dw/dk$ is positive at $k=k_+$ and negative at $k=k_-$.)  A scattering solution may be written outside the sample as
\beq
    \Psi(n) = 
    \begin{cases}
        \Psi_L^+  e^{ik_+ (n-1)}u_{k_+} +\Psi_L^-  e^{-ik_- (n-1)} u_{k_-} & n \le 1,\\
        \\
        \Psi_R^+ e^{ik_+(n-N-1)} u_{k_+} +\Psi_R^- e^{-ik_-(n-N-1)}u_{k_-} & n \ge N+1.
    \end{cases}
\eeq
Note that Eq.~\eqref{eq:scattering wavefn DTQW Tarasinski leads expanded form} is a special case with (assuming $k>0$) $k_\pm = \pm k$, $u_k=(1 , 0)$, and $u_{-k}= ( 0 , 1)$.

To bring this scattering problem to the form to which our setup applies, we define (as in Sec.~\ref{sec:Anderson model with diagonal disorder}) a matrix $\Lambda$ that converts position-space amplitudes to scattering amplitudes:
\beq
    \Lambda = 
    \bpmat
        u_{k_+,\uparrow} & u_{k_-,\uparrow}\\
        u_{k_+,\downarrow} & u_{k_-,\downarrow}
    \epmat.
\eeq
Then
\beq
    \Psi(1) = \Lambda
    \bpmat
        \Psi_L^+\\
        \Psi_L^-
    \epmat,\qquad
    \Psi_(N+1)=\Lambda
    \bpmat
        \Psi_R^+\\
        \Psi_R^-
    \epmat.
\eeq
Multiplying transfer matrices then yields
\beq
    \bpmat
        \Psi_R^+\\
        \Psi_R^-
    \epmat
    = \tilde{T}_{1\dots N}
    \bpmat
        \Psi_L^+\\
        \Psi_L^-
    \epmat,
\eeq
where
\beq
    \tilde{\mathcal{T}}_{1\dots N} = \Lambda^{-1} \mathcal{M}_N \dots \mathcal{M}_1 \Lambda
\eeq
is the scattering transfer matrix for the sample in this setup.  We then read off $\tilde{\mathcal{T}}_n = \Lambda^{-1}\mathcal{M}_n\Lambda$.

While we could now apply our results to the problem with these more general leads, it is more convenient to instead show that the probability distribution of the minus logarithm of the transmission coefficient is the same for all choices of leads.  To do this, let us write the sample transmission coefficient as $T_{1\dots N}$ for the problem with Tarasinski leads and as $\tilde{T}_{1\dots N}$ for the problem with more general leads.  Our goal is to show that for large samples, $-\ln T_{1\dots N}$ and $-\ln \tilde{T}_{1\dots N}$ have the same probability distribution.

The claim follows immediately from the relation
\beq
    \tilde{\mathcal{T}}_{1\dots N} = \Lambda^{-1}\mathcal{T}_{1 \dots N}\Lambda,\label{eq:relation between Tarasinski and general}
\eeq
in which the only difference between the two scattering transfer matrices is multiplication by boundary terms, which have negligible effect as $N\to\infty$.  To see this explicitly, we parametrize $\mathcal{T}_{1\dots N}$ in terms of the transmission coefficient and transmission and reflection phases, and then we expand in small $T_{1 \dots N}$:
\bseq
\begin{align}
    \mathcal{T}_{1\dots N} &=
    \frac{1}{\sqrt{T_{1\dots N}} }
    \bpmat
         e^{i \phi_{t_{1\dots N} }} & - \sqrt{1 - T_{1\dots N}} e^{i ( \phi_{t_{1\dots N}} - \phi_{r_{1\dots N}})} \\
         -\sqrt{1 - T_{1\dots N}} e^{i ( \phi_{r_{1\dots N}} - \phi_{t_{1\dots N}'})}& e^{-i \phi_{t_{1\dots N}' }}
    \epmat\\
    &= \frac{1}{\sqrt{T_{1\dots N}}}
    \bpmat
         e^{i \phi_{t_{1\dots N} }} & - e^{i ( \phi_{t_{1\dots N}} - \phi_{r_{1\dots N}})} \\
         - e^{i ( \phi_{r_{1\dots N}} - \phi_{t_{1\dots N}'})}& e^{-i \phi_{t_{1\dots N}' }}
    \epmat
    + O(T_{1\dots N}^{1/2}).\label{eq:scattering transfer matrix expansion}
\end{align}
\eseq
By the assumption that localization occurs, the error term is exponentially small for large $N$.  Eq.~\eqref{eq:relation between Tarasinski and general} [with $\tilde{\mathcal{T}}_{1\dots N}$ also expanded as in Eq.~\eqref{eq:scattering transfer matrix expansion}] then yields
\begin{multline}
    -\ln \tilde{T}_{1\dots N} = -\ln T_{1\dots N} - 2 \ln |u_{k_+,\uparrow}u_{k_-,\downarrow} -u_{k_-,\uparrow}u_{k_+,\downarrow}| + 2 \ln | u_{k_+,\uparrow}- e^{-i \phi_{r_{1\dots N}}} u_{k_+,\downarrow} | \\
    + 2 \ln | e^{i(\phi_{r_{1\dots N}} -\phi_{t_{1\dots N}'})}u_{k_-,\uparrow} + e^{i\phi_{t_{1\dots N}}} u_{k_-,\downarrow} |
    + \dots,
\end{multline}
in which the second, third, and fourth terms on the right-hand side are all $O(N^0)$ as $N\to\infty$ (all may be bounded by $N$-independent constants).  The omitted terms are exponentially small in $N$.  Thus, we confirm that $-\ln \tilde{T}_{1\dots N}$ and $-\ln T_{1\dots N}$ are asymptotically equal, and hence they have the same probability distribution.

\section{Toy recursion relation}\label{sec:Toy recursion relation}
To help illustrate our calculation of the joint probability distribution, we apply the same method to a toy example: a random walk with variable step lengths.  This example will also explain the comment below Eq.~\eqref{eq:phi integration identity} regarding the convenience of working with a variable that does not increase on average.

Consider a toy recursion relation given by
\beq
    s_{1\dots N+1} = s_{1\dots N} + a_{N+1},
\eeq
where the random variables $a_n$ are i.i.d. and where some initial value $s_1$ is given.  For large $N$ the probability distribution of $s$ [$P_{N}(s)\equiv \langle \delta(s_{1\dots N} - s)\rangle_{1\dots N}$] goes to a Gaussian with mean $\langle a_n\rangle_n N + O(N^0)$ and variance $N (\langle a_n^2\rangle_n - \langle a_n\rangle_n^2) + O(N^0)$.  We proceed next to recover this result by applying the approach we used in Sec.~\ref{sec:Joint probability distribution}.

We start by defining a shifted variable $\tilde{s}_{1\dots N} - \alpha$, where $\alpha$ is any constant.  The recursion relation for $\tilde{s}$ is readily found to be $\tilde{s}_{1\dots N+1} = \tilde{s}_{1\dots N} + a_{N+1} - \alpha$, from which we obtain the recursion relation for the probability distribution of $\tilde{s}$ [$\tilde{P}_N(\tilde{s})\equiv \langle \delta(\tilde{s}_{1\dots N} - \tilde{s})\rangle_{1\dots N}$]
\bseq
\begin{align}
    \tilde{P}_{N+1}(\tilde{s}) &= \int d\tilde{s}'\ \tilde{P}_N(\tilde{s}')\delta( \tilde{s}' + a_n - \alpha - \tilde{s} )\rangle_n\label{eq:prob dist recursion toy recursion}\\
    &\equiv \tilde{\mathcal{F}}[N,\tilde{s};\{\tilde{P}_N\}],
\end{align}
\eseq
where we note that $\tilde{\mathcal{F}}$ is a linear functional in its last argument.  To solve this equation, we will make an ansatz $\tilde{P}_N^{(\text{ansatz})}(\tilde{s})$ that is proportional to $1/\sqrt{N}$ and show 
\beq
    \tilde{\mathcal{F}}[\tilde{s};\{\tilde{P}_N^{(\text{ansatz})}\}] = \tilde{P}_{ N+1}^{(\text{ansatz})}(\tilde{s}) + O(1/N^2).\label{eq:goal toy recursion}
\eeq
Our ansatz is a Gaussian with mean and variance that both scale linearly with $N$, i.e., 
\beq
    \tilde{P}_N^{(\text{ansatz})}(\tilde{s}) = \frac{1}{\sqrt{4\pi c N}} e^{-\frac{1}{2}(\tilde{s}- b N)^2/(2 c N)}, 
\eeq
where $b$ and $c$ are constants to be determined by requiring Eq.~\eqref{eq:goal toy recursion} to hold.  Equation~\eqref{eq:goal toy recursion} will also hold if the mean and variance in the ansatz have arbitrary $N$-independent constants added; in the terminology of Sec.~\ref{sec:Calculation}, these constants parametrize a family of trajectories that satisfy the recursion relation up to small error.

Converting to Fourier space, we obtain
\bseq
\begin{align}
    \tilde{\mathcal{F}}[q;\{\tilde{P}_N\}] &= \langle e^{-i q(a_n-\alpha)}\rangle_n \tilde{P}_N(q)\\
    &=\left[ 1- i q(\langle a_n\rangle_n -\alpha) -\frac{1}{2}q^2 \langle(a_n-\alpha)^2\rangle_n + O(q^3)\right] \tilde{P}_N(q),\label{eq:mathcalF expansion toy recursion}
\end{align}
\eseq
and
\beq
    \tilde{P}_N^{(\text{ansatz})}(q) = e^{-i b N q}e^{-c N q^2}.
\eeq

The error estimate~\eqref{eq:FT error estimate} from the main text, with $\tilde{F}_N(q)$ replaced by  $\tilde{P}_N^{(\text{ansatz})}(q)$, still holds because the constant $b$ only appears in the overall phase.  Thus, we can neglect terms of the form $q^j\tilde{P}_N^{(\text{ansatz})}(q)$ (with $j\ge 3$) in $\tilde{\mathcal{F}}[q;\{\tilde{P}_N^{(\text{ansatz})}\}]$, since they are $O(1/N^2)$ in $\tilde{s}$ space.

We next do a Taylor expansion of $\tilde{P}_N^{(\text{ansatz})}(q)$ in $N$.  Unless the constant $b$ is zero (as in the main text), we must include the second derivative term in this expansion; the key relation is
\bseq
\begin{align}
    \tilde{P}_{N+1}^{(\text{ansatz})}(q) &= \tilde{P}_{N}^{(\text{ansatz})}(q) +  \frac{\pd}{\pd N} \tilde{P}_{N}^{(\text{ansatz})}(q) + \frac{1}{2}\frac{\pd^2}{\pd N^2} \tilde{P}_{N}^{(\text{ansatz})}(q) + \dots\label{eq:2nd order Taylor expansion general toy recursion}\\
    &=\left[ 1 - (i b q + c q^2) - \frac{1}{2}b^2 q^2 + O(q^3)\right] \tilde{P}_{N}^{(\text{ansatz})}(q),\label{eq:2nd order Taylor expansion explicit toy recursion}
\end{align}
\eseq
from which Eq.~\eqref{eq:mathcalF expansion toy recursion} yields
\beq
    \tilde{P}_{N+1}^{(\text{ansatz})}(q)-\tilde{\mathcal{F}}[q;\{\tilde{P}_{N}^{(\text{ansatz})}\}] =  \\
    i (-b +\langle a_n\rangle_n - \alpha)q\tilde{P}_N^{(\text{ansatz})}(q) - \left(c +\frac{1}{2}b^2 -\frac{1}{2} \langle(a_n-\alpha)^2\rangle_n\right)q^2 \tilde{P}_N^{(\text{ansatz})}(q) + \dots
\eeq
Thus, requiring Eq.~\eqref{eq:goal toy recursion} to hold for large $N$ uniquely determines the constants $b$ and $c$:
\bseq
\begin{align}
    b&= \langle a_n \rangle_n - \alpha,\\
    c&= \frac{1}{2}\left( \langle a_n^2\rangle_n - \langle a_n\rangle_n^2\right),
\end{align}
\eseq
which are the correct values.  Note that if we choose $\alpha=\langle a_n\rangle_n$, then we need not include the second derivative term in Eq.~\eqref{eq:2nd order Taylor expansion general toy recursion}, since $b=0$; this is essentially what is done in the main text. 

In contrast, we show next that a naive calculation based on finding a Fokker-Planck equation for $\tilde{P}_N(\tilde{s})$ yields the variance of $s$ depending on $\alpha$, which is incorrect unless a particular value of $\alpha$ is chosen.

Starting from the recursion relation~\eqref{eq:prob dist recursion toy recursion}, we consider $a_n-\alpha$ to be a small parameter [more precisely, we replace $a_n-\alpha\to \lambda(a_n -\alpha)$ and expand in $\lambda$].  Expanding the delta function on the right-hand side yields
\beq
    \tilde{P}_{N+1}(\tilde{s}) = \langle \tilde{P}_N(\tilde{s}) - (\langle a_n\rangle_n - \alpha)\frac{\pd}{\pd\tilde{s}} \tilde{P}_N(\tilde{s}) + \frac{1}{2} \langle(a_n-\alpha)^2 \frac{\pd^2}{\pd \tilde{s}^2}\tilde{P}_N(\tilde{s}) \rangle_n + O((a_n-\alpha)^3).
\eeq
Taking $N$ large we replace (although this step turns out not to be correct in general)
\beq
    \tilde{P}_{N+1}(\tilde{s}) - \tilde{P}_N(\tilde{s})\to\frac{\pd}{\pd N}\tilde{P}_N(\tilde{s}),\label{eq:1st order Taylor expansion toy recursion}
\eeq
and we thus obtain the Fokker-Planck equation:
\beq
    \frac{\pd}{\pd N} \tilde{P}_N(\tilde{s}) = -(\langle a_n\rangle_n - \alpha)\frac{\pd}{\pd\tilde{s}} \tilde{P}_N(\tilde{s}) + \frac{1}{2} \langle(a_n-\alpha)^2\rangle_n \frac{\pd^2}{\pd \tilde{s}^2}\tilde{P}_N(\tilde{s})+ O((a_n-\alpha)^3).\label{eq:incorrect Fokker-Planck eqn toy recursion}
\eeq
The initial condition $s_1$ does not affect the slope in $N$ of the mean or variance, but for definiteness let us take $s_1 = \alpha$ so that $\tilde{s}_1=0$ and hence $\tilde{P}_1(\tilde{s})=\delta(\tilde{s})$.  Then the solution of Eq.~\eqref{eq:incorrect Fokker-Planck eqn toy recursion} (dropping the third-order error term) is the Gaussian with mean $N(\langle a_n\rangle_n - \alpha)$ and variance $N \langle(a_n-\alpha)^2\rangle_n$.  Shifting back to the original variable $s$, we see that the mean is correct, but the variance is wrong unless we choose $\alpha=\langle a_n\rangle_n$.

The problem with the calculation just presented is that the expansion~\eqref{eq:1st order Taylor expansion toy recursion} needs to be done to one more order, as in Eqs.~\eqref{eq:2nd order Taylor expansion general toy recursion} and~\eqref{eq:2nd order Taylor expansion explicit toy recursion}. In other words, it is turns out that the solution Eq.~\eqref{eq:incorrect Fokker-Planck eqn toy recursion} is such that the replacement~\eqref{eq:1st order Taylor expansion toy recursion} is not consistent (in that the second derivative term is not in fact negligible).  If $\alpha=\langle a_n\rangle_n$ is chosen, then $b=0$, so the second derivative term in~\eqref{eq:2nd order Taylor expansion explicit toy recursion} vanishes and the replacement~\eqref{eq:1st order Taylor expansion toy recursion} is correct.

\end{widetext}

\bibliography{ms}

\end{document}